\pdfoutput=1
\documentclass[prd,showpacs,letterpaper,10pt,aps,notitlepage,nofootinbib,superscriptaddress,fleqn]{revtex4-2}

\usepackage{amsfonts,amsmath,graphicx}
\usepackage{hyperref}
\usepackage{placeins}
\usepackage{diagbox}
\usepackage[capitalize]{cleveref}
\usepackage{bm}
\usepackage{braket}
\usepackage{ulem}
\usepackage[pdftex,dvipsnames]{xcolor}
\usepackage{array}
\usepackage{tikz}
\usetikzlibrary{calc,positioning}

\usepackage{xargs}

\hypersetup{colorlinks, linkcolor = [rgb]{0,0.0,0.75}, citecolor=[rgb]{0,0.0,0.75}, urlcolor = [rgb]{0,0.0,0.75}}

\renewcommand{\arraystretch}{1.2} 

\newcommand{\ltapprox}{\raisebox{-0.5ex}{$\,\stackrel{<}{\scriptstyle\sim}\,$}}

\newcolumntype{C}[1]{>{\centering\arraybackslash}m{#1}}

\begin{document}

\title{Lattice gauge theory computation of \texorpdfstring{$1/m_Q$}{1/mQ} and \texorpdfstring{$1/m_Q^2$}{1/mQ2} corrections to the quarkonium potential}

\author{Michael Eichberg}
\affiliation{Goethe-Universit\"at Frankfurt am Main, Institut f\"ur Theoretische Physik, Max-von-Laue-Stra{\ss}e 1, D-60438 Frankfurt am Main, Germany}
\affiliation{Helmholtz Research Academy Hesse for FAIR, Campus Riedberg, Max-von-Laue-Stra{\ss}e 12, D-60438 Frankfurt am Main, Germany}

\author{Marc Wagner}
\affiliation{Goethe-Universit\"at Frankfurt am Main, Institut f\"ur Theoretische Physik, Max-von-Laue-Stra{\ss}e 1, D-60438 Frankfurt am Main, Germany}
\affiliation{Helmholtz Research Academy Hesse for FAIR, Campus Riedberg, Max-von-Laue-Stra{\ss}e 12, D-60438 Frankfurt am Main, Germany}

\date{\today}

\begin{abstract}
We compute the static quark-antiquark potential and its $1/m_Q$ and $1/m_Q^2$ corrections, i.e.\ the $1/m_Q$ potential and all $1/m_Q^2$ potentials, which can be expressed in terms of correlators of two field strength insertions, in SU(3) lattice gauge theory and provide parameterizations for quark-antiquark separations $0.2 \, \text{fm} \ltapprox r \ltapprox 0.7 \, \text{fm}$.
The field strength insertions in generalized Wilson loops are renormalized using gradient flow, which also leads to small statistical uncertainties.
All potentials are extrapolated to zero flow time and, in contrast to previous lattice studies, to the continuum, using three lattice spacings and four flow times.
The spin dependent potentials are matched to the $\overline{\text{MS}}$ scheme at NLO and provided at $\mu = 3 \, \text{GeV}$.
They can easily be converted to specific heavy quark masses, e.g.\ for Born-Oppenheimer effective field theory studies of quarkonium.
\end{abstract}

\maketitle


\section{Introduction}

Heavy quarkonium has been a prime testing ground for QCD since the discovery of the $J/\psi$, and the large number of conventional and exotic heavy quarkonium-like states observed in recent years has further increased the experimental and theoretical interest \cite{Brambilla:2010cs,Brambilla:2019esw}.
Theoretical predictions based on QCD are, therefore, essential.
In principle, lattice QCD allows to compute the spectrum of such states from first principles \cite{HadronSpectrum:2012gic}.
In practice, however, such computations are expensive, and states above strong decay thresholds have to be treated as resonances, which requires a rather involved scattering analysis.
For hybrid mesons, where the gluons contribute to the quantum numbers and which can decay via several channels into multi-particle states with two, three or more hadrons, a rigorous treatment within full lattice QCD is currently out of reach.

A promising alternative is to exploit the large mass of the heavy quarks.
Integrating out the hard and soft scales leads to the effective field theories NRQCD and potential NRQCD (pNRQCD) (see Ref.\ \cite{Brambilla:2004jw} for a review) and, more generally, to the Born-Oppenheimer effective field theory (BOEFT) \cite{Brambilla:2017uyf}, which describes conventional quarkonium as well as exotic states such as hybrid mesons within a common framework (see also Refs.\ \cite{Braaten:2014qka,Berwein:2015vca,Oncala:2017hop,Brambilla:2018pyn,Soto:2020xpm} as well as Ref.\ \cite{Berwein:2024ztx} for a recent comprehensive study).
Within these effective field theories, heavy quarkonium is described by a Schr\"odinger equation with potentials, which are systematically organized in powers of the inverse heavy quark mass $1/m_Q$ and which can be expressed in terms of Wilson loops with field strength insertions \cite{Brambilla:2000gk,Pineda:2000sz}.
Similar Born-Oppenheimer approaches based on lattice QCD potentials have also been used to study heavy quarkonium resonances \cite{Bicudo:2019ymo,Bicudo:2020qhp,Bicudo:2022ihz}.
The potentials are, however, nonperturbative input, which the effective field theories do not provide themselves.
At small quark-antiquark separations they can be computed in perturbation theory \cite{Peset:2015vvi,Peset:2017wef,Peset:2018jkf,Anzai:2018eua}, while at large separations effective string theory \cite{Luscher:2004ib} provides constraints \cite{Perez-Nadal:2008wtr,Brambilla:2014eaa}.
At intermediate separations, which are most relevant for the observed quarkonium spectrum, they have to be computed with lattice gauge theory.
This is the aim of the present work.

We compute the static potential $V^{(0)}$, the $1/m_Q$ potential $V^{(1)}$, the spin independent $1/m_Q^2$ potentials $V_{p^2}^{(2,0)}$, $V_{p^2}^{(1,1)}$, $V_{L^2}^{(2,0)}$ and $V_{L^2}^{(1,1)}$ as well as the spin dependent $1/m_Q^2$ potentials $V_{LS}^{(2,0)}$, $V_{L_2S_1}^{(1,1)}$, $V_{S_{12}}^{(1,1)}$ and $V_{S^2}^{(1,1)}$ with lattice gauge theory.
The remaining two $1/m_Q^2$ potentials, $V_r^{(2,0)}$ and $V_r^{(1,1)}$, require Wilson loops with three and four field strength insertions and are not considered in this work.
Lattice computations of the $1/m_Q$ and $1/m_Q^2$ potentials face two main technical difficulties.
First, correlators of field strength insertions in Wilson loops exhibit a poor signal-to-noise ratio.
Second, the lattice field strength operators require renormalization and, for the spin dependent potentials, matching to QCD.
The pioneering work of Ref.\ \cite{Bali:1997am} addressed the latter, but has comparatively large statistical uncertainties, while Refs.\ \cite{Koma:2006fw,Koma:2006si,Koma:2007jq,Koma:2009ws,Koma:2012bc} achieved very small statistical uncertainties using the multilevel algorithm, but did not carry out a matching.
In this work we address both difficulties using gradient flow \cite{Narayanan:2006rf,Luscher:2010iy,Luscher:2011bx}.
Gradient flow can be interpreted as a smearing technique, which suppresses ultraviolet fluctuations and, thus, leads to small statistical uncertainties.
Moreover, correlators of flowed field strength operators are ultraviolet finite, i.e.\ at fixed flow time they can be extrapolated to the continuum.
Combined with the matching coefficient from the gradient flow scheme to the $\overline{\text{MS}}$ scheme, which is known at NLO \cite{Brambilla:2023vwm} and has recently been computed at NNLO \cite{Wang:2026toy}, a subsequent extrapolation to zero flow time yields renormalized spin dependent potentials in the $\overline{\text{MS}}$ scheme, which we provide at $\mu = 3 \, \text{GeV}$ and which can easily be converted to specific heavy quark masses.
For these extrapolations we use three lattice spacings and four flow times. In particular, a continuum extrapolation has not been carried out in previous lattice studies.
Preliminary results have been presented at recent conferences \cite{Eichberg:2023trq,Eichberg:2024svw}.

The paper is structured as follows: in Sec.\ \ref{sec:potential_definitions} we summarize the definitions of the potentials in terms of field strength correlators, in Sec.\ \ref{SEC_lattice_setup} we describe the lattice setup, gradient flow and the extrapolation strategy, in Sec.\ \ref{SEC_results} we present and discuss our results, and in Sec.\ \ref{SEC_conclusions} we conclude.


\newpage

\section{\label{sec:potential_definitions}The quarkonium potential up to order \texorpdfstring{$1/m_Q^2$}{1/mQ2}}

The quarkonium potential $V(r)$ has been derived in pNRQCD up to order $1/m_Q^2$ \cite{Pineda:2000sz} (in the following
$m_Q = m_1$ and $m_Q = m_2$ denote the heavy quark masses; $r = |\bm{r}|$ and $\bm{r} = \bm{r}_1 - \bm{r}_2$, where
$\bm{r}_1$ and $\bm{r}_2$ are the positions of the heavy quark and antiquark). It depends on the momenta $\bm{p}_1$ and $\bm{p}_2$
of the heavy quark and antiquark
and their spins $\bm{S}_1 = \bm{\sigma}_1/2$ and $\bm{S}_2 = \bm{\sigma}_2/2$, where $\bm{\sigma} = (\sigma_1,
	\sigma_2, \sigma_3)$ are the Pauli matrices. In the center of mass frame, where $\bm{p}_1 = -\bm{p}_2 = \bm{p}$,
it is given by
\begin{eqnarray}
\nonumber & & \hspace{-0.7cm} V(r) = U^{(0)} + V^{(0)}(r) + \frac{m_1 + m_2}{m_1 m_2} \Big(U^{(1)} + V^{(1)}(r)\Big)
\\
\nonumber & & \hspace{0.7cm} + \frac{1}{m_1^2} V^{(2,0)}(r, m_1,\bm{p},\bm{L},\bm{S}_1) + \frac{1}{m_2^2} V^{(0,2)}(r,m_2,\bm{p},\bm{L},\bm{S}_2) + \frac{1}{m_1 m_2} V^{(1,1)}(r,m_1,m_2,\bm{p},\bm{L},\bm{S}_1,\bm{S_2})
\\
\label{eq:VpNRQCD} & & \hspace{0.7cm} + \mathcal{O} (1 / m_Q^3)
\end{eqnarray}
with
\begin{eqnarray}
\nonumber & & \hspace{-0.7cm} V^{(2, 0)} (r, m_j, \bm{p}, \bm{L}, \bm{S}) = V^{(0, 2)} (r, m_j, \bm{p}, \bm{L}, \bm{S}) = \\
\label{eq:V20} & & \hspace{0.7cm} = \frac{1}{2} \left\lbrace \bm{p}^2, U_{p^2}^{(2,0)} + V_{p^2}^{(2,0)}(r) \right\rbrace
+ \frac{\bm{L}^2}{r^2} V_{L^2}^{(2,0)}(r) + V_r^{(2,0)}(r, m_j)
+ \bm{L}\bm{S}_1 V_{LS}^{(2,0)}(r, m_j) \\
\nonumber & & \hspace{-0.7cm} V^{(1, 1)}(r, m_1, m_2, \bm{p}, \bm{L}, \bm{S}_1, \bm{S_2})  =
\frac{1}{2} \left\lbrace \bm{p}^2, V_{p^2}^{(1, 1)}(r) \right\rbrace
+ \frac{\bm{L}^2}{r^2} V_{L^2}^{(1, 1)}(r)
+ V_r^{(1, 1)}(r, m_1, m_2) + \bm{L}\bm{S}_1 V_{L_2S_1}^{(1, 1)}(r, m_1) \\
& & \hspace{0.7cm} + \bm{L}\bm{S}_2 V_{L_1S_2}^{(1, 1)}(r, m_2) + 4 \Big(3 (\bm{e}_r \bm{S}_1) (\bm{e}_r \bm{S}_2) - \bm{S}_1 \bm{S}_2\Big) V_{S_{12}}^{(1, 1)}(r, m_1, m_2) + \bm{S}_1 \bm{S}_2 V_{S^2}^{(1, 1)}(r, m_1, m_2)
\label{eq:V11}
\end{eqnarray}
($\bm{L} = \bm{r} \times \bm{p}$ is the quarkonium orbital angular momentum and $\{ \ldots \, , \, \ldots \}$ denotes the anticommutator). The quarkonium masses are given by $M = m_1 + m_2 + E$, where $E$ are the eigenvalues of the Hamiltonian $\bm{p}^2/(2 m_\text{red}) + V(r)$ with the reduced mass $m_\text{red} = m_1 m_2/(m_1 + m_2)$ \cite{Pineda:2000sz}. The static potential $V^{(0)}(r)$ as well as the $1/m_Q$ correction $V^{(1)}(r)$ neither depend on momentum nor spin. The quark mass dependence of the potentials $V_X(r,m_j)$ in Eqs.\ (\ref{eq:V20}) and (\ref{eq:V11}) enters exclusively via NRQCD matching coefficients (for the potentials computed in this work see Eqs.\ (\ref{eq:cf_V_LS20}) to (\ref{eq:cf_V_Ssq})), while the potentials defined in terms of Wilson loops with field strength insertions do not depend on the quark masses. 
$U^{(0)}$, $U^{(1)}$ and $U_{p^2}^{(2,0)}$ are finite constants, which do not depend on the quark masses, once a scheme for the definition of the heavy quark masses has been chosen (i.e.\ the same constants apply e.g.\ to charmonium and bottomonium). $U^{(0)}$ and $U^{(1)}$ are independent and shift a quarkonium spectrum as a whole, but do not affect mass differences between its states. $U_{p^2}^{(2,0)}$ is fixed by $U^{(0)}$ via the BBMP relation (\ref{eq:bbmp2}), $U_{p^2}^{(2,0)} = -U^{(0)}/4$. Its role, however, is different. It multiplies $\bm{p}^2/m_j^2$ in Eq.\ (\ref{eq:V20}), i.e.\ it is a correction to the kinetic energy, which also affects mass differences between quarkonium states. All three constants have to be distinguished from $r$ independent, regulator dependent contributions to the potentials, which arise in a cutoff regularization like lattice gauge theory and diverge when the regulator is removed, e.g.\ the self-energy contribution to the static potential. In our analysis these divergent contributions are absorbed by fit parameters depending on $a$ and $t_f$ (see Secs.~\ref{sec:result_static_potential} and \ref{SEC_spin_independent}), while $U^{(0)}$, $U^{(1)}$ and $U_{p^2}^{(2,0)}$ are finite. In pNRQCD \cite{Brambilla:2004jw} and in the Born-Oppenheimer effective field theory \cite{Berwein:2024ztx}, these constants and the heavy quark masses can be defined consistently in a single scheme, where the quark masses are taken as external input. Determining the constants in such a scheme from our lattice results would, however, require substantial additional effort and is beyond the scope of this work, i.e.\ we do not provide results for these constants. Alternatively, the two independent constants $U^{(0)}$ and $U^{(1)}$, which also fix $U_{p^2}^{(2,0)} = -U^{(0)}/4$, (and, if desired, also the heavy quark masses) can be determined by matching the spectrum resulting from a Born-Oppenheimer calculation to a selected set of experimentally known quarkonium masses, similar to the tuning of quark masses to experimental hadron masses in lattice QCD. The remaining quarkonium masses can then be predicted.


\subsection{\label{SEC_notation}Notation}

In this paper we carry out a lattice computation of the potentials $V^{(0)}(r)$, $V^{(1)}(r)$, $V_{p^2}^{(2, 0)}(r)$, $V_{p^2}^{(1, 1)}(r)$, $V_{L^2}^{(2, 0)}(r)$, $V_{L^2}^{(1, 1)}(r)$, $V_{LS}^{(2, 0)}(r, m_j)$, $V_{L_2S_1}^{(1, 1)}(r, m_j)$, $V_{S_{12}}^{(1, 1)}(r, m_1, m_2)$ and $V_{S^2}^{(1, 1)}(r, m_1, m_2)$. 
We use gradient flow and extrapolate to lattice spacing $a = 0$ as well as to flow time $t_f = 0$, where the spin dependent potentials $V_{LS}^{(2, 0)}$, $V_{L_2S_1}^{(1, 1)}$, $V_{S_{12}}^{(1, 1)}$ and $V_{S^2}^{(1, 1)}$ require renormalization and matching, either to physical heavy quark masses $m_Q$ or to another scale defined by the chosen renormalization scheme.
In this work we use the lattice results for spin dependent potentials to determine renormalized potentials in the $\overline{\text{MS}}$ scheme at scale $\mu$, which can be converted in a simple and straightforward way to the quark mass dependent potentials $V_{LS}^{(2, 0)}(r,m_j)$, $V_{L_2S_1}^{(1, 1)}(r,m_j)$, $V_{S_{12}}^{(1, 1)}(r,m_1,m_2)$ and $V_{S^2}^{(1, 1)}(r,m_1,m_2)$ appearing in Eqs.\ (\ref{eq:V20}) and (\ref{eq:V11}).

Throughout this paper we use the following notation, where quantities are not only described by their names, but also by their arguments:
\begin{itemize}
\item $V_X(r, t_f, a)$: lattice results for the potentials at finite $a$ and $t_f$ (here and in the following $X$ represents both the lower and upper labels of the potentials, e.g.\ $X \in \{ {}^{(0)} , {}^{(1)} , {}_{p^2}^{(2,0)} , \ldots \}$).

\item $V_X(r, t_f)$: potentials in the continuum, but at finite $t_f$, i.e.\ $V_X(r,t_f) = \lim_{a \rightarrow 0} V_X(r,t_f,a)$.

\item For spin independent potentials $V^{(0)}$, $V^{(1)}$, $V_{p^2}^{(2, 0)}$, $V_{p^2}^{(1, 1)}$, $V_{L^2}^{(2, 0)}$, $V_{L^2}^{(1, 1)}$:
\begin{itemize}
\item $V_X(r)$: potentials in the continuum and at $t_f = 0$, i.e.\ $V_X(r) = \lim_{t_f \rightarrow 0} V_X(r, t_f)$ (for $V^{(0)}$, $V^{(1)}$ and $V_{p^2}^{(2,0)}$ up to $r$ independent, regulator dependent contributions; see the discussion after Eq.\ (\ref{eq:V11}) and Secs.\ \ref{sec:result_static_potential} and \ref{SEC_spin_independent}). 
These potentials appear in Eqs.\ (\ref{eq:VpNRQCD}) to (\ref{eq:V11}) and can be used for Born-Oppenheimer studies of quarkonium.
\end{itemize}

\item For spin dependent potentials $V_{LS}^{(2, 0)}$, $V_{L_2S_1}^{(1, 1)}$, $V_{S_{12}}^{(1, 1)}$, $V_{S^2}^{(1, 1)}$:
\begin{itemize}
\item $V_X(r, \mu)$: potentials in the continuum and at $t_f = 0$ in the $\overline{\text{MS}}$ scheme at scale $\mu$ (how to convert $V_X(r,t_f)$ to $V_X(r,\mu)$ is discussed in Sec.\ \ref{sec:gradient_flow}).

\item $V_X(r, m_j)$ and $V_X(r, m_1, m_2)$: potentials in the continuum and at $t_f = 0$ for specific quark masses (how to convert $V_X(r, \mu)$ to $V_X(r, m_j)$ or $V_X(r, m_1, m_2)$, is discussed in Sec.\ \ref{sec:potential_defs}). 
These potentials appear in Eqs.\ (\ref{eq:V20}) to (\ref{eq:V11}) and can be used for Born-Oppenheimer studies of quarkonium.
\end{itemize}
\end{itemize}


\subsection{\label{sec:potential_defs}Potentials expressed in terms of field strength correlators and Wilson loops with field strength insertions}

The $1/m_Q^0$ potential $V^{(0)}(r)$ in Eq.\ (\ref{eq:VpNRQCD}) is the well-known static quark-antiquark potential.
The potentials appearing in order $1/m_Q$ and $1/m_Q^2$ in Eqs.\ (\ref{eq:VpNRQCD}) to (\ref{eq:V11}) have been derived using Wilson loops with chromoelectric and chromomagnetic field strength insertions in Refs.\ \cite{Brambilla:2000gk,Pineda:2000sz}. 
They can be expressed in terms of field strength correlators, the spin independent potentials as
\begin{eqnarray}
& & \hspace{-0.7cm} V^{(1)}(r) = -\frac{1}{2} \int_0^\infty \mathrm{d}t \, t \left( 2 \bra{0 , \Sigma_g^+} g E_x^0(t) g E_x^0(0) \ket{0 , \Sigma_g^+}(r) + \bra{0 , \Sigma_g^+} g E_z^0(t) g E_z^0(0) \ket{0 , \Sigma_g^+}_c(r) \right)
\label{eq:V_1}
\\
& & \hspace{-0.7cm} V_{p^2}^{(2,0)}(r) = \frac{1}{2} \int_0^\infty \mathrm{d}t \, t^2 \bra{0 , \Sigma_g^+} g E_z^0(t) g E_z^0(0) \ket{0 , \Sigma_g^+}_c(r)
\label{eq:V_psq20} 
\\
& & \hspace{-0.7cm} V_{p^2}^{(1,1)}(r) = \int_0^\infty \mathrm{d}t \, t^2 \bra{0 , \Sigma_g^+} g E_z^r(t) g E_z^0(0) \ket{0 , \Sigma_g^+}_c(r)
\label{eq:V_psq11} 
\\
& & \hspace{-0.7cm} V_{L^2}^{(2,0)}(r) = \frac{1}{2} \int_0^\infty \mathrm{d}t \, t^2 \left( \bra{0 , \Sigma_g^+} g E_x^0(t) g E_x^0(0) \ket{0 , \Sigma_g^+}(r) - \bra{0 , \Sigma_g^+} g E_z^0(t) g E_z^0(0) \ket{0 , \Sigma_g^+}_c(r) \right)
\label{eq:V_Lsq20} 
\\
& & \hspace{-0.7cm} V_{L^2}^{(1,1)}(r) = \int_0^\infty \mathrm{d}t \, t^2 \left( \bra{0 , \Sigma_g^+} g E_x^r(t) g E_x^0(0) \ket{0 , \Sigma_g^+}(r) - \bra{0 , \Sigma_g^+} g E_z^r(t) g E_z^0(0) \ket{0 , \Sigma_g^+}_c(r) \right)
\label{eq:V_Lsq11} 
\end{eqnarray}
and the spin dependent potentials as
\begin{eqnarray}
& & \hspace{-0.7cm} V_{LS}^{(2,0)}(r, m_j) = c_F^{(m_j)}(\mu) V_{LS}^{(2, 0)}(r,\mu) - \frac{1}{2r} V^{(0)\prime}(r)
\label{eq:cf_V_LS20} 
\\
& & \hspace{-0.7cm} V_{L_2S_1}^{(1,1)}(r, m_j) = c_F^{(m_j)}(\mu) V_{L_2S_1}^{(1, 1)}(r,\mu)
\label{eq:cf_V_LS11} 
\\
& & \hspace{-0.7cm} V_{S_{12}}^{(1,1)}(r, m_1, m_2) = c_F^{(m_1)}(\mu) c_F^{(m_2)}(\mu) V_{S_{12}}^{(1,1)}(r,\mu)
\label{eq:cf_V_ST} 
\\
& & \hspace{-0.7cm} V_{S^2}^{(1,1)}(r, m_1, m_2) = c_F^{(m_1)}(\mu) c_F^{(m_2)}(\mu) V_{S^2}^{(1,1)}(r,\mu) - 4(d_{sv} + d_{vv} C_F) \delta^{(3)}(\bm{r})
\label{eq:cf_V_Ssq}
\,,
\end{eqnarray}
where
\begin{eqnarray}
& & \hspace{-0.7cm} V_{LS}^{(2,0)}(r,\mu) = -\frac{2}{r} \int_0^\infty \mathrm{d}t \, t \bra{0 , \Sigma_g^+} g B_x^0(t) g E_y^0(0) \ket{0 , \Sigma_g^+}(r) + \frac{1}{r} V^{(0)\prime}(r)
\label{eq:V_LS20}
\\
& & \hspace{-0.7cm} V_{L_2S_1}^{(1,1)}(r,\mu) = -\frac{2}{r} \int_0^\infty \mathrm{d}t \, t \bra{0 , \Sigma_g^+} g B_x^r(t) g E_y^0(0) \ket{0 , \Sigma_g^+}(r)
\label{eq:V_LS11} 
\\
& & \hspace{-0.7cm} V_{S_{12}}^{(1,1)}(r,\mu) = \frac{1}{6} \int_0^\infty \mathrm{d}t \,\Big( \bra{0 , \Sigma_g^+} g B_z^r(t) g B_z^0(0) \ket{0 , \Sigma_g^+}(r) - \bra{0 , \Sigma_g^+} g B_x^r(t) g B_x^0(0) \ket{0 , \Sigma_g^+}(r) \Big)
\label{eq:V_ST} 
\\
& & \hspace{-0.7cm} V_{S^2}^{(1,1)}(r,\mu) = \frac{2}{3} \int_0^\infty \mathrm{d}t \, \Big( 2\bra{0 , \Sigma_g^+} g B_x^r(t) g B_x^0(0) \ket{0 , \Sigma_g^+}(r) + \bra{0 , \Sigma_g^+} g B_z^r(t) g B_z^0(0) \ket{0 , \Sigma_g^+}(r) \Big)
\label{eq:V_Ssq}
\end{eqnarray}
are renormalized quantities and $c_F^{(m_j)}(\mu)$ are NRQCD matching coefficients computed for the $\overline{\text{MS}}$ scheme (see below).
Note that, in contrast to Ref.\ \cite{Pineda:2000sz}, we have included the term $V^{(0)\prime}(r)/r$ in the definition (\ref{eq:V_LS20}) of $V_{LS}^{(2,0)}(r,\mu)$. Then $V_{LS}^{(2,0)}(r,\mu)$ is renormalized multiplicatively by a single matching coefficient, as all other spin dependent potentials (see Sec.\ \ref{sec:gradient_flow}). Using $c_S = 2 c_F^{(m_j)} - 1$ \cite{Manohar:1997qy}, where $c_S$ is the NRQCD matching coefficient of the static force term in Eq.\ (42) of Ref.\ \cite{Pineda:2000sz}, Eqs.\ (\ref{eq:cf_V_LS20}) and (\ref{eq:V_LS20}) are equivalent to that equation.

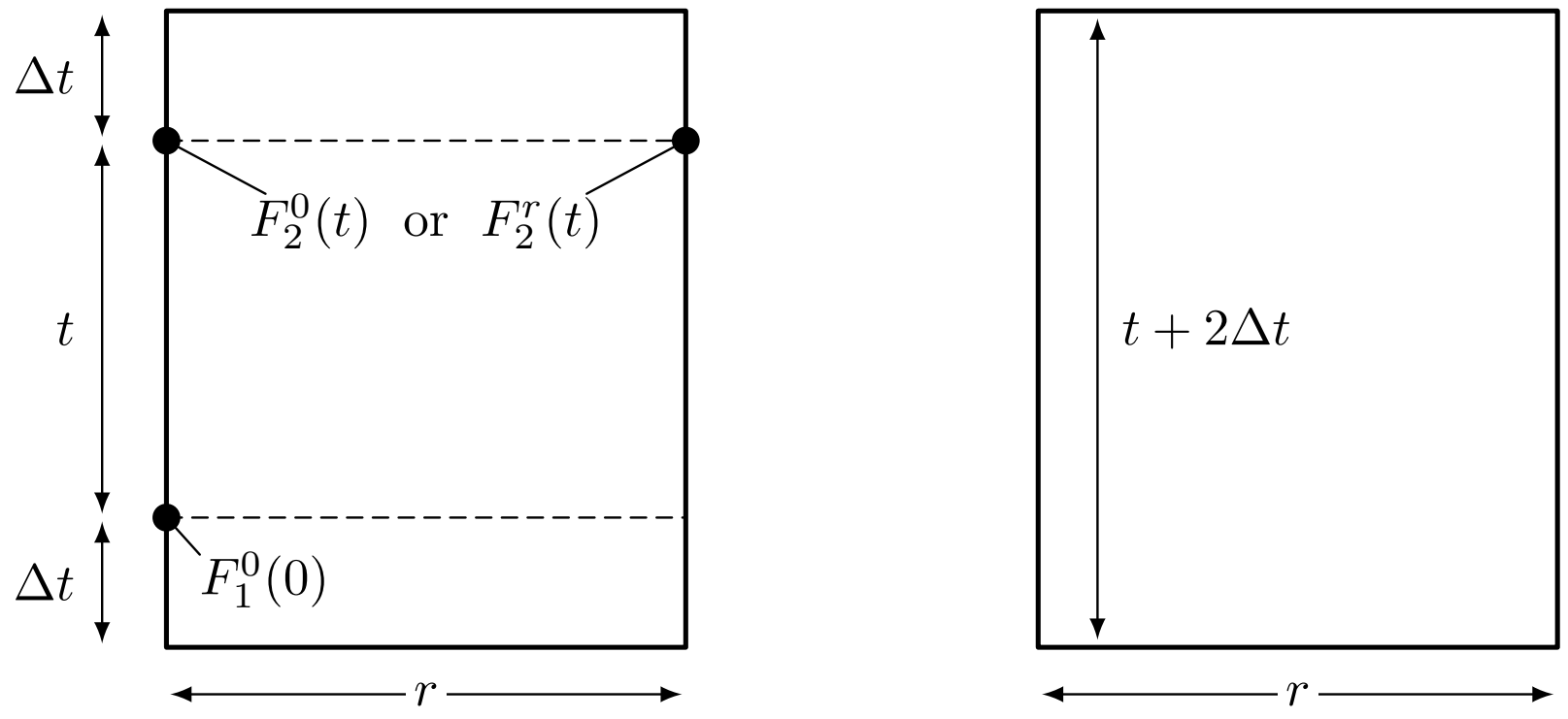
\begin{figure}[ht]
\centering
\begin{tikzpicture}[
x=0.047\textwidth,
y=0.047\textwidth,
font=\normalfont\normalsize,
draw=black,
text=black,
line cap=round,
line join=round,
contour/.style={line width=0.95pt},
guide/.style={line width=0.45pt, dash pattern=on 3pt off 2pt},
insertion/.style={circle, fill=black, draw=black, minimum size=5pt, inner sep=0pt},
leader/.style={line width=0.45pt},
dimension/.style={line width=0.45pt, <->, >=latex},
operator/.style={fill=white, inner sep=1pt}
]
\draw[contour] (0,0) rectangle (4.2,5.15);
\draw[guide] (0,1.05) -- (4.2,1.05);
\draw[guide] (0,4.10) -- (4.2,4.10);
\node[insertion] (Fone)     at (0,1.05) {};
\node[insertion] (FtwoZero) at (0,4.10) {};
\node[insertion] (FtwoR)    at (4.2,4.10) {};
\node[operator, anchor=north west] (FoneLabel) at (0.22,0.82) {$F_1^{0}(0)$};
\node[operator, anchor=north] (FtwoLabel) at (2.1,3.72) {$F_2^{0}(t)$\; or \;$F_2^{r}(t)$};
\draw[leader] (Fone) -- ($(FoneLabel.north west)+(0.05,-0.07)$);
\draw[leader] (FtwoZero) -- ($(FtwoLabel.north west)+(0.18,-0.05)$);
\draw[leader] (FtwoR) -- ($(FtwoLabel.north east)+(-0.18,-0.05)$);
\draw[dimension] (-0.52,0.03) -- node[left=2pt] {$\Delta t$} (-0.52,1.02);
\draw[dimension] (-0.52,1.08) -- node[left=2pt] {$t$} (-0.52,4.07);
\draw[dimension] (-0.52,4.13) -- node[left=2pt] {$\Delta t$} (-0.52,5.12);
\draw[dimension] (0.03,-0.38) -- node[fill=white, inner sep=1.5pt] {$r$} (4.17,-0.38);
\begin{scope}[shift={(7.05,0)}]
\draw[contour] (0,0) rectangle (4.2,5.15);
\draw[dimension] (0.48,0.06) -- node[right=3pt, fill=white, inner sep=1.5pt] {$t+2\Delta t$} (0.48,5.09);
\draw[dimension] (0.03,-0.38) -- node[fill=white, inner sep=1.5pt] {$r$} (4.17,-0.38);
\end{scope}
\end{tikzpicture}
\caption{\label{fig:skizze}Wilson loop and generalized Wilson loop with two field strength insertions.
}
\end{figure}

For Eqs.\ (\ref{eq:V_1}) to (\ref{eq:V_Lsq11}) and (\ref{eq:V_LS20}) to (\ref{eq:V_Ssq}) and all following related equations we have separated the heavy quark and antiquark along the $z$ axis.
The superscripts $0$ and $r$ of the chromoelectric and chromomagnetic fields denote their spatial positions, i.e.\ the position of the quark, $(0,0,0)$, or of the antiquark, $(0,0,r)$. The matrix elements in these equations are gauge invariant, since the temporal Wilson lines representing the static quark and antiquark, into which the field strengths are inserted, are implicitly contained in them. This is particularly obvious, when expressing them in terms of Wilson loops with field strength insertions (see Eq.\ (\ref{EQN_W_ratio}) and Fig.\ \ref{fig:skizze}). Note that, in a cutoff regularization like lattice gauge theory or gradient flow, the right hand sides of Eqs.\ (\ref{eq:V_1}) and (\ref{eq:V_psq20}), as well as the Wilson loop expression for the static potential, Eq.\ (\ref{EQN_V_standard}), contain $r$ independent contributions in addition to $V^{(1)}(r)$, $V_{p^2}^{(2,0)}(r)$ and $V^{(0)}(r)$, i.e.\ the constants $U^{(1)}$, $U_{p^2}^{(2,0)}$ and $U^{(0)}$ as well as regulator dependent, divergent contributions (see the discussion after Eq.\ (\ref{eq:V11}) and Secs.\ \ref{sec:result_static_potential} and \ref{SEC_spin_independent}). For simplicity, we nevertheless use the same notation.
$\ket{0 , \Sigma_g^+}$ denotes the ground state in the sector with a heavy quark antiquark pair at separation $r$ and
gluons in a $\Sigma_g^+$ configuration (i.e.\ a gluon configuration invariant under rotations around the axis of
separation [indicated by $\Sigma$], invariant under a combined parity and charge conjugation operation [indicated by
		$g$] and invariant under spatial reflections along an axis perpendicular to the axis of separation [indicated by $+$];
see e.g.\ Ref.\ \cite{Schlosser:2021wnr} for details).
$V^{(0)\prime}(r) = \mathrm{d} V^{(0)}(r) / \mathrm{d}r$ in Eq.\ (\ref{eq:V_LS20}) denotes the negative static force.
Note that $E_z$ has a non-vanishing expectation value in the $\ket{0 , \Sigma_g^+}$ state. The subscript $c$ denotes
that this expectation value has to be subtracted, i.e.\
\begin{eqnarray}
\label{eq:def_connected_correlator}
& & \hspace{-0.7cm} \bra{0, \Sigma_g^+} g E_z^{0,r}(t) g E_z^0(0) \ket{0, \Sigma_g^+}_c(r) = \bra{0, \Sigma_g^+} g E_z^{0,r}(t) g E_z^0(0) \ket{0, \Sigma_g^+}(r) - \Big( \bra{0, \Sigma_g^+} g E_z \ket{0, \Sigma_g^+}(r) \Big)^2
\,.
\end{eqnarray}

The static potential can be extracted from Wilson loops via
\begin{eqnarray}
& & \hspace{-0.7cm}
V^{(0)}(r) = \lim_{T \rightarrow \infty} \frac{1}{a} \ln \bigg(\frac{W(r, T)}{W(r, T + a)}\bigg)
\,,
\label{EQN_V_standard}
\end{eqnarray}
where $a$ is an arbitrary time difference (in lattice computations, $a$ typically corresponds to the lattice spacing and the quantity appearing to the right of $\lim_{T\rightarrow \infty}$ is referred to as effective potential).
Similarly, the field strength correlators on the right hand sides of Eqs.\ (\ref{eq:V_1}) to (\ref{eq:V_Lsq11}) and (\ref{eq:V_LS20}) to (\ref{eq:V_Ssq}) can be expressed as
\begin{eqnarray}
\label{EQN_W_ratio} & & \hspace{-0.7cm} \bra{0 , \Sigma_g^+} g F_2^{0, r}(t) g F_1^{0}(0) \ket{0 , \Sigma_g^+}
=
\lim_{\Delta t \rightarrow \infty} \frac{W^{F_1^{0} F_2^{0, r}}(r, t, \Delta t)}{W(r,T = t + 2 \Delta t)} \,.
\end{eqnarray}
$W(r, T)$ represents the Euclidean path integral expectation value of a standard Wilson loop with spatial extent $r$ and temporal extent $T$, while $W^{F_1^0 F_2^{0, r}}(r,t,\Delta t)$ is an analogous expectation value of a generalized Wilson loop with two field strength insertions $F_1^{0}$ and $F_2^{0, r}$ as sketched in Fig.~\ref{fig:skizze}. 

$F_1^0$ and $F_2^{0, r}$ are separated in the temporal direction by $t$ ($F_1^0$ at the earlier, $F_2^{0, r}$ at the later time).
$F_1^0$ is always located at the position of the quark (which we choose for simplicity as $\bm{r}_1 = (0,0,0)$), while $F_2^{0, r}$ is either located at the same position or at the position of the antiquark (which is $\bm{r}_2 = (0,0,r)$), as indicated by the superscript.
For example $\bra{0 , \Sigma_g^+} B_x^r(t) E_y^0(0) \ket{0 , \Sigma_g^+}$ corresponds to $F_1^0 = E_y^0(0) = E_y(\bm{r} = (0,0,0),0)$ and $F_2^{0, r}(t) = B_x^r(t) = B_x(\bm{r} = (0,0,r),t)$).

In addition to their $1/m_Q^2$ prefactors, spin dependent potentials $V_{LS}^{(2, 0)}(r, m_j)$,  $V_{L_2S_1}^{(1, 1)}(r, m_j)$, $V_{S_{12}}^{(1, 1)}(r, m_1, m_2)$ and $V_{S^2}^{(1, 1)}(r, m_1, m_2)$ exhibit an additional quark mass dependence due to the NRQCD matching coefficients $c_F^{(m_j)}(\mu)$, see Eqs.\ (\ref{eq:cf_V_LS20}) to (\ref{eq:cf_V_Ssq}).
As stated above, the integrated correlators $V_X(r,\mu)$ in Eqs.\ (\ref{eq:V_LS20}) to (\ref{eq:V_Ssq}) are renormalized quantities
in the $\overline{\text{MS}}$ scheme
depending on the scale $\mu$.
$c_F^{(m_j)}(\mu)$ has been calculated perturbatively in the $\overline{\text{MS}}$ scheme in Feynman gauge (for $n_f=0$)
in next-to-leading order (NLO) \cite{Eichten:1990vp,Manohar:1997qy},
\begin{eqnarray}
\label{EQN753} & & \hspace{-0.7cm} c_F^{(m_j)}(\mu) = 1 + \frac{\alpha_s(\mu)}{2 \pi} \bigg(C_F + C_A \bigg(1 - \ln\bigg(\frac{m_j}{\mu}\bigg)\bigg)\bigg) + \mathcal{O}(\alpha_s^2) \, ,
\end{eqnarray}
and in next-to-leading logarithms (NLL) \cite{Amoros:1997rx,Czarnecki:1997dz,Peset:2018jkf},
\begin{eqnarray}
\label{eq:cf_msbar} & & \hspace{-0.7cm} c_F^{(m_j)}(\mu) = 
\bigg(\frac{\alpha_s(\mu)}{\alpha_s(m_j)}\bigg)^{-3/11}
\bigg(1 + \frac{\alpha_s(m_j)}{2 \pi} \Big(C_F + C_A\Big) 
+ \frac{17 C_A}{363\pi} \Big(\alpha_s(m_j) - \alpha_s(\mu)\Big)
+ \mathcal{O}(\alpha_s^2)\bigg) \,,
\end{eqnarray}
where $\alpha_s$ is the strong coupling in the $\overline{\text{MS}}$ scheme (see Sec.\ \ref{sec:basic_parameterization} for details on its determination) and $C_F = 4/3$ and $C_A = 3$ are the Casimir charges of SU(3) in the fundamental and the adjoint representation. 
At NLO, the renormalized $\overline{\text{MS}}$ potentials $V_X(r,\mu)$ containing chromomagnetic fields depend logarithmically on the scale $\mu$. This dependence is compensated by the corresponding logarithms in the matching coefficients (either Eq.\ (\ref{EQN753}) or Eq.\ (\ref{eq:cf_msbar})), such that the spin dependent potentials $V_{LS}^{(2, 0)}(r, m_j)$, $V_{L_2S_1}^{(1, 1)}(r, m_j)$, $V_{S_{12}}^{(1, 1)}(r, m_1, m_2)$ and $V_{S^2}^{(1, 1)}(r, m_1, m_2)$ are independent of $\mu$ up to higher orders in $\alpha_s$. 
Eq.\ (\ref{eq:cf_V_Ssq}) contains two further matching coefficients $d_{sv}$ and $d_{vv}$, which are not relevant in the context of this work, where we aim to compute the $t$ integrals over renormalized field strength correlators with lattice gauge theory ($d_{sv}$ and $d_{vv}$ appear only in the last term on the right hand side of Eq.\ (\ref{eq:cf_V_Ssq}), which does not contain a field strength correlator; see Ref.\ \cite{Pineda:1998kj} for details).

In this work, we carry out lattice computations of integrated field strength correlators using gradient flow at several values of the lattice spacing and flow time. 
Correlators not containing chromomagnetic fields (Eqs.\ (\ref{eq:V_1}) to (\ref{eq:V_Lsq11})) can be extrapolated in a straightforward way to the continuum and flow time zero.
However, for the integrated field strength correlators relevant for the spin dependent potentials (\ref{eq:V_LS20}) to (\ref{eq:V_Ssq})), which contain chromomagnetic fields, the continuum and zero flow time extrapolation is more complicated. 
In Sec.\ \ref{sec:gradient_flow}, we discuss, how these extrapolations can be done to arrive at corresponding results in the $\overline{\text{MS}}$ scheme. Throughout this work we use $\mu = 3 \,\text{GeV}$, which is somewhere between the charm and bottom quark mass. 
In contrast to a direct matching to specific heavy quark masses, we consider this approach to provide more flexibility, when using our results for Born-Oppenheimer studies of quarkonium systems:
the quantities, $V_{LS}^{(2, 0)}(r,\mu)$, $V_{L_2S_1}^{(1, 1)}(r,\mu)$, $V_{S_{12}}^{(1, 1)}(r,\mu)$, $V_{S^2}^{(1, 1)}(r,\mu)$, can be easily converted to quark mass dependent potentials, $V_{LS}^{(2, 0)}(r,m_j)$, $V_{L_2S_1}^{(1, 1)}(r,m_j)$, $V_{S_{12}}^{(1, 1)}(r,m_1,m_2)$, $V_{S^2}^{(1, 1)}(r,m_1,m_2)$, using Eqs.\ (\ref{eq:cf_V_LS20}) to (\ref{eq:cf_V_Ssq}) and Eq.\ (\ref{EQN753}) or (\ref{eq:cf_msbar}).
The resulting quark mass dependent potentials are independent of $\mu$ up to corrections of order $\alpha_s^2$. We have verified numerically that the residual $\mu$ dependence is of the order of our statistical uncertainties by repeating our analysis for $\mu = 2 \, \text{GeV}$ and $\mu = 4 \, \text{GeV}$ (see Sec.~\ref{sec:discussion_mu}).

To summarize, our main goal in this work is to provide numerical results, based on first principles lattice gauge theory, for
\begin{itemize}
\item $V^{(0)}(r)$, $V^{(1)}(r)$, $V_{p^2}^{(2, 0)}(r)$, $V_{p^2}^{(1, 1)}(r)$, $V_{L^2}^{(2, 0)}(r)$ and $V_{L^2}^{(1, 1)}(r)$
\end{itemize}
as well as for
\begin{itemize}
\item $V_{LS}^{(2, 0)}(r,\mu)$, $V_{L_2S_1}^{(1, 1)}(r,\mu)$, $V_{S_{12}}^{(1, 1)}(r,\mu)$ and $V_{S^2}^{(1, 1)}(r,\mu)$
\end{itemize}
in the $\overline{\text{MS}}$ scheme at scale $\mu$.
This is the necessary non-perturbative lattice gauge theory input needed for quarkonium studies in pNRQCD \cite{Brambilla:2004jw,Berwein:2024ztx,Brambilla:2000gk,Pineda:2000sz}.

We finally note that there are two potentials on the right hand sides of Eqs.\ (\ref{eq:V20}) and (\ref{eq:V11}),
$V_r^{(2,0)}(r,m_j)$ and $V_r^{(1,1)}(r,m_1,m_2)$, for which we did not provide expressions in terms of field strength
correlators similar to Eqs.\ (\ref{eq:V_1}) to (\ref{eq:V_Lsq11}) and (\ref{eq:V_LS20}) to (\ref{eq:V_Ssq}). In this work we do not further consider these two
potentials. The reason is that their computation requires not only generalized Wilson loops with two field strength
insertions, but also with three and four field strength insertions. Computing such loops with lattice gauge theory, in
particular for large temporal separations between all three or four field strength insertions, is technically extremely
difficult and beyond the scope of this paper. $V_r^{(2,0)}(r,m_j)$ and $V_r^{(1,1)}(r,m_1,m_2)$ were, however, calculated in perturbation theory in NLO and NLL \cite{Anzai:2018eua,Peset:2018jkf} as well as in effective string theory
\cite{Brambilla:2014eaa}. The results obtained within the latter approach depend on just a single parameter, the string tension, which can be computed with lattice
gauge theory in a straightforward way rather precisely. Thus, for future Born-Oppenheimer predictions of quarkonium
spectra and properties a promising strategy could be to approximate the two potentials $V_r^{(2,0)}(r,m_j)$ and
$V_r^{(1,1)}(r,m_1,m_2)$ using perturbative expressions for small separations and effective string theory expressions
for large separations, while all other $1/m_Q^2$ potentials are provided more rigorously by lattice gauge theory.


\subsection{Gromes and BBMP relations}

Some of the $1/m_Q^2$ potentials introduced in Eqs.\ (\ref{eq:V20}) and (\ref{eq:V11}) are related to the static
potential $V^{(0)}(r)$ and the static force $-V^{(0)\prime}(r)$. These relations were derived using Poincar\'e invariance by
Gromes \cite{Gromes:1984ma},
\begin{eqnarray}
\label{eq:gromes_a} & & \hspace{-0.7cm} V_{LS}^{(2,0)}(r, m_j) - V_{L_2S_1}^{(1,1)}(r, m_j) = -\frac{1}{2r} V^{(0)\prime}(r) \, ,
\end{eqnarray}
and Barchielli, Brambilla, Montaldi and Prosperi (BBMP) \cite{Barchielli:1988zp},
\begin{eqnarray}
\label{eq:bbmp1} & & \hspace{-0.7cm} 2 V_{L^2}^{(2,0)}(r) - V_{L^2}^{(1,1)}(r) = -\frac{r}{2} V^{(0)\prime}(r) \\
\label{eq:bbmp2} & & \hspace{-0.7cm} 4 \Big(U_{p^2}^{(2,0)} + V_{p^2}^{(2,0)}(r)\Big) - 2 V_{p^2}^{(1,1)}(r) = -\Big(U^{(0)} + V^{(0)}(r)\Big) + r V^{(0)\prime}(r) \, ,
\end{eqnarray}
(see also Refs.\ \cite{Brambilla:2001xk,Brambilla:2003nt,Berwein:2018fos}) and are valid in
the continuum. In lattice gauge theory these relations are only approximately fulfilled, because of discretization errors. The splitting of the potentials into $r$ independent constants $U$ and $r$ dependent parts $V(r)$ is fixed by our parameterizations of $V^{(0)}(r)$, $V_{p^2}^{(1,1)}(r)$ and $V_{p^2}^{(2,0)}(r)$ (see Eqs.\ (\ref{EQN_Cornell}), (\ref{EQN_V_p_square_11_para}) and (\ref{EQN_V_p_square_20_para}) in Sec.\ \ref{SEC_results}). These parameterizations do not contain constant terms and fulfill Eq.\ (\ref{eq:bbmp2}) without the constants $U^{(0)}$ and $U_{p^2}^{(2,0)}$, i.e.\ $4 V_{p^2}^{(2,0)}(r) - 2 V_{p^2}^{(1,1)}(r) = -V^{(0)}(r) + r V^{(0)\prime}(r)$. Consequently, Eq.\ (\ref{eq:bbmp2}) implies $U_{p^2}^{(2,0)} = -U^{(0)}/4$.

Note that the Gromes relation can also be expressed in terms of $\mu$ dependent potentials in the $\overline{\text{MS}}$ scheme. Inserting Eqs.\ (\ref{eq:cf_V_LS20}) and (\ref{eq:cf_V_LS11}) into Eq.\ (\ref{eq:gromes_a}) leads to
\begin{eqnarray}
\label{eq:gromes_b} & & \hspace{-0.7cm} 
V_{LS}^{(2, 0)}(r,\mu) = V_{L_2 S_1}^{(1, 1)}(r,\mu) \, ,
\end{eqnarray}
where we have used $c_S = 2 c_F^{(m_j)} - 1$ (see the text below Eq.\ (\ref{eq:V_Ssq})), which is already incorporated in Eq.\ (\ref{eq:cf_V_LS20}).


\newpage

\section{\label{SEC_lattice_setup}Lattice setup and techniques}


\subsection{{\label{sec:ensemble_generation}}Gauge link ensembles}

For this work we generated three ensembles of pure SU(3) gauge link configurations using the CL2QCD software package~\cite{Sciarra:2021cl2qcd,Philipsen:2014mra}. 
The gauge action is the standard Wilson plaquette action and the ensembles mainly differ in their gauge coupling $\beta$, i.e.\ the lattice spacing $a$, while their volume in physical units is almost the same.

To introduce physical units, we set the Sommer scale to $r_0 = 0.5\,\text{fm}$~\cite{Sommer:1993ce,Necco:2001xg}. 
This choice for $r_0$ is standard in pure gauge theory, even though it slightly differs from values obtained in full QCD~\cite{Sommer:2014mea}.
We determine $r_0 / a$ and as a direct consequence the lattice spacing $a$ in fm from the static potential in Sec.\ \ref{sec:result_static_potential}. 
The extent of the periodic spatial dimensions $L$ is around $1.4 \, \text{fm}$ for all three ensembles. 

For each ensemble $N_\text{conf}$ thermalized gauge link configurations were generated to evaluate standard and generalized Wilson loops. We carried out a sufficient number of heatbath sweeps between consecutively generated configurations, such that  autocorrelations are negligible.
The ensembles were generated in parallel using $N_\text{run}$ independent simulations, each contributing $N_\text{conf} / N_\text{run}$ configurations.

Details of the ensembles are summarized in Tab.\ \ref{tab:ensembles}.

\begin{table}[htb]
	\centering
	\begin{tabular}{cccccccc} \hline \hline
		ensemble & $\beta$ & $a\,[\text{fm}]$ & $(L/a)^3 \times T/a$ & $N_\text{conf}$  & $N_\text{run}$
		\\
		\hline
		A        & $6.284$ & $0.060109(82)$   & $24^3 \times 48$     & 10000            & 10
		\\
		B        & $6.451$ & $0.047810(81)$   & $30^3 \times 60$     & \hphantom{0}4800 & 10
		\\
		C        & $6.594$ & $0.039520(80)$   & $36^3 \times 72$     & \hphantom{0}2400 & \hphantom{0}5
		\\
		\hline \hline
	\end{tabular}
	\caption{\label{tab:ensembles}Ensembles of gauge link configurations.}
\end{table}


\subsection{{\label{sec:analysis_details}}Computation and analysis of standard and generalized Wilson loops}

The field strength insertions in the generalized Wilson loops (see Fig.\ \ref{fig:skizze}) were discretized according to the clover leaf definition,
\begin{eqnarray}
& & \hspace{-0.7cm} gB_j = \frac{\epsilon_{jkl}}{2 i} \Big( \Pi_{kl} - \Pi_{kl}^\dagger \Big) \quad , \quad
gE_j = \frac{1}{2i} \Big( \Pi_{j0} - \Pi_{j0}^\dagger \Big) \quad , \quad
\Pi_{\mu\nu} = \frac{1}{4} \Big( P_{\mu, \nu} + P_{\nu, -\mu} + P_{-\mu, -\nu} + P_{-\nu, \mu} \Big) ,
\end{eqnarray}
where $P_{\mu, \nu}$ denotes the plaquette of gauge links in the $\mu$ and $\nu$ direction.

For the spatial parallel transporters in both standard and generalized Wilson loops we used APE-smeared links to enhance the corresponding ground state overlaps. Consequently, the limits $T \rightarrow \infty$ and $\Delta t \rightarrow \infty$ in Eqs.\ (\ref{EQN_V_standard}) and (\ref{EQN_W_ratio}) are effectively reached at smaller temporal extents $T$ and $\Delta t$, which is numerically highly advantageous.
As smearing parameters we used $\alpha_\text{APE} = 0.5$ and $N_\text{APE} = 75 \, , \, 100 \, , \, 100$ for ensembles A, B and C, respectively. 
These parameters refer to Eq.\ (23) in Ref.\ \cite{Jansen:2008si}.

To compute statistical errors and to determine autocorrelation times, we used \texttt{pyerrors} \cite{Joswig:2022qfe},
which implements the Gamma method \cite{Wolff:2003sm}, including the automatic windowing procedure \cite{Schaefer:2010hu}, and propagates errors linearly using automatic differentiation \cite{Ramos:2020scv}. 
As a cross-check we also carried out Jackknife analyses assuming negligible autocorrelations. The corresponding results turned out to be consistent within statistical errors.


\subsection{\label{sec:gradient_flow}Gradient flow and continuum and flow time extrapolations}

The computation of potential corrections from generalized Wilson loops, as defined in Eqs.\ (\ref{eq:V_1}) to (\ref{eq:V_Ssq}), is numerically quite difficult. 
One reason is that observables involving field strength insertions, like generalized Wilson loops, exhibit poor signal-to-noise ratios. 
Previous studies have addressed this issue using techniques such as link integration \cite{Bali:1997am} and the multilevel algorithm \cite{Koma:2006fw}.
Another difficulty arises from $r$ independent ultraviolet divergences associated with the lattice definition of field strength insertions, which require renormalization. 
Earlier approaches employ multiplicative Huntley-Michael coefficients \cite{Huntley:1986de,Bali:1996cj,Bali:1997am}, eliminating such divergent contributions only up to $\mathcal{O}(g^4)$ or
$\mathcal{O}(g^6)$, where $g$ is the gauge coupling (see Ref.\ \cite{Bali:1997am} for a detailed discussion).

In this work we instead use gradient flow \cite{Narayanan:2006rf,Luscher:2010iy,Luscher:2011bx} (see Refs.\ \cite{Luscher:2013vga,Shindler:2022tlx} for reviews), which provides a rigorous framework to regularize and renormalize.
Gradient flow simultaneously reduces ultraviolet fluctuations, thereby improving statistical precision.

Applying gradient flow amounts to evolving the gauge field $A_\mu$ along an extra dimension, the so-called flow time $t_f$, according to the flow equation
\begin{eqnarray}
\label{eq:flow_equation} & & \hspace{-0.7cm} \frac{\mathrm{d}}{\mathrm{d} t_f} B_\mu = D_\nu G_{\nu\mu} + \alpha_0 D_\mu \partial_\nu B_\nu \, .
\end{eqnarray}
$B_\mu$ is the flowed gauge field with initial conditions
\begin{eqnarray}
& & \hspace{-0.7cm} B_\mu \Big|_{t_f = 0} = A_\mu
\end{eqnarray}
and $G_{\mu\nu}$ and $D_\mu$ are the field strength tensor and the covariant derivative with respect to $B_\mu$, i.e.\
\begin{eqnarray}
\label{EQN_G_D} & & \hspace{-0.7cm} G_{\mu\nu} = \partial_\mu B_\nu - \partial_\nu B_\mu + [B_\mu, B_\nu] \quad , \quad D_\mu \ldots = \partial_\mu + [B_\mu, \ldots] .
\end{eqnarray}
Here and in Eq.\ (\ref{eq:flow_equation}) we use the convention common in lattice gauge theory, where the gauge coupling $g$ is absorbed into the gauge fields, i.e.\ $A_\mu$, $B_\mu$ and $G_{\mu\nu}$ correspond to $g A_\mu$, $g B_\mu$ and $g G_{\mu\nu}$ in the canonical normalization used in the remainder of this work.
$\alpha_0$ is a parameter related to gauge fixing. We use $\alpha_0 = 1$, corresponding to Feynman gauge, which is a common choice.
Note that in the context of gradient flow one conventionally uses traceless anti-hermitian instead of traceless hermitian generators for the gauge group. 
As a consequence there are no factors $i$ in Eq.\ (\ref{EQN_G_D}) in front of the commutator terms appearing in the field strength tensor and the covariant derivative. 

Gradient flow can be interpreted as a smearing technique. The corresponding smearing or flow radius $r_f$ is related to the flow time via $r_f = \sqrt{8 t_f}$.
The discretization of the gradient flow equations (\ref{eq:flow_equation}) to (\ref{EQN_G_D}) follows Ref.\ \cite{Luscher:2010iy}. 
It is reminiscent of the Wilson plaquette action and the corresponding flow is commonly referred to as Wilson flow. 
We computed Wilson loops and generalized Wilson loops at several flow times $t_f$ corresponding to the same flow radii in physical units for all three ensembles, 
$r_f = \sqrt{8 t_f} \approx 0.087 \, \text{fm} \, , \, 0.100 \, \text{fm} \, , \, 0.110 \, \text{fm} \, , \, 0.118 \, \text{fm}$ \footnote{For ensemble~C we used the dimensionless flow times $t_f / a^2 = 0.60 \, , \, 0.80 \, , \, 0.96 \, , \, 1.12$, which can easily be converted to their counterparts for ensembles A and B using the lattice spacings provided in Tab.\ \ref{tab:ensembles}.}.

The bare lattice gauge theory results for potentials, $V_X(r,t_f,a)$, depend on the lattice spacing $a$ and the gradient flow time $t_f$. Keeping the flow time $t_f$ fixed, they can be extrapolated in a straightforward way to the continuum,
\begin{eqnarray}
\label{EQN_lim_a} & & \hspace{-0.7cm} V_X(r,t_f) = \lim_{a \rightarrow 0} V_X(r,t_f,a) .
\end{eqnarray}
Spin independent potentials $V^{(0)}$, $V^{(1)}$, $V_{p^2}^{(2, 0)}$, $V_{p^2}^{(1, 1)}$, $V_{L^2}^{(2, 0)}$ and $V_{L^2}^{(1, 1)}$ can then be extrapolated to flow time zero in a similarly simple way,
\begin{eqnarray}
\label{EQN_lim_tf} & & \hspace{-0.7cm} V_X(r) = \lim_{t_f \rightarrow 0} V_X(r,t_f)
\end{eqnarray}
for $V^{(0)}$, $V^{(1)}$ and $V_{p^2}^{(2,0)}$ up to $r$ independent, regulator dependent contributions (see the discussion after Eq.\ (\ref{eq:V11}) and Secs.\ \ref{sec:result_static_potential} and \ref{SEC_spin_independent}).
The spin dependent potentials $V_{LS}^{(2, 0)}$, $V_{L_2S_1}^{(1, 1)}$, $V_{S_{12}}^{(1, 1)}$ and $V_{S^2}^{(1, 1)}$ at finite flow time, however, depend logarithmically on $t_f$, which prevents a naive extrapolation to $t_f = 0$. This dependence is compensated by a matching coefficient (see also Ref.\ \cite{Brambilla:2026mvc}). As discussed in Secs.\ \ref{SEC_notation} and \ref{sec:potential_defs} we use matching coefficients computed for the $\overline{\text{MS}}$ scheme, i.e.\ 
\begin{eqnarray}
\label{EQN_lim_tf_with_cF} & & \hspace{-0.7cm} V_X(r,\mu) = \lim_{t_f \rightarrow 0} \Big(c_F(\mu,t_f)\Big)^k V_X(r,t_f)
\end{eqnarray}
with $k$ denoting the number of chromomagnetic fields appearing in the corresponding field strength correlator, i.e.\ $k = 1$ for $V_{LS}^{(2, 0)}$ and $V_{L_2S_1}^{(1, 1)}$ and $k = 2$ for $V_{S_{12}}^{(1, 1)}$ and $V_{S^2}^{(1, 1)}$.
The matching coefficient is known in NLO \cite{Brambilla:2023vwm},
\begin{eqnarray}
& & \hspace{-0.7cm} c_F(\mu,t_f) =
1 - \frac{C_A\alpha_s(\mu)}{4\pi} \ln\Big(2\mu^2t_f e^{\gamma_E}\Big) + \mathcal{O}(\alpha_s^2) \, ,
\label{eq:cf_flow}
\end{eqnarray}
where $\alpha_s(\mu)$ is the strong coupling in the $\overline{\text{MS}}$ scheme (see Sec.\ \ref{sec:basic_parameterization} for details on its determination).
Recently, $c_F(\mu,t_f)$ has also been computed at NNLO \cite{Wang:2026toy}, which we became aware of only after completing our analysis. In the example studied in Ref.\ \cite{Wang:2026toy} ($n_f = 3$, $\mu = 2 \, \text{GeV}$), the change of $c_F(\mu,t_f)$ from NLO to NNLO is below about $4 \, \%$. Although this cannot be directly transferred to our setup, we do not expect that using the NNLO expression would change our results significantly within uncertainties. A consistent NNLO analysis, which would also require the NNLO expression for the NRQCD matching coefficient $c_F^{(m_j)}(\mu)$ \cite{Czarnecki:1997dz}, is left for future work.


\subsection{\label{sec:tree_level}Tree level improvement for the static potential and field strength correlators}

To reduce lattice artefacts and allow for more controlled continuum and flow time extrapolations, we apply tree level improvement.
This technique is applied to both the static potential and the integrated field strength correlators (\ref{eq:V_1}) to (\ref{eq:V_Lsq11}) and (\ref{eq:V_LS20}) to (\ref{eq:V_Ssq}). 
It is based on comparing tree level results in continuum perturbation theory and lattice perturbation theory at both zero and finite flow time $t_f$. 
Common approaches include the definition of an improved distance $r$ \cite{Necco:2001xg,Koma:2006fw}, performing a global fit which includes the  difference between tree level results \cite{Bali:1996cj,Bali:1997am,Schlosser:2021wnr}, or multiplying lattice data with the ratio of tree level results \cite{Brambilla:2023fsi}. In this work we mainly use the second approach, i.e.\ global fits.

At tree level, the static potential, as well as field strength correlators, are described by one gluon exchange diagrams \cite{Bali:1996cj,Bali:1997am}. 
Concrete details on how we use tree level improvement for $V^{(0)}(r)$ and the potential corrections in Eqs.\ (\ref{eq:V_1}) to (\ref{eq:V_Ssq}) are given in Secs.\ \ref{sec:result_static_potential} and \ref{sec:basic_parameterization}.
Further technical aspects concerning tree level improvement are provided in App.\ \ref{sec:apdx_tree_level}.


\newpage

\section{\label{SEC_results}Results}


\subsection{{\label{sec:potentials_at_a}}Potentials at finite flow time and lattice spacing}


\subsubsection{\label{SEC467}Static potential}

In pNRQCD the static potential $V^{(0)}(r)$ constitutes the leading-order term (of order $1/m_Q^0$) in the heavy-quark mass expansion.
Moreover, we use its derivative $V^{(0)\prime}(r)$ for scale setting, see Sec.\ \ref{sec:ensemble_generation} and Sec.\ \ref{sec:result_static_potential}. 
The potential corrections $V_{p^2}^{(2, 0)}(r)$, $V_{p^2}^{(1, 1)}(r)$, $V_{L^2}^{(2, 0)}(r)$ and $V_{L^2}^{(1, 1)}(r)$ are also related to $V^{(0)}(r)$ and $V^{(0)\prime}(r)$ via the BBMP relations (\ref{eq:bbmp1}) and (\ref{eq:bbmp2}). Consequently, computing $V^{(0)}(r)$ is also necessary for using these relations to stabilize fits of suitable parameterizations to lattice data for these potentials (see Secs.\ \ref{SEC_spin_independent_p_square} and \ref{SEC_spin_independent_L_square}). Moreover, $V^{(0)\prime}(r)$ appears in the definition of $V_{LS}^{(2, 0)}(r,\mu)$, Eq.\ (\ref{eq:V_LS20}) (see also Sec.\ \ref{SEC_spin_orbit}).

Following Eq.\ (\ref{EQN_V_standard}) lattice data points for the static potential $V^{(0)}(r, t_f, a)$ are extracted for each ensemble (characterized by the lattice spacing $a$), flow time $t_f$ and quark-antiquark separation $r$ using a strategy similar to that described in Ref.\ \cite{Schlosser:2021wnr}. 
For each data point we perform a correlated fit of a constant to the effective mass
\begin{eqnarray}
& & \hspace{-0.7cm} a V_\text{eff}^{(0)}(r, T, t_f, a) = \ln\bigg(\frac{W(r, T, t_f, a)}{W(r, T+a, t_f, a)}\bigg)
\label{eq:V_eff}
\end{eqnarray}
for $T \in [T_1, T_2]$ in plateau regions, which we select in the following way:
\begin{enumerate}

\item
$T_1 \geq 2 r_f$, where $r_f = \sqrt{8 t_f}$ is the flow radius \footnote{In this way, we exclude Wilson loops, in which the parallel transporters in the spatial direction, smeared by the application of gradient flow, have a sizable overlap. Such an overlap is expected to significantly alter the Wilson loop expectation value, which in turn makes a controlled and precise $t_f \rightarrow 0$ extrapolation difficult.}.

\item
For $T_1$ the minimum value is chosen, for which condition 1.\ and
\begin{eqnarray}
& & \hspace{-0.7cm} \Big\vert V_\text{eff}^{(0)}(r, T_1, t_f, a) - V_\text{eff}^{(0)}(r, T_1+a, t_f, a)\Big\vert < 2 \Delta\Big(V_\text{eff}^{(0)}(r, T_1, t_f, a) - V_\text{eff}^{(0)}(r, T_1+a, t_f, a)\Big)
\end{eqnarray}
are satisfied, where $\Delta(X)$ is the statistical error of $X$.

\item
$T_2$ is chosen, such that the correlated $\chi^2/\mathrm{DOF} \leq 1$ and $T_2 - T_1 > 3 a$ is as large as possible.

\item
If no such fit range exists, we use the fit range with $T_2 - T_1 > 3 a$ with the smallest correlated $\chi^2/\mathrm{DOF}$.
\end{enumerate}


\subsubsection{\label{SEC_potential_corrections}Potential corrections}

The potential corrections are defined in terms of integrals of field strength correlators (Eqs.\ (\ref{eq:V_1}) to (\ref{eq:V_Lsq11}) and (\ref{eq:V_LS20}) to (\ref{eq:V_Ssq})).

Lattice data points for these field strength correlators are obtained from ratios of generalized and standard Wilson loops according to Eq.\ (\ref{EQN_W_ratio}) (see also Fig.\ \ref{fig:skizze}), where the limit $\Delta t \rightarrow \infty$ is approximated by a sufficiently large but finite $\Delta t$.
This is more restrictive than the corresponding limit $T \rightarrow \infty$ for standard Wilson loops in Sec.\ \ref{SEC467}: the temporal extent of the generalized Wilson loops, $t + 2 \Delta t$, grows twice as fast with $\Delta t$, and it has to accommodate in addition the separation $t$ of the field strength insertions, which is the integration variable in Eqs.\ (\ref{eq:V_1}) to (\ref{eq:V_Lsq11}) and (\ref{eq:V_LS20}) to (\ref{eq:V_Ssq}) defining the potential corrections.
Therefore, we do not use a plateau selection as in Sec.\ \ref{SEC467}, but a fixed $\Delta t = 4a \, , \, 5a \, , \, 6a$ for ensembles A, B and C, respectively, corresponding to $\Delta t \approx 0.24 \, \text{fm}$.
We have checked using the effective potentials (\ref{eq:V_eff}) of standard Wilson loops that excited states are essentially suppressed at this temporal separation.
Moreover, we have repeated the computation with $\Delta t = 5a \, , \, 6a \, , \, 7a$ and found consistent field strength correlators within statistical errors.

For the numerical integration from $0$ to $\infty$, we split these integrals into two parts,
\begin{eqnarray}
\label{EQN_I_split} & & \hspace{-0.7cm} 
I^{(s)} (0, \infty)
 =
I^{(s)} (0, \tau) + I^{(s)}(\tau, \infty) \, ,
\end{eqnarray}
where we have introduced the shorthand notation
\begin{eqnarray}
& & \hspace{-0.7cm} I^{(s)} (t_1, t_2)
 =
\int_{t_1}^{t_2} \mathrm{d}t\, t^s \bra{0, \Sigma_g^+} gF_{\mu\nu}^{0, r}(t) gF_{\rho \sigma}^{0}(0) \ket{0, \Sigma_g^+}(r) \, .
\label{eq:general_correlator_integral}
\end{eqnarray}
To minimize statistical uncertainties we use different methods to compute the small $t$ integrals $I^{(s)} (0, \tau)$ and the large $t$ integrals $I^{(s)}(\tau, \infty)$ as detailed in the following. 
Throughout this paper we use $\tau / a = 8 \, , \, 10 \, , \, 12$ for ensemble A, B and C, respectively, corresponding to $\tau \approx 0.48 \, \text{fm}$, which turned out to be a numerically efficient choice.


\vspace{6pt}
{\noindent \underline{Small $t$ integral}:}

\vspace{3pt}
Lattice data points for field strength correlators $\bra{0, \Sigma_g^+} gF_{\mu\nu}^{0, r}(t) gF_{\rho \sigma}^{0}(0) \ket{0, \Sigma_g^+}(r)$ at small $t$ can be computed rather precisely, i.e.\ with small statistical errors. 
Thus, we can compute $I^{(s)}(0, \tau)$ numerically in a straightforward way using the composite Simpson's rule, which is defined as
\begin{eqnarray}
& & \hspace{-0.7cm} I_\text{Simpson}^{(s)} (t_1, t_2)
 =
a \sum_{j = t_1 / a}^{t_2 / a} \omega_j (j a)^s
\bra{0, \Sigma_g^+} gF_{\mu\nu}^{0, r}(j a) gF_{\rho \sigma}^{0}(0) \ket{0, \Sigma_g^+}(r) =
I^{(s)} (t_1, t_2) + \mathcal{O}(a^4) \, .
\label{eq:integral_simpson}
\end{eqnarray}
For an odd number of data points (i.e.\ $(t_2 -t_1) / a$ is even), we use exclusively the composite Simpson's-$1/3$ rule with coefficients $\omega_j = 1/3 \, , \, 4/3 \, , \, 2/3 \, , \, 4/3 \, , \ldots , \, 4/3 \, , \, 1/3$. 
For an even number of data points (i.e.\ $(t_2 -t_1) / a$ is odd), we use the composite Simpson's-$1/3$ rule in the range $t_1 \leq t \leq t_2 - 3a$ and Simpson's-$3/8$ in the range $t_2 - 3a \leq t \leq t_2$, where the latter has coefficients $\omega_j = 3/8 \, , \, 9/8 \, , \, 9/8 \, , \, 3/8$. 
By proceeding in that way, errors due to the numerical integration are proportional to $a^4$, i.e.\ strongly suppressed compared to the $\mathcal{O}(a^2)$ errors already present in the data lattice points due to the chosen discretization (Wilson plaquette action, clover leaf definition for field strength insertions).


\vspace{6pt}
{\noindent \underline{Large $t$ integral:}}

\vspace{3pt}
Lattice data points for field strength correlators $\bra{0, \Sigma_g^+} gF_{\mu\nu}^{0, r}(t) gF_{\rho \sigma}^{0}(0) \ket{0, \Sigma_g^+}(r)$ at large $t$ have rapidly growing statistical errors. 
As a consequence, standard numerical integration techniques to compute or approximate $I^{(s)}(\tau, \infty)$ introduce unacceptably large uncertainties.

Therefore, we adopt a different strategy and write the field strength correlators as spectral decompositions \cite{Bali:1996cj,Bali:1997am},
\begin{eqnarray}
& & \hspace{-0.7cm} \bra{0, \Sigma_g^+} gF_{\mu\nu}^{0, r}(t) gF_{\rho\sigma}^0(0) \ket{0, \Sigma_g^+}(r) =
\sum_{n, \Lambda_\eta^\epsilon}
D_{n, \Lambda_\eta^\epsilon}(r, t_f, a)
e^{-\Delta V_{n, \Lambda_\eta^\epsilon}(r, t_f, a)t}
\label{eq:spectral_decomposition}
\end{eqnarray}
with
\begin{eqnarray}
& & \hspace{-0.7cm} D_{n, \Lambda_\eta^\epsilon}(r, t_f, a) =
\bra{0, \Sigma_g^+} gF_{\mu\nu}^{0, r} \ket{n, \Lambda_\eta^\epsilon}(r)
\bra{n, \Lambda_\eta^\epsilon} gF_{\rho\sigma}^0 \ket{0, \Sigma_g^+}(r) \\
& & \hspace{-0.7cm} \Delta V_{n, \Lambda_\eta^\epsilon}(r, t_f, a) =
V_{n, \Lambda_\eta^\epsilon}(r, t_f, a) - V_{0,\Sigma_g^+}(r, t_f, a) .
\label{eq:spectrum_amplitude}
\end{eqnarray}
$\Lambda_\eta^\epsilon = \Sigma_g^+$ refers to the sector, where the ground state corresponds to the ordinary static potential. All other $\Lambda_\eta^\epsilon$ combinations denote sectors related to hybrid static potentials ($\Lambda = \Sigma (\equiv 0) \ , \ \Pi (\equiv 1) \ , \ \Delta (\equiv 2) \ , \ \ldots$ is the total angular momentum with respect to the quark-antiquark separation axis, $\eta = g (\equiv +) \ , \ u (\equiv -)$ describes the even (g) or odd (u) behavior under the combined parity
and charge conjugation operation and $\epsilon = + , -$ is the eigenvalue of a spatial reflection along an axis perpendicular to the quark-antiquark separation axis; see e.g.\ Ref.\ \cite{Schlosser:2021wnr} for a detailed discussion).
$n = 0, 1, 2 \dots$ label the ground state, first excitation, second excitation, etc.\ in the given $\Lambda_\eta^\epsilon$ sector.
One can show that for field strength correlators with $B_x$,
$B_y$, $E_x$ or $E_y$ the sum over $\Lambda_\eta^\epsilon$ is restricted to $\Lambda_\eta^\epsilon = \Pi_u , \Pi_g$,
while for $B_z$ one has $\Lambda_\eta^\epsilon = \Sigma_g^-,\Sigma_u^-$ and for $E_z$ $\Lambda_\eta^\epsilon = \Sigma_g^+,\Sigma_u^+$.
Moreover, one can show that all matrix elements $D_{n, \Lambda_\eta^\epsilon}(r, t_f, a)$ are real.

At large $t$ the lattice data points for $\bra{0, \Sigma_g^+} gF_{\mu\nu}^{0, r}(t) gF_{\rho \sigma}^{0}(0) \ket{0,
		\Sigma_g^+}(r)$ can be parameterized accurately by a truncated spectral decomposition. In Refs.\
\cite{Bali:1996cj,Bali:1997am,Koma:2006fw} the relevant matrix elements $D_{n, \Lambda_\eta^\epsilon}(r, t_f, a)$ and
potential differences $\Delta V_{n, \Lambda_\eta^\epsilon}(r, t_f, a)$ were determined by a fit at sufficiently large but not too large
$t$, where statistical errors of lattice data points are still on a moderate level. Once a parameterization had been fixed by such a fit, it was inserted into the integrand of the corresponding large-$t$ integral $I^{(s)}(\tau,\infty)$, which could then be evaluated analytically.

In our present work, we use a slightly different approach. Before determining the matrix elements $D_{n, \Lambda_\eta^\epsilon}(r, t_f, a)$ and potential differences $\Delta V_{n, \Lambda_\eta^\epsilon}(r, t_f, a)$, we integrate the truncated spectral decomposition analytically,
\begin{eqnarray}
& & \hspace{-0.7cm} I_\text{spectral}^{(0)}(t_1, t_2)
=
-\sum_{n, \Lambda_\eta^\epsilon} \frac{D_{n, \Lambda_\eta^\epsilon}}{\Delta V_{n, \Lambda_\eta^\epsilon}}
e^{-\Delta V_{n, \Lambda_\eta^\epsilon}t} \bigg\vert_{t_1}^{t_2}
\label{eq:rootfct_t0}
\\
& & \hspace{-0.7cm} I_\text{spectral}^{(1)}(t_1, t_2)
=
-\sum_{n, \Lambda_\eta^\epsilon}
\frac{D_{n, \Lambda_\eta^\epsilon} (1 + \Delta V_{n, \Lambda_\eta^\epsilon} t)}
{(\Delta V_{n, \Lambda_\eta^\epsilon})^2}
e^{-\Delta V_{n, \Lambda_\eta^\epsilon}t} \bigg\vert_{t_1}^{t_2}
\label{eq:rootfct_t1}
\\
& & \hspace{-0.7cm} I_\text{spectral}^{(2)}(t_1, t_2)
=
-\sum_{n, \Lambda_\eta^\epsilon}
\frac{D_{n, \Lambda_\eta^\epsilon} (2 + 2\Delta V_{n, \Lambda_\eta^\epsilon} t + (\Delta V_{n, \Lambda_\eta^\epsilon} t)^2)}
{(\Delta V_{n, \Lambda_\eta^\epsilon})^3}
e^{-\Delta V_{n, \Lambda_\eta^\epsilon}t}\bigg\vert_{t_1}^{t_2} \, ,
\label{eq:rootfct_t2}
\end{eqnarray}
where we have suppressed the arguments $(r, t_f, a)$ of $D_{n, \Lambda_\eta^\epsilon}$ and $\Delta V_{n, \Lambda_\eta^\epsilon}$.
Then we fit these expressions to numerical integrals of lattice data points of field strength correlators, i.e.\ to $I_\text{Simpson}^{(s)} (t_1, t_2)$ with fixed $t_1$ and $t_1 + 2 a \leq t_2 \leq t_{2, \text{max}}$, to determine the parameters $D_{n, \Lambda_\eta^\epsilon}(r, t_f, a)$ and $\Delta V_{n, \Lambda_\eta^\epsilon}(r, t_f, a)$ (we restrict the summations in Eqs.\ (\ref{eq:rootfct_t0}) to (\ref{eq:rootfct_t2}) to two states, the ground states in the two contributing $\Lambda_\eta^\epsilon$ sectors; an exception are $E_z$ correlators, where a term with $\Sigma_g^+$ and $n=0$ does not appear and $n=1$ is used instead).
The technical details on how we select $t_1$ and $t_{2, \text{max}}$ can be found in App.\ \ref{apdx:correlator_fits}.
We use correlated fits and stabilize them using Gaussian priors for $\Delta V_{n, \Lambda_\eta^\epsilon}(r, t_f, a)$, for which we use mean values and statistical uncertainties for (hybrid) static potentials provided by Ref.\ \cite{Capitani:2018rox} (for further details we refer to App.\ \ref{apdx:priors}). Once the parameters are known, we just have to set $t_1 = \tau$ and $t_2 \rightarrow \infty$ in Eqs.\ (\ref{eq:rootfct_t0}) to (\ref{eq:rootfct_t2}) to get the results for the large $t$ integrals $I^{(s)}(\tau, \infty)$ needed in Eq.\ (\ref{EQN_I_split}).

The main advantage of using the antiderivatives (\ref{eq:rootfct_t0}) to (\ref{eq:rootfct_t2}) is their obvious
relation to $I^{(s)} (\tau, \infty)$, i.e.\
\begin{eqnarray}
& & \hspace{-0.7cm} I^{(s)} (\tau, \infty) = \lim_{t_2 \rightarrow \infty} I_\text{spectral}^{(s)} (\tau, t_2) \,,
\label{EQN633}
\end{eqnarray}
which implies that the data points $I_\text{Simpson}^{(s)} (t_1, t_2)$ form a plateau at sufficiently large $t_2$.
For example plotting these data points as well as the righthand side of Eq.\ (\ref{EQN633}), $\lim_{t_2 \rightarrow \infty} I_\text{spectral}^{(s)} (\tau, t_2)$, provides a
helpful visualization of the quality of the data and the numerical precision of the resulting large $t$ integral.


\subsection{\label{sec:result_static_potential}Static potential and scale setting}

The extrapolation of lattice data points $V^{(0)}(r, t_f, a)$ to the continuum and zero flow time is carried out consistently with the conceptual discussion in Sec.\ \ref{sec:gradient_flow}.
To derive an appropriate fit ansatz, we start from
\begin{eqnarray}
\label{EQN_V_starting_point_fit_ansatz} & & \hspace{-0.7cm} V^{(0)}(r) =
V^{(0)}(r, t_f, a) - \mu^{(0)}(t_f, a) - \Delta V^{(0)}(r, t_f, a)
\, .
\end{eqnarray}
$V^{(0)}(r)$, which was introduced in Eq.\ (\ref{eq:VpNRQCD}), is the static potential in the continuum and at zero flow time. $\mu^{(0)}(t_f, a)$ accounts for the $r$ independent self-energy of the quark-antiquark pair, i.e.\ $\mu^{(0)}(t_f, a)$ represents a set of independent constants, one for each ensemble and flow time.
This self-energy diverges when the regulator is removed and is not part of $V^{(0)}(r)$. Since $\mu^{(0)}(t_f, a)$ are fit parameters, $V^{(0)}(r)$ can be determined only up to a finite additive constant, which cannot be separated from $\mu^{(0)}(t_f, a)$ in our analysis. This constant, $U^{(0)}$ in Eq.\ (\ref{eq:VpNRQCD}), is independent of $t_f$ and $a$ and is not determined in this work.
$\Delta V^{(0)}(r, t_f, a)$ comprises discretization errors as well as flow effects, which vanish in the limit $a \rightarrow 0$ and $t_f \rightarrow 0$.

Solving Eq.\ (\ref{EQN_V_starting_point_fit_ansatz}) with respect to $V^{(0)}(r, t_f, a)$ and replacing both $V^{(0)}(r)$ and $\Delta V^{(0)}(r, t_f, a)$ by suitable parameterizations (which carry the additional label ``fit'') leads to the fit ansatz
\begin{eqnarray}
& & \hspace{-0.7cm} V_\text{fit}^{(0)}(r, t_f, a) = V_\text{fit}^{(0)}(r) + \mu^{(0)}(t_f, a) + \Delta V_\text{fit}^{(0)}(r, t_f, a) \, .
\end{eqnarray}
For the static potential in the continuum and at zero flow time we use the commonly used Cornell parameterization,
\begin{eqnarray}
\label{EQN_Cornell} & & \hspace{-0.7cm} V_\text{fit}^{(0)}(r) = -\frac{\alpha^{(0)}}{r} + \sigma r \, .
\end{eqnarray}
For the discretization errors and flow effects we checked numerically that the perturbative tree level result
\begin{eqnarray}
& & \hspace{-0.7cm} \Delta V_\text{fit}^{(0)}(r, t_f, a) = V^{(0),\text{LO}}(r, t_f, a) - V^{(0),\text{LO}}(r)
\end{eqnarray}
with
\begin{eqnarray}
\label{SEC_V_LO} & & \hspace{-0.7cm} V^{(0),\text{LO}}(r) = -C_F \alpha_s \frac{1}{r}
\quad , \quad
V^{(0),\text{LO}}(r, t_f, a) = -4\pi C_F \alpha_s G(r, t_f, a)
\end{eqnarray}
is well-suited as parameterization, when replacing the strong coupling $\alpha_s$ by a fit parameter $\tilde{\alpha}$ in addition to the parameters $\alpha^{(0)}$, $\sigma$ and $\mu^{(0)}(t_f, a)$. Thus, $\tilde{\alpha}$ plays the role of $\alpha_s$, but is not determined by a perturbative formula; it is a pure fit parameter, which effectively accounts for discretization errors and flow effects. For detailed derivations and discussions of the tree level expressions (\ref{SEC_V_LO}), in particular the quantity $G(r, t_f, a)$, we refer to App.\ \ref{sec:apdx_tree_level} (for $G(r, t_f, a)$ see App.\ \ref{sec:apdx_tree_level_lattice}).

\begin{figure}[htb]
	\centering
	\includegraphics[width=0.495\textwidth]{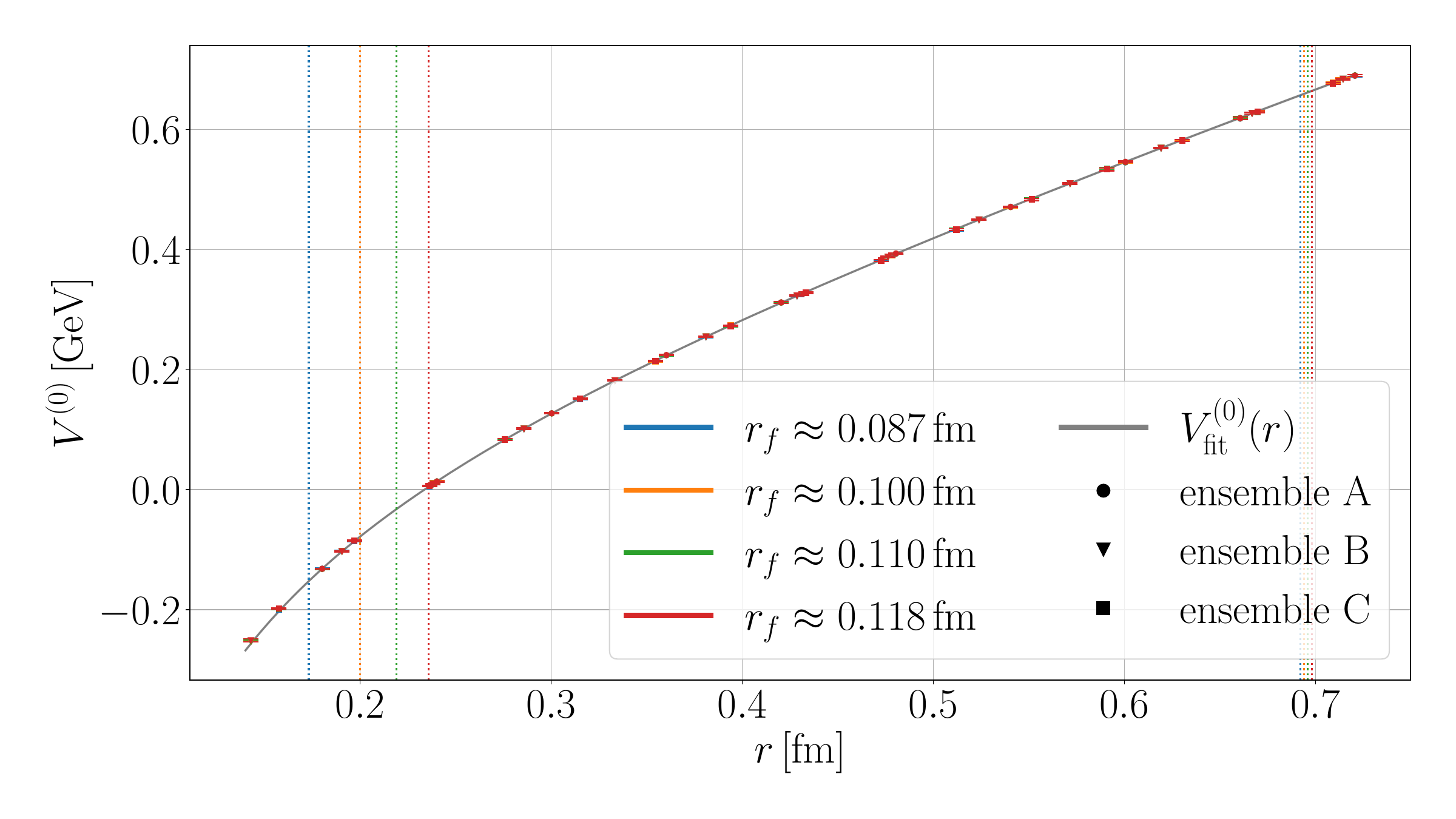}
	\caption{\label{fig:V0plot}Results for the static potential $V^{(0)}$. The curve represents the resulting parameterization in the continuum and at zero flow time (\ref{EQN_Cornell}).
    Data points have been corrected by subtracting all terms representing discretization errors and flow time effects as well as the constants $\mu^{(0)}(t_f, a)$. Fit ranges for each flow time are indicated by vertical lines in the corresponding color.}
	
\end{figure}

In summary the fit ansatz for the static potential is
\begin{eqnarray}
& & \hspace{-0.7cm} V_\text{fit}^{(0)}(r, t_f, a)
 =
-\frac{\alpha^{(0)}}{r} + \sigma r
+ \mu^{(0)}(t_f, a)
- C_F \tilde{\alpha} \bigg(4\pi G(r, t_f, a) - \frac{1}{r}\bigg) \, ,
\label{eq:ansatz_cornell_global}
\end{eqnarray}
which we use to carry out an uncorrelated global fit across all ensembles and flow times in the range $2 r_f \leq r < L / 2$  \footnote{The restriction $2 r_f \leq r$
excludes Wilson loops, in which the parallel transporters in the temporal direction, smeared by the application of gradient flow, have a sizable overlap. Such an overlap is expected to significantly alter the Wilson loop expectation value, which in turn makes a controlled and precise $t_f \rightarrow 0$ extrapolation difficult. See also Sec.\ \ref{SEC467}, where Wilson loops with temporal extent smaller than $2 r_f$ were excluded for the same reason.}. Results are shown in Fig.\ \ref{fig:V0plot} and the physically relevant parameters of the Cornell parameterization $\alpha^{(0)}$ and $\sigma$ are collected in Tab.\ \ref{tab:v0_results}, together with the corresponding $\chi^2/\text{DOF}$, which is of order $1$ and indicates a good fit. Moreover, we obtain $\tilde{\alpha} = 0.169(19)$, which seems to be a reasonable value.

\begin{table}[htb]
	\centering
	\begin{tabular}{c|c|c}
		$\alpha^{(0)}$    & $\sigma\,[\text{GeV}^2]$ & $\chi^2/\text{DOF}$
		\\
		\hline
		$0.29638(95)$ & $0.21152(32)$            & $1.15$
	\end{tabular}
	\caption{\label{tab:v0_results}Fit results for the parameterization of the static potential (\ref{EQN_Cornell}). Correlations between all fit parameters in Tabs.~\ref{tab:v0_results} to \ref{tab:v_sq_results} are given in App.~\ref{apdx:correlations}.}
\end{table}

We also compared to results from an analogous correlated fit and found agreement well within statistical errors, although with larger $\chi^2/\text{DOF} = 2.86$. 
A correlated fit requires a covariance matrix, which in the present case is rather large due to the use of multiple ensembles and flow times. A reliable estimate of such large covariance matrices requires substantially more statistical samples than available \cite{Michael:1993yj,Michael:1994sz}. Consequently, the correlated $\chi^2/\text{DOF}$ value larger than 1, should be interpreted with caution and not necessarily be regarded as an indicator of a poor fit.
Since some of the potential corrections discussed in the following subsections do not even allow correlated fits due to limited statistical accuracy, we prefer for consistency reasons uncorrelated fits for all potentials and quote e.g.\ in Tab.\ \ref{tab:v0_results} the corresponding uncorrelated fit results.

We tested these results also for stability with respect to a variation of the lower and upper bound of the fit range by $2 r_f \leq r \rightarrow 3 r_f \leq r$ and $r < L / 2 \rightarrow r < 3 L / 8$, respectively. 
The continuum and zero flow time result $-\alpha^{(0)}/r + \sigma r$ is again stable under these variations within statistical uncertainties.

To introduce physical units, we determined the Sommer scale $r_0 / a$ for each of our three ensembles listed in Tab.\ \ref{tab:ensembles}. For this we evaluate
$V^{(0)\prime}(r + a/2, t_f, a) = (V^{(0)}(r + a, t_f, a) - V^{(0)}(r, t_f, a)) / a$ in the range $0.4\,\text{fm} \ltapprox r \ltapprox 0.6\,\text{fm}$.  
Again we use tree level improvement, this time, however, via improved distances $r_\text{imp}$ derived from the lattice gluon propagator at finite flow time, in close analogy to the approach described in Ref.\ \cite{Necco:2001xg}.
The derivative of the Cornell ansatz expressed in terms of dimensionless lattice quantities,
\begin{eqnarray}
& & \hspace{-0.7cm} \Big(a^2 V_\text{fit}^{(0)\prime}(r/a)\Big)
 =
\frac{\alpha^{(0)}}{(r/a)^2} + (a^2 \sigma) ,
\end{eqnarray}
is fitted for each ensemble and flow time to the lattice data points $a^2 V^{(0)\prime}(r + a/2, t_f, a)$ at the improved separations $r_\text{imp}$.
The Sommer scale in units of the lattice spacing is then computed for each ensemble and flow time via $r_0 / a = \sqrt{(1.65 - \alpha^{(0)}) / (a^2 \sigma)}$.
We then carried out linear extrapolations of $r_0/a$ at fixed $a$ to $t_f = 0$. The resulting slopes are consistent with zero, as expected, because the used flow radii are significantly smaller than the considered quark antiquark separations. From the flow time extrapolated results we read off the lattice spacing $a$ as a function of $\beta$ by setting $r_0 = 0.5 \, \text{fm}$.
The resulting lattice spacings for the three ensembles are listed in Tab.\ \ref{tab:ensembles}. They are consistent with scale setting results from Ref.\ \cite{Necco:2001xg} within the cited uncertainty of $1\%$.


\subsection{\label{sec:basic_parameterization}Method to determine parameterizations of potential corrections}

Lattice data points for potential corrections (the integrated field strength correlators on the right hand sides of Eqs.\ (\ref{eq:V_1}) to (\ref{eq:V_Lsq11}) and (\ref{eq:V_LS20}) to (\ref{eq:V_Ssq}); see also Sec.\ \ref{SEC_potential_corrections}), denoted as $V_X(r, t_f, a)$, have to be extrapolated to the continuum and then to zero flow time.
For spin independent potentials the relevant Eqs.\ are (\ref{EQN_lim_a}) and (\ref{EQN_lim_tf}).
For spin dependent potentials one has to include matching coefficients, to compensate for the logarithmic dependence on $t_f$. In this work we provide spin dependent potentials in the $\overline{\text{MS}}$ scheme at scale $\mu$, where the relevant Eqs.\ are (\ref{EQN_lim_a}) and (\ref{EQN_lim_tf_with_cF}) and matching coefficients $c_F(\mu,t_f)$ are given by Eq.\ (\ref{eq:cf_flow}).
We note that the order of the two limits is important: to avoid unphysical $r$ independent divergences originating from the lattice discretization, one has to extrapolate first to the continuum ($a \rightarrow 0$) at finite flow time and then to flow time zero ($t_f \rightarrow 0$).

In the following, we discuss the general strategy we use to determine all nine potential corrections, 
$V^{(1)}(r)$, $V_{p^2}^{(2, 0)}(r)$, $V_{p^2}^{(1, 1)}(r)$, $V_{L^2}^{(2, 0)}(r)$, $V_{L^2}^{(1, 1)}(r)$, $V_{LS}^{(2, 0)}(r, \mu)$, $V_{L_2S_1}^{(1, 1)}(r, \mu)$, $V_{S_{12}}^{(1, 1)}(r, \mu)$ and $V_{S^2}^{(1, 1)}(r, \mu)$. The statements and equations in this subsection both apply to spin independent and spin dependent potentials. 
For the former one has to set $k=0$, i.e.\ matching coefficients $c_F(\mu, t_f)$ do not appear, and one can ignore the $\mu$ argument, i.e.\ $V_X(r, \mu) \rightarrow V_X(r)$.

Numerically, we implement the two extrapolations -- first to the continuum and then to flow time zero -- by means of a single fit to the lattice data points obtained from three ensembles (see Tab.~\ref{tab:ensembles}) and at four flow times (as discussed in Sec.~\ref{sec:gradient_flow}).
The starting point to construct a suitable fit function is conceptually the same as for the static potential, Eq.\ (\ref{EQN_V_starting_point_fit_ansatz}),
the relation between a potential in the continuum and at zero flow time, $V_X(r,\mu)$, and the corresponding lattice data points, $V_X(r,t_f,a)$,
\begin{eqnarray}
\label{eq:extrapolation_steps} & & \hspace{-0.7cm} V_X(r, \mu)
=
\Big(c_{F}(\mu, t_f)\Big)^k \underbrace{\Big(V_X(r, t_f, a) - \Delta V_{X,\text{lattice}}(r, t_f, a)\Big)}_{= V_X(r, t_f)} - \Delta V_{X,\text{flow}}(r, \mu, t_f) \, .
\end{eqnarray}
The logic on the right hand side of this equation is to first subtract lattice artifacts
\begin{eqnarray}
& & \hspace{-0.7cm} \Delta V_{X,\text{lattice}}(r, t_f, a) = \Delta V_{X,\text{lattice}}^{\text{LO}}(r, t_f, a) + \Delta V_{X,\text{lattice}}^{\text{beyond LO}}(r, t_f, a)
\end{eqnarray}
(which we separate into a tree level contribution and a higher-order contribution) from the lattice data points at finite $t_f$, $V_X(r, t_f, a)$, which yields corresponding results in the continuum, but at finite flow time $t_f$, i.e.\ $V_X(r, t_f)$. For spin dependent potentials, these are then multiplied by an appropriate power of the matching coefficients $(c_F(\mu, t_f))^k$ ($k$ denotes the number of chromomagnetic fields appearing in the corresponding field strength correlator) to compensate logarithmic divergences in $t_f$. 
Finally, flow effects
\begin{eqnarray}
& & \hspace{-0.7cm} \Delta V_{X,\text{flow}}(r,\mu,t_f) = \Delta V_{X,\text{flow}}^{\text{LO}}(r, t_f) + \Delta V_{X,\text{flow}}^{\text{beyond LO}}(r, \mu, t_f)
\end{eqnarray}
(which we also separate into a tree level contribution and a higher-order contribution) are subtracted to arrive at $V_X(r,\mu)$.

Since $c_{F}(\mu, t_f) = 1 + \mathcal{O}(\alpha_s)$, the LO gradient flow effects $\Delta V_{X,\text{flow}}^{\text{LO}}(r, t_f)$ can also be absorbed into the bracket multiplied by $c_{F}(\mu, t_f)$ and combined with $\Delta V_{X,\text{lattice}}^{\text{LO}}(r, t_f, a)$ into a single term $\Delta V_X^{\text{LO}}(r, t_f, a)$  that describes both lattice discretization and flow effects, i.e.\
\begin{eqnarray}
\label{EQN_starting_point} & & \hspace{-0.7cm} V_X(r, \mu)
=
\Big(c_{F}(\mu, t_f)\Big)^k \Big(V_X(r, t_f, a) - \Delta V_X^{\text{LO}}(r, t_f, a) - \Delta V_{X,\text{lattice}}^{\text{beyond LO}}(r, t_f, a)\Big) - \Delta V_{X,\text{flow}}^{\text{beyond LO}}(r, \mu, t_f) \, .
\end{eqnarray}

Solving Eq.\ (\ref{EQN_starting_point}) with respect to $V_X(r, t_f, a) - \Delta V_X^{\text{LO}}(r, t_f, a)$ and providing suitable parameterizations for the potential in the continuum and at zero flow time as well as for lattice artifacts and flow effects beyond LO (denoted in the following by $V_{X, \text{fit}}(r, \mu)$,  $\Delta V_{X, \text{lattice},\text{fit}}^{\text{beyond LO}}(r, t_f, a)$ and $\Delta V_{X, \text{flow},\text{fit}}^{\text{beyond LO}}(r, \mu, t_f)$) leads to the fit function
\begin{eqnarray}
& & \hspace{-0.7cm} \mathcal{U}_X(r, t_f, a) =
\Big(c_F(\mu, t_f) \Big)^{-k}
\Big(V_{X, \text{fit}}(r, \mu)
+ \Delta V_{X, \text{flow},\text{fit}}^{\text{beyond LO}}(r, \mu, t_f)\Big)
+ \Delta V_{X, \text{lattice},\text{fit}}^{\text{beyond LO}}(r, t_f, a)
\label{eq:ansatz_general}
\,,
\end{eqnarray}
which has to be fitted to data points
\begin{eqnarray}
& & \hspace{-0.7cm} V_X(r, t_f, a) - \Delta V_X^{\text{LO}}(r, t_f, a) \, .
\end{eqnarray}
The reason for subtracting $\Delta V_X^{\text{LO}}(r, t_f, a)$ from the lattice data points $V_X(r, t_f, a)$, before performing the fit with $\mathcal{U}_X(r, t_f, a)$, is to ensure a straightforward and consistent treatment of error propagation. The tree level correction $\Delta V_X^{\text{LO}}(r, t_f, a)$ can be calculated analytically, except for the value of the coupling (see App.\ \ref{sec:apdx_tree_level}). For the latter we use the fit parameter $\tilde{\alpha}$ instead of $\alpha_s$, which has already been determined in Sec.\ \ref{sec:result_static_potential}, when fitting the parameterization (\ref{eq:ansatz_cornell_global}) to the highly precise data for the static potential. Consequently, $\Delta V_X^{\text{LO}}(r, t_f, a)$ is not part of the fit ansatz, as it does not contain any free parameters, but rather a correction applied to the lattice data points $V_X(r, t_f, a)$. Its statistical uncertainty, arising from the uncertainty in $\tilde{\alpha}$, must therefore be combined consistently with the statistical uncertainties of $V_X(r, t_f, a)$, namely at the level of the individual samples used in our statistical analysis.

For the terms on the right hand side of Eq.\ (\ref{eq:ansatz_general}) we use or select expressions as follows:
\begin{enumerate}
\item \textbf{Matching coefficients:} \\
For $c_F(\mu, t_f)$ we use the perturbative result (\ref{eq:cf_flow}).
The $\overline{\text{MS}}$ coupling $\alpha_s(\mu)$ of pure SU(3) gauge theory ($n_f = 0$) is computed with the five-loop $\beta$ function as implemented in CRunDec \cite{Chetyrkin:2000yt,Schmidt:2012az,Herren:2017osy}, using the central value $r_0 \Lambda_{\overline{\text{MS}}}^{(n_f=0)} = 0.647$ from Ref.\ \cite{FlavourLatticeAveragingGroupFLAG:2024oxs} and $r_0 = 0.5 \, \text{fm}$ (the uncertainty of $r_0 \Lambda_{\overline{\text{MS}}}^{(n_f=0)}$ changes $\alpha_s$ by less than $1\%$ at the relevant scales, which is negligible in our context).
Throughout this work, $\alpha_s$ is always determined in this way, in particular also for the NRQCD matching coefficients $c_F^{(m_j)}(\mu)$ in Sec.\ \ref{sec:discussion_mu}.

\item \textbf{Potentials in the continuum and at zero flow time:} \\
We use physically motivated parameterizations $V_{X, \text{fit}}(r, \mu)$, which are consistent with the lattice data points, i.e.\ where the fit leads to an acceptable $\chi^2 / \text{DOF}$. The parameterizations are discussed individually in the following sections \ref{SEC_spin_independent} to \ref{sec:result_spin_spin}.

\item 
\textbf{Lattice discretization errors beyond LO:} \\
Lattice discretization errors  at finite flow time $t_f$ beyond LO are modeled by
\begin{eqnarray}
\label{eq:ansatz_general_lattice} & & \hspace{-0.7cm} \Delta V_{X, \text{lattice},\text{fit}}^{\text{beyond LO}}(r, t_f, a) = \frac{1}{r^{d_X}} \bigg(\beta_{X, \text{lattice}}^{1}(t_f/r^2) \frac{a^2}{r^2} + \beta_{X, \text{lattice}}^{2}(t_f/r^2) \frac{a^2}{t_f} \bigg) \, ,
\end{eqnarray}
where we primarily aim to account for errors proportional to $a^2 / r^2$ and $a^2 / t_f$. $d_X$ is the mass dimension of the corresponding potential. For the dimensionless coefficients $\beta_{X, \text{lattice}}^{1,2}$ we allow for the spin dependent potentials 
$V_{LS}^{(2, 0)}$, $V_{L_2S_1}^{(1, 1)}$, $V_{S_{12}}^{(1, 1)}$ and $V_{S^2}^{(1, 1)}$
a simple logarithmic dependency on $t_f/r^2$ motivated by and consistent with the lattice data, i.e.\
\begin{eqnarray}
& & \hspace{-0.7cm} \beta_{X, \text{lattice}}^{1}(t_f/r^2)
=
\beta_{X, \text{lattice}}^{1a} + \beta_{X, \text{lattice}}^{1b} \ln(t_f / r^2)
\,,
\label{eq:beta_term_with_r_log}
\\
& & \hspace{-0.7cm} \beta_{X, \text{lattice}}^{2}(t_f/r^2)
=
\beta_{X, \text{lattice}}^{2a} + \beta_{X, \text{lattice}}^{2b} \ln(t_f / r^2)
\label{eq:beta_term_with_tf_log}
\end{eqnarray}
(logarithms typically appear in NLO of perturbation theory, e.g.\ when calculating the static potential, which further motivates the $\ln(t_f / r^2)$ terms on the right hand sides of Eqs.\ (\ref{eq:beta_term_with_r_log}) and (\ref{eq:beta_term_with_tf_log})). For the spin independent potentials $V^{(1)}$, $V_{p^2}^{(2, 0)}$, $V_{p^2}^{(1, 1)}$, $V_{L^2}^{(2, 0)}$ and $V_{L^2}^{(1, 1)}$, however, the logarithmic terms do not lead to notably smaller $\chi^2 / \text{DOF}$ and, thus, are not needed, i.e.\ we use Eqs.\ (\ref{eq:beta_term_with_r_log}) and (\ref{eq:beta_term_with_tf_log}) with $\beta_{X, \text{lattice}}^{1b} = \beta_{X, \text{lattice}}^{2b} = 0$ in these cases.

\item \textbf{Gradient flow effects beyond LO:} \\ 
Gradient flow effects in the continuum beyond LO are modeled by
\begin{eqnarray}
\label{eq:ansatz_general_flow_effects} & & \hspace{-0.7cm} \Delta V_{X, \text{flow},\text{fit}}^{\text{beyond LO}}(r, \mu, t_f) = \frac{1}{r^{d_X}} 
\beta_{X, \text{flow}}(\mu) \frac{t_f}{r^2} \, ,
\end{eqnarray}
where  we primarily aim to account for errors proportional to $t_f / r^2$.
\end{enumerate}

All fit results for the potential corrections presented in the following sections \ref{SEC_spin_independent}, \ref{SEC_spin_orbit} and \ref{sec:result_spin_spin} were tested for stability with respect to a variation of the lower and upper bound of the fit range by $2 r_f \leq r \rightarrow r_f + 2a \leq r$ as well as $r < L / 2 \rightarrow r < L/2 - 2 a$, respectively. The continuum and zero flow time results turned out to be stable
under these variations within statistical uncertainties.


\subsection{\label{SEC_spin_independent}Spin independent potentials}

The spin independent potentials $V^{(1)}$, $V_{p^2}^{(2, 0)}$, $V_{p^2}^{(1, 1)}$, $V_{L^2}^{(2, 0)}$ and $V_{L^2}^{(1, 1)}$ are related to integrated field strength correlators containing two chromoelectric, but no chromomagnetic insertions (see Eqs.\ (\ref{eq:V_1}) to (\ref{eq:V_Lsq11})). Thus, matching coefficients $c_F$ are not required for the extrapolation to the continuum and zero flow time, i.e.\ one has to set $k = 0$ in the general equations provided in Sec.\ \ref{sec:basic_parameterization} and can ignore the $\mu$ argument.


\subsubsection{\label{SEC_spin_independent_p}Potential \texorpdfstring{$V^{(1)}$}{V(1)}}

For the potential $V^{(1)}(r)$ in the continuum and at zero flow time we use the parameterization
\begin{eqnarray}
\label{EQN_V_1_para} & & \hspace{-0.7cm} V_{\text{fit}}^{(1)}(r) = -\frac{9 \big(\alpha^{(1)}\big)^2}{16 r^2} + \frac{\Lambda^{(1)}}{2 \pi} \ln\Big((r / r_0)^2\Big) \, ,
\end{eqnarray}
where the first term is motivated by perturbation theory \cite{Brambilla:2000gk,Peset:2015vvi,Peset:2018jkf,Anzai:2018eua} and the second term by effective string theory
\cite{Perez-Nadal:2008wtr,Brambilla:2014eaa}.
Here, $\alpha^{(1)}$ and $\Lambda^{(1)}$ (as well as $\alpha_{p^2}$, $\alpha_{L^2}$, $\Lambda_{L^2}$, $\alpha_{LS}(\mu)$, $\Lambda_{LS}(\mu)$, $\alpha_{S_{12}}(\mu)$, $\alpha_{S_{12}}^\prime(\mu)$, $\alpha_{S^2}(\mu)$, $\alpha_{S^2}^\prime(\mu)$ and $\Lambda_{S^2}(\mu)$ in Secs.~\ref{SEC_spin_independent} to \ref{sec:result_spin_spin}) are fit parameters.
Their names and the numerical prefactors are chosen in analogy to the perturbative expressions motivating the parameterizations, e.g.\ $\alpha^{(1)}$ corresponds at leading order to $C_F \alpha_s$ (as $\alpha^{(0)}$ in the Cornell parameterization (\ref{EQN_Cornell})), such that Eq.\ (\ref{EQN_V_1_para}) reproduces the perturbative result of Ref.\ \cite{Brambilla:2000gk} for $\alpha^{(1)} = C_F \alpha_s$.
Since our lattice data points are not in the perturbative regime, perturbation theory only serves as a guideline for a suitable fit function, and the fitted values should not be interpreted as the strong coupling.

Just like the static potential, the potential $V^{(1)}$ also contains an $r$ independent but regulator dependent contribution. 
At leading non-vanishing order in perturbation theory (order $\alpha_s^2$) and finite flow time, this contribution is proportional to $\alpha_s^2/t_f$, i.e.\ it diverges for $t_f \rightarrow 0$. We account for it by adding $\mu^{(1)}(t_f, a)$ to the fit function (\ref{eq:ansatz_general}), where $\mu^{(1)}(t_f, a)$ represents a set of independent constants, one for each ensemble and flow time. Consequently, instead of $r_0 = 0.5 \, \text{fm}$, any other constant with units of length could be used in the logarithm in Eq.\ (\ref{EQN_V_1_para}), since changing $r_0$ would merely result in an additive constant in $V_{\text{fit}}^{(1)}(r)$, which can be absorbed into $\mu^{(1)}(t_f, a)$. Since $\mu^{(1)}(t_f, a)$ are fit parameters, $V_\text{fit}^{(1)}(r)$ can be determined only up to a finite additive constant, which cannot be separated from $\mu^{(1)}(t_f, a)$ in our analysis. This constant, $U^{(1)}$ in Eq.\ (\ref{eq:VpNRQCD}), is independent of $t_f$ and $a$. It is expected that the regulator dependent contribution is related to the $r$ independent $\mathcal{O}(\Lambda_\text{QCD}^2/m)$ renormalon of $V^{(1)}$ and that the latter cancels the corresponding renormalon of the heavy quark pole mass \cite{Ayala:2019hkn,Agemura:2023xqq}. In a consistent scheme the regulator dependent contribution could then be absorbed into the definition of the heavy quark mass.

Since the tree level correction $\Delta V^{(1), \text{LO}}(r, t_f, a)$ is $r$ independent (see App.\ \ref{sec:apdx_tree_level}), it can also be absorbed into $\mu^{(1)}(t_f, a)$ and, thus, is not considered explicitly in this subsection.

\begin{figure}[htb]
\centering
\includegraphics[width=0.495\textwidth]{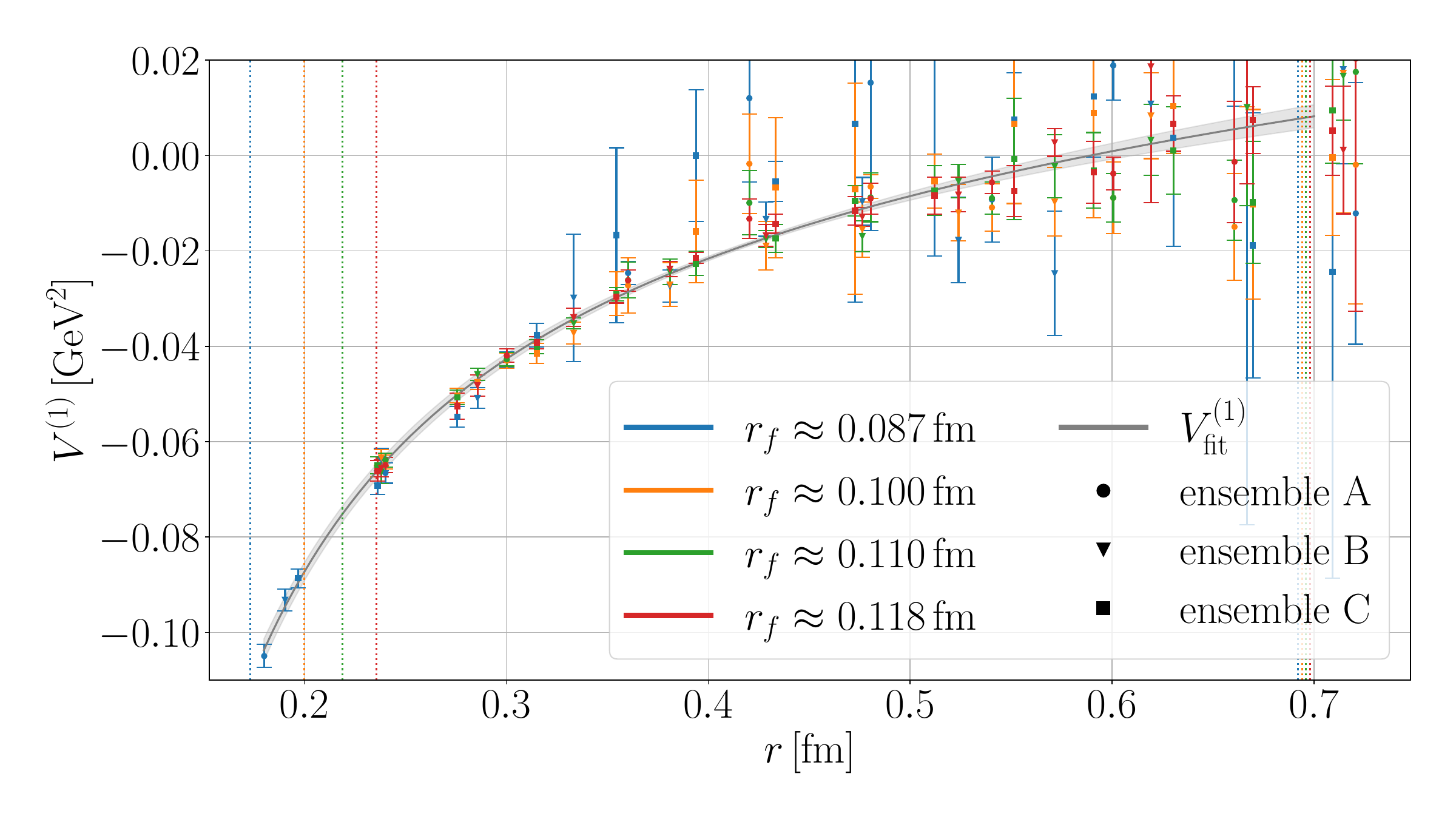}
\caption{
\label{fig:extrapolated_V_1}
Results for the spin independent potential $V^{(1)}$. The curve represents the resulting parameterization in the continuum and at flow
time zero (\ref{EQN_V_1_para}). Data points have been corrected by subtracting all terms representing discretization errors and flow time effects as well as the constants $\mu^{(1)}(t_f, a)$. Fit ranges for each flow time are indicated by vertical lines in the corresponding
color.}
\end{figure}

To summarize, we fit
\begin{eqnarray}
\label{eq:ansatz_v_1} & & \hspace{-0.7cm} \mathcal{U}^{(1)}(r,t_f,a) = -\frac{9 \big(\alpha^{(1)}\big)^2}{16 r^2} + \frac{\Lambda^{(1)}}{2 \pi} \ln\Big((r / r_0)^2\Big) + \mu^{(1)}(t_f, a) + \frac{1}{r^2} \beta_{\text{flow}}^{(1)} \frac{t_f}{r^2} + \frac{1}{r^2} \bigg(\beta_{\text{lattice}}^{(1),1a} \frac{a^2}{r^2} + \beta_{\text{lattice}}^{(1),2a} \frac{a^2}{t_f} \bigg)
\end{eqnarray}
to data points
\begin{eqnarray}
& & \hspace{-0.7cm} V^{(1)}(r, t_f, a) .
\end{eqnarray}
The fit parameters are $\alpha^{(1)}$ and $\Lambda^{(1)}$ and for each ensemble and flow time one $\mu^{(1)}(t_f, a)$ as well as three common $\beta$ parameters, describing discretization errors and flow effects.
Results from an uncorrelated fit across all ensembles and flow times in the range $2 r_f \leq r < L/2$ are shown in Fig.~\ref{fig:extrapolated_V_1}. The physically relevant parameters $\alpha^{(1)}$ and $\Lambda^{(1)}$ are collected in Tab.\ \ref{tab:v_spin_independent_results}, together with the corresponding $\chi^2 / \text{DOF}$, which is of order 1 and indicates a good fit.

\begin{table}[htb]
	\centering
	\begin{tabular}{c|c|c|c|c|c}
		$\alpha^{(1)}$ & $\Lambda^{(1)}\,[\text{GeV}^2]$ & $\alpha_{p^2}$ & $\alpha_{L^2}$ & $\Lambda_{L^2}\,[\text{GeV}^2]$ & $\chi^2/\text{DOF}$
		\\
		\hline
		$0.311(16)$ & $0.117(19)$ & & & & $0.87$
		\\
		& & $0.144(9)$ & & & $3.17$
		\\
		& & & $0.607(47)$ & $0.313(16)$ & $2.26$
	\end{tabular}
    \caption{\label{tab:v_spin_independent_results}Fit results for the parameterizations of the spin independent potentials (\ref{EQN_V_1_para}), (\ref{EQN_V_p_square_11_para}), (\ref{EQN_V_p_square_20_para}), (\ref{EQN_V_L_square_11_para}) and (\ref{EQN_V_L_square_20_para}). The parameters $\alpha^{(0)}$ and $\sigma$ appearing in the parameterizations (\ref{EQN_V_p_square_20_para}) and (\ref{EQN_V_L_square_20_para}) have already been determined in the context of the static potential and their values can be found in Tab.\ \ref{tab:v0_results}. Correlations between all fit parameters in Tabs.~\ref{tab:v0_results} to \ref{tab:v_sq_results} are given in App.~\ref{apdx:correlations}.}
\end{table}


\subsubsection{\label{SEC_spin_independent_p_square}Potentials \texorpdfstring{$V_{p^2}^{(2, 0)}$}{Vp2(2,0)} and \texorpdfstring{$V_{p^2}^{(1, 1)}$}{Vp2(1,1)}}

For the potential $V_{p^2}^{(1, 1)}(r)$ in the continuum and at zero flow time we use the parameterization
\begin{eqnarray}
\label{EQN_V_p_square_11_para} & & \hspace{-0.7cm} V_{p^2,\text{fit}}^{(1, 1)}(r) = -\frac{\alpha_{p^2}}{r} \, ,
\end{eqnarray}
which is motivated by perturbation theory \cite{Peset:2015vvi,Peset:2018jkf,Anzai:2018eua} and consistent with effective string theory
\cite{Perez-Nadal:2008wtr,Brambilla:2014eaa}.
Since $V_{p^2}^{(1, 1)}$ and $V_{p^2}^{(2, 0)}$ are related by the BBMP relation (\ref{eq:bbmp2}), the parameterization of the latter in the continuum and at zero flow time is fixed by Eq.\ (\ref{EQN_V_p_square_11_para}),
\begin{eqnarray}
\label{EQN_V_p_square_20_para} & & \hspace{-0.7cm} 
V_{p^2,\text{fit}}^{(2, 0)}(r) = 
\frac{1}{2} V_{p^2,\text{fit}}^{(1, 1)}(r) - \frac{1}{4} V_\text{fit}^{(0)}(r) + \frac{r}{4} V_\text{fit}^{(0)\prime}(r) = -\frac{\alpha_{p^2}}{2 r} + \frac{\alpha^{(0)}}{2 r} \, .
\end{eqnarray}
Just like the potentials $V^{(0)}$ and $V^{(1)}$, the potential $V_{p^2}^{(2,0)}$ also contains an $r$ independent but regulator dependent contribution.
We account for it by adding $\mu_{p^2}^{(2,0)}(t_f, a)$ to the fit function (\ref{eq:ansatz_general}), where $\mu_{p^2}^{(2,0)}(t_f, a)$ represents a set of independent constants, one for each ensemble and flow time. Since $\mu_{p^2}^{(2,0)}(t_f, a)$ are fit parameters, $V_{p^2,\text{fit}}^{(2,0)}(r)$ can be determined only up to a finite additive constant, which cannot be separated from $\mu_{p^2}^{(2,0)}(t_f, a)$ in our analysis. This constant, $U_{p^2}^{(2,0)}$ in Eq.\ (\ref{eq:V20}), is independent of $t_f$ and $a$ and, in the continuum, related to the constant of the static potential via $U_{p^2}^{(2,0)} = -U^{(0)}/4$.
Because the two parameterizations (\ref{EQN_V_p_square_11_para}) and (\ref{EQN_V_p_square_20_para}) contain the same parameter $\alpha_{p^2}$, we have to carry out a combined fit.

As discussed in Sec.\ \ref{sec:basic_parameterization}, we do not include tree level corrections in the fit functions, but subtract them from the lattice data points so that we can perform a straightforward and consistent error propagation. 
Their calculation can be found in App.\ \ref{sec:apdx_tree_level}. 
Since $\Delta V_{p^2}^{(2,0), \text{LO}}(r, t_f, a)$ is $r$ independent, it can be absorbed into $\mu_{p^2}^{(2,0)}(t_f, a)$ and, thus, is not considered explicitly in this subsection.
Similarly, we subtract $\alpha^{(0)} / 2 r$, which is part of the parameterization (\ref{EQN_V_p_square_20_para}).

\begin{figure}[htb]
\centering
\includegraphics[width=0.495\textwidth, page=1]{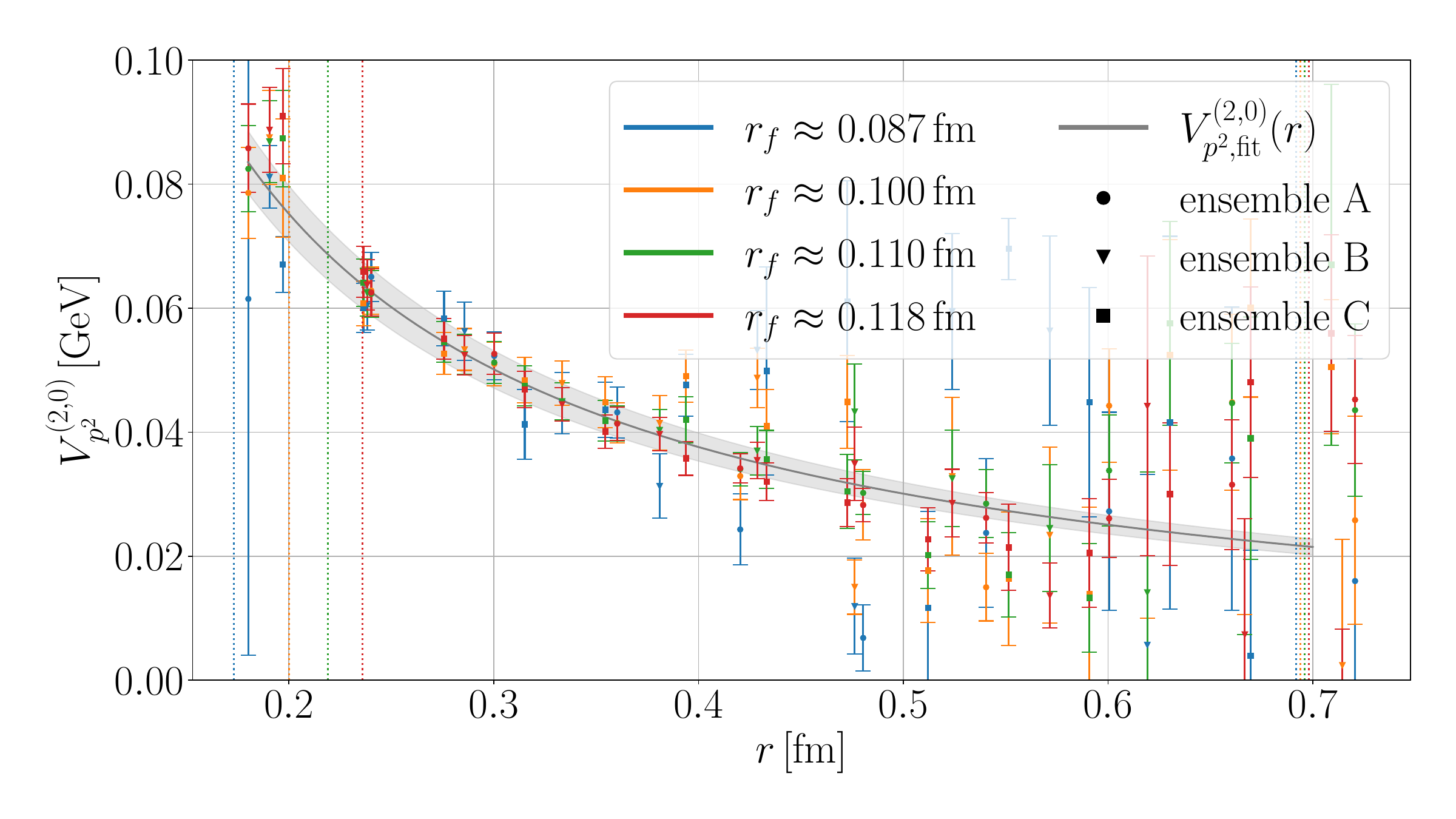}
\includegraphics[width=0.495\textwidth, page=2]{figs/momentum_dependent.pdf}
\caption{
\label{fig:extrapolated_V_p_squared}
Results for the spin independent potentials $V_{p^2}^{(2, 0)}$ and $V_{p^2}^{(1, 1)}$. The curves represent the resulting parameterizations in the continuum and at flow
time zero (\ref{EQN_V_p_square_20_para}) and (\ref{EQN_V_p_square_11_para}). Data points have been corrected by subtracting all terms representing discretization errors and flow time effects as well as the constants $\mu_{p^2}^{(2,0)}(t_f,a)$. Fit ranges for each flow time are indicated by vertical lines in the corresponding
color.}
\end{figure}

To summarize, we fit
\begin{eqnarray}
\label{eq:ansatz_v_p_square_20} & & \hspace{-0.7cm} \mathcal{U}_{p^2}^{(2,0)}(r,t_f,a) = -\frac{\alpha_{p^2}}{2 r} + \mu_{p^2}^{(2,0)}(t_f,a) + \frac{1}{r} \beta_{p^2, \text{flow}}^{(2, 0)} \frac{t_f}{r^2} + \frac{1}{r} \bigg(\beta_{p^2, \text{lattice}}^{(2,0),1a} \frac{a^2}{r^2} + \beta_{p^2, \text{lattice}}^{(2,0),2a} \frac{a^2}{t_f} \bigg)
\end{eqnarray}
to data points
\begin{eqnarray}
& & \hspace{-0.7cm} V_{p^2}^{(2,0)}(r, t_f, a) - \frac{\alpha^{(0)}}{2 r}
\end{eqnarray}
and
\begin{eqnarray}
\label{eq:ansatz_v_p_square_11} & & \hspace{-0.7cm} \mathcal{U}_{p^2}^{(1,1)}(r,t_f,a) = -\frac{\alpha_{p^2}}{r} + \frac{1}{r} \beta_{p^2, \text{flow}}^{(1,1)} \frac{t_f}{r^2} + \frac{1}{r} \bigg(\beta_{p^2, \text{lattice}}^{(1,1),1a} \frac{a^2}{r^2} + \beta_{p^2, \text{lattice}}^{(1,1),2a} \frac{a^2}{t_f}\bigg)
\end{eqnarray}
to data points
\begin{eqnarray}
& & \hspace{-0.7cm} V_{p^2}^{(1,1)}(r, t_f, a) - \Delta V_{p^2}^{(1, 1), \text{LO}}(r, t_f, a) .
\end{eqnarray}
The fit parameters are $\alpha_{p^2}$ and for each ensemble and flow time one $\mu_{p^2}^{(2,0)}(t_f,a)$ as well as three $\beta$ parameters for each potential, describing discretization errors and flow effects.
Results from an uncorrelated fit across all ensembles and flow times in the range $2 r_f \leq r < L/2$ are shown in Fig.~\ref{fig:extrapolated_V_p_squared}. The physically relevant parameter $\alpha_{p^2}$ is collected in Tab.\ \ref{tab:v_spin_independent_results}, together with the corresponding $\chi^2 / \text{DOF}$, which is rather large. This is, however not due to inadequately chosen fit functions (\ref{eq:ansatz_v_p_square_20}) and (\ref{eq:ansatz_v_p_square_11}), but rather due to the partially poor data quality, which manifests itself in individual outlier data points (points that deviate from the majority of data points while simultaneously having small uncertainties). If these outliers, which account for approximately 5\% of all data points, are removed, one obtains $\chi^2 / \text{DOF} = 1.13$, while the resulting continuum and zero flow time parameterizations (\ref{EQN_V_p_square_11_para}) and (\ref{EQN_V_p_square_20_para}) remain essentially unchanged.

\subsubsection{\label{SEC_spin_independent_L_square}Potentials \texorpdfstring{$V_{L^2}^{(2, 0)}$}{VL2(2,0)} and \texorpdfstring{$V_{L^2}^{(1, 1)}$}{VL2(1,1)}}

For the potential $V_{L^2}^{(1, 1)}(r)$ in the continuum and at zero flow time we use the parameterization
\begin{eqnarray}
\label{EQN_V_L_square_11_para} & & \hspace{-0.7cm} V_{L^2,\text{fit}}^{(1, 1)}(r) = \frac{\alpha_{L^2}}{2 r} + \frac{\Lambda_{L^2} r}{6} \, ,
\end{eqnarray}
where the first term is motivated by perturbation theory \cite{Peset:2015vvi,Peset:2018jkf,Anzai:2018eua} and the second term by effective string theory
\cite{Perez-Nadal:2008wtr,Brambilla:2014eaa}.
Since $V_{L^2}^{(1, 1)}$ and $V_{L^2}^{(2, 0)}$ are related by the BBMP relation (\ref{eq:bbmp1}), the parameterization of the latter in the continuum and at zero flow time is fixed by Eq.\ (\ref{EQN_V_L_square_11_para}),
\begin{eqnarray}
\label{EQN_V_L_square_20_para}& & \hspace{-0.7cm} 
V_{L^2,\text{fit}}^{(2, 0)}(r) = 
\frac{1}{2} V_{L^2,\text{fit}}^{(1, 1)}(r) - \frac{r}{4} V_\text{fit}^{(0)\prime}(r) = \frac{\alpha_{L^2}}{4 r} + \frac{\Lambda_{L^2} r}{12} - \bigg(\frac{\alpha^{(0)}}{4 r} + \frac{\sigma r}{4}\bigg) \, .
\end{eqnarray}
Because the two parameterizations (\ref{EQN_V_L_square_11_para}) and (\ref{EQN_V_L_square_20_para}) contain the same parameters $\alpha_{L^2}$ and $\Lambda_{L^2}$, we have to carry out a combined fit.

As discussed in Sec.\ \ref{sec:basic_parameterization}, we do not include tree level corrections in the fit functions, but subtract them from the lattice data points so that we can perform a straightforward and consistent error propagation. 
Their calculation can be found in App.\ \ref{sec:apdx_tree_level}. 
Since $\Delta V_{L^2}^{(2,0), \text{LO}}(r, t_f, a) = 0$, it will not appear in the following.
Similarly, we subtract $-(\alpha^{(0)} / 4 r + \sigma r / 4)$, which is part of the parameterization (\ref{EQN_V_L_square_20_para}).

\begin{figure}[htb]
\centering
\includegraphics[width=0.495\textwidth, page=1]{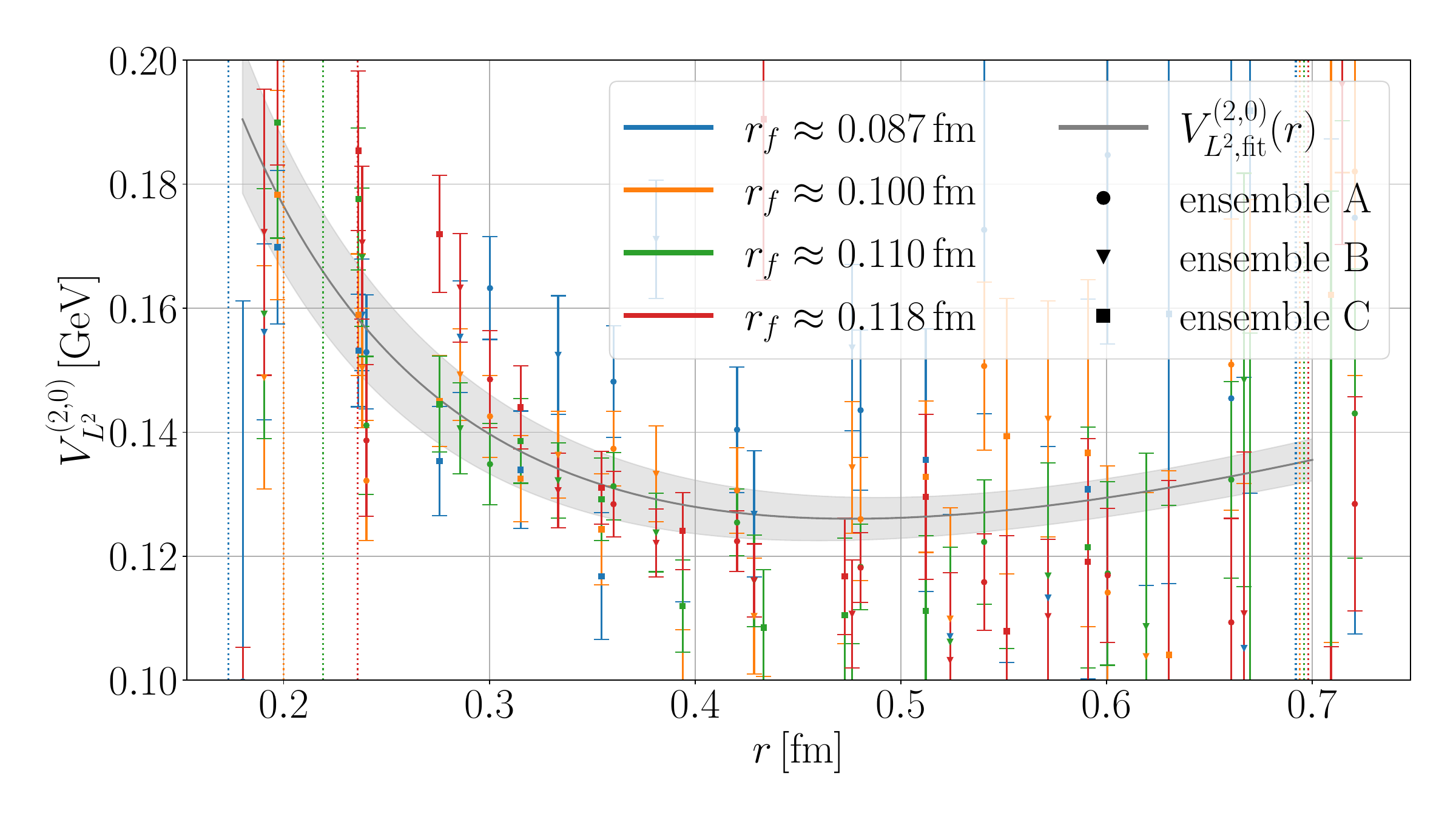}
\includegraphics[width=0.495\textwidth, page=2]{figs/angular_momentum_dependent.pdf}
\caption{
\label{fig:extrapolated_V_L_squared}
Results for the spin independent potentials $V_{L^2}^{(2, 0)}$ and $V_{L^2}^{(1, 1)}$. The curves represent the resulting parameterizations in the continuum and at flow
time zero (\ref{EQN_V_L_square_20_para}) and (\ref{EQN_V_L_square_11_para}). Data points have been corrected by subtracting all terms representing discretization errors and flow time effects. Fit ranges for each flow time are indicated by vertical lines in the corresponding
color.}
\end{figure}

To summarize, we fit
\begin{eqnarray}
\label{eq:ansatz_v_L_square_20} & & \hspace{-0.7cm} \mathcal{U}_{L^2}^{(2,0)}(r,t_f,a) = \frac{\alpha_{L^2}}{4 r} + \frac{\Lambda_{L^2} r}{12} + \frac{1}{r} \beta_{L^2, \text{flow}}^{(2, 0)} \frac{t_f}{r^2} + \frac{1}{r} \bigg(\beta_{L^2, \text{lattice}}^{(2,0),1a} \frac{a^2}{r^2} + \beta_{L^2, \text{lattice}}^{(2,0),2a} \frac{a^2}{t_f} \bigg)
\end{eqnarray}
to data points
\begin{eqnarray}
& & \hspace{-0.7cm} 
V_{L^2}^{(2,0)}(r, t_f, a) + \bigg(\frac{\alpha^{(0)}}{4 r} + \frac{\sigma r}{4}\bigg)
\end{eqnarray}
and
\begin{eqnarray}
\label{eq:ansatz_v_L_square_11} & & \hspace{-0.7cm} \mathcal{U}_{L^2}^{(1,1)}(r,t_f,a) = \frac{\alpha_{L^2}}{2 r} + \frac{\Lambda_{L^2} r}{6} + \frac{1}{r} \beta_{L^2, \text{flow}}^{(1,1)} \frac{t_f}{r^2} + \frac{1}{r} \bigg(\beta_{L^2, \text{lattice}}^{(1,1),1a} \frac{a^2}{r^2} + \beta_{L^2, \text{lattice}}^{(1,1),2a} \frac{a^2}{t_f}\bigg)
\end{eqnarray}
to data points
\begin{eqnarray}
& & \hspace{-0.7cm} V_{L^2}^{(1,1)}(r, t_f, a) - \Delta V_{L^2}^{(1, 1), \text{LO}}(r, t_f, a) .
\end{eqnarray}
The fit parameters are $\alpha_{L^2}$ and $\Lambda_{L^2}$ as well as three $\beta$ parameters for each potential, describing discretization errors and flow effects.
Results from an uncorrelated fit across all ensembles and flow times in the range $2 r_f \leq r < L/2$ are shown in Fig.~\ref{fig:extrapolated_V_L_squared}. The physically relevant parameters $\alpha_{L^2}$ and $\Lambda_{L^2}$ are collected in Tab.\ \ref{tab:v_spin_independent_results}. As in the case of the potentials $V_{p^2}^{(2, 0)}$ and $V_{p^2}^{(1, 1)}$, the corresponding $\chi^2 / \text{DOF} = 2.26$ is large, but can be significantly reduced by removing a small percentage of outlier data points before carrying out the fit.   

\subsection{\label{SEC_spin_orbit}Spin-orbit potentials \texorpdfstring{$V_{LS}^{(2, 0)}$}{VLS(2,0)} and \texorpdfstring{$V_{L_2 S_1}^{(1, 1)}$}{VL2S1(1,1)}}

The spin-orbit potentials $V_{LS}^{(2, 0)}$ and $V_{L_2 S_1}^{(1, 1)}$ are related to integrated field strength correlators containing one chromoelectric and one chromomagnetic insertion (see Eqs.\ (\ref{eq:V_LS20}) and (\ref{eq:V_LS11})). Due to the latter, matching coefficients $c_F$ are required for the extrapolation to the continuum and zero flow time and one has to set $k = 1$ in the general equations provided in Sec.\ \ref{sec:basic_parameterization}.
As already stated in Sec.\ \ref{sec:potential_defs}, we provide continuum and flow time extrapolated potentials in the $\overline{\text{MS}}$ scheme at fixed scale $\mu$, i.e.\ the relevant matching coefficients are $c_F(\mu,t_f)$ (see Eq.\ (\ref{eq:cf_flow})). We use $\mu = 3 \, \text{GeV}$, which is somewhere between the charm and bottom quark mass. To convert the $\overline{\text{MS}}$ potentials to quark mass dependent potentials, which can be used for Born-Oppenheimer studies of quarkonium, i.e.\ potentials $V_{LS}^{(2, 0)}(r,m_j)$ and $V_{L_2 S_1}^{(1, 1)}(r,m_1,m_2)$ appearing in Eqs.\ (\ref{eq:V20}) and (\ref{eq:V11}), one has to multiply the $\overline{\text{MS}}$ results with matching coefficients $c_F^{(m_j)}(\mu)$ (see Eq.\ (\ref{eq:cf_msbar})) and, for $V_{LS}^{(2,0)}$, add the static force term (see Eq.\ (\ref{eq:cf_V_LS20})).

For the potential $V_{L_2 S_1}^{(1, 1)}(r,\mu)$ in the continuum and at zero flow time we use the parameterization
\begin{eqnarray}
\label{EQN_V_L2S1_para} & & \hspace{-0.7cm} V_{L_2 S_1,\text{fit}}^{(1, 1)}(r,\mu) = \frac{\alpha_{LS}(\mu)}{r^3} - \frac{\Lambda_{LS}(\mu)}{r^2} \, ,
\end{eqnarray}
where the first term is motivated by perturbation theory \cite{Peset:2015vvi,Peset:2018jkf,Anzai:2018eua} and the second term by effective string theory \cite{Perez-Nadal:2008wtr,Brambilla:2014eaa}. Since $V_{L_2 S_1}^{(1, 1)}(r,\mu) = V_{LS}^{(2, 0)}(r,\mu)$ due to the Gromes relation (\ref{eq:gromes_b}), the parameterization of the latter in the continuum and at zero flow time is identical to Eq.\ (\ref{EQN_V_L2S1_para}),
\begin{eqnarray}
\label{EQN_V_LS_para} & & \hspace{-0.7cm} V_{LS,\text{fit}}^{(2, 0)}(r,\mu) = \frac{\alpha_{LS}(\mu)}{r^3} - \frac{\Lambda_{LS}(\mu)}{r^2} \, ,
\end{eqnarray}
and we have to carry out a combined fit.

As discussed in Sec.\ \ref{sec:basic_parameterization}, we do not include tree level corrections in the fit functions, but subtract them from the lattice data points so that we can perform a straightforward and consistent error propagation. 
Their calculation can be found in App.\ \ref{sec:apdx_tree_level}.
Note that, according to Eq.\ (\ref{eq:V_LS20}), $V_{LS}^{(2,0)}(r,t_f,a)$ also contains the term $V^{(0)\prime}/r$, for which we use the continuum and zero flow time parameterization of the static potential from Sec.\ \ref{sec:result_static_potential}, i.e.\ $(\alpha^{(0)}/r^2 + \sigma)/r$.
Since the tree level contribution of the field strength correlator in Eq.\ (\ref{eq:V_LS20}) vanishes, $\Delta V_{L S}^{(2,0), \text{LO}}(r, t_f, a) = 0$ and it will not appear in the following.

\begin{figure}[htb]
\centering
\includegraphics[width=0.495\textwidth, page=1]{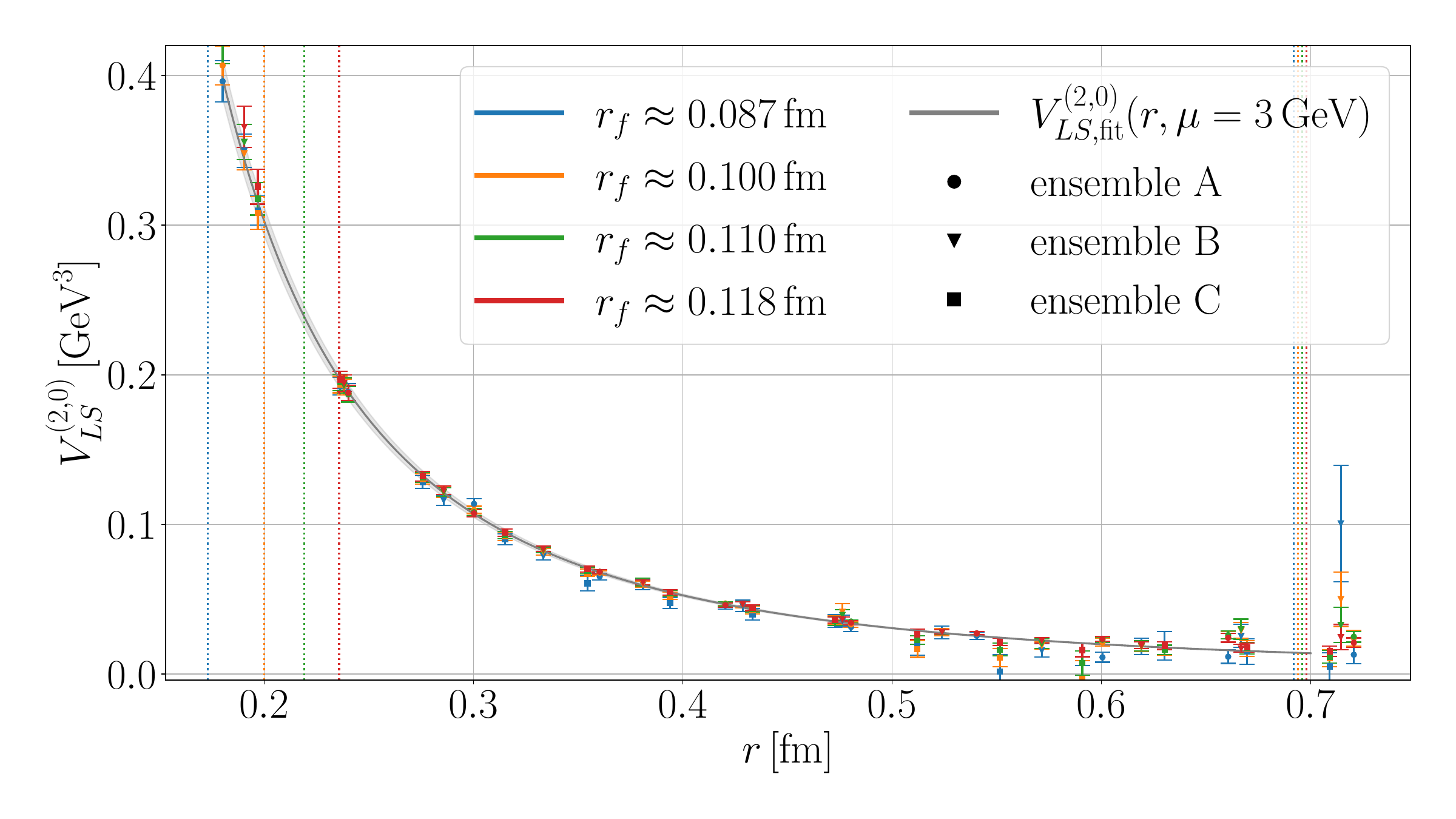}
\includegraphics[width=0.495\textwidth, page=2]{figs/spin_orbit.pdf}
\caption{
\label{fig:extrapolated_V_spinorbit}
Results for the spin-orbit potentials $V_{LS}^{(2, 0)}$ and $V_{L_2 S_1}^{(1, 1)}$ in the $\overline{\text{MS}}$ scheme at scale $\mu = 3 \, \text{GeV}$. The curves represent the resulting parameterizations in the continuum and at flow
time zero (\ref{EQN_V_LS_para}) and (\ref{EQN_V_L2S1_para}). 
Data points have been corrected by subtracting all terms representing discretization errors and flow time effects and multiplied with $c_F(\mu, t_f)$. Fit ranges for each flow time are indicated by vertical lines in the corresponding
color.
}
\end{figure}

To summarize, we fit
\begin{eqnarray}
\nonumber & & \hspace{-0.7cm} \mathcal{U}_{LS}^{(2,0)}(r,t_f,a) =
\Big(c_F(\mu, t_f)\Big)^{-1} \bigg(
\frac{\alpha_{LS}(\mu)}{r^3} - \frac{\Lambda_{LS}(\mu)}{r^2} + \frac{1}{r^3} \beta_{LS, \text{flow}}^{(2, 0)}(\mu) \frac{t_f}{r^2}
\bigg) \\
\label{eq:ansatz_v_ls_20} & &  \hspace{0.675cm} + \frac{1}{r^3} \bigg(\beta_{LS, \text{lattice}}^{(2,0),1}(t_f/r^2) \frac{a^2}{r^2} + \beta_{LS, \text{lattice}}^{(2,0),2}(t_f/r^2) \frac{a^2}{t_f} \bigg)
\end{eqnarray}
to data points
\begin{eqnarray}
& & \hspace{-0.7cm} V_{LS}^{(2,0)}(r, t_f, a)
\label{eq:datapoints_ls_20}
\end{eqnarray}
and
\begin{eqnarray}
\nonumber & & \hspace{-0.7cm} \mathcal{U}_{L_2 S_1}^{(1,1)}(r,t_f,a) =
\Big(c_F(\mu, t_f)\Big)^{-1} \bigg(
\frac{\alpha_{LS}(\mu)}{r^3} - \frac{\Lambda_{LS}(\mu)}{r^2} + \frac{1}{r^3} \beta_{L_2 S_1, \text{flow}}^{(1,1)}(\mu) \frac{t_f}{r^2}
\bigg) \\
\label{eq:ansatz_v_ls_11} & &  \hspace{0.675cm} + \frac{1}{r^3} \bigg(\beta_{L_2 S_1, \text{lattice}}^{(1,1),1}(t_f/r^2) \frac{a^2}{r^2} + \beta_{L_2 S_1, \text{lattice}}^{(1,1),2}(t_f/r^2) \frac{a^2}{t_f} \bigg)
\end{eqnarray}
to data points
\begin{eqnarray}
& & \hspace{-0.7cm} V_{L_2 S_1}^{(1,1)}(r, t_f, a) - \Delta V_{L_2 S_1}^{(1, 1), \text{LO}}(r, t_f, a) .
\end{eqnarray}
The fit parameters are $\alpha_{LS}(\mu)$ and $\Lambda_{LS}(\mu)$ as well as five $\beta$ parameters for each potential, describing discretization errors and flow effects. The argument $\mu$ of $\alpha_{LS}(\mu)$ and $\Lambda_{LS}(\mu)$ (as well as of $\alpha_{S_{12}}(\mu)$, $\alpha_{S_{12}}^\prime(\mu)$, $\alpha_{S^2}(\mu)$, $\alpha_{S^2}^\prime(\mu)$ and $\Lambda_{S^2}(\mu)$ in Sec.~\ref{sec:result_spin_spin}) indicates that the corresponding potentials are given in the $\overline{\text{MS}}$ scheme at fixed scale $\mu = 3 \, \text{GeV}$.
Results from an uncorrelated fit across all ensembles and flow times in the range $2 r_f \leq r < L/2$ are shown in Fig.~\ref{fig:extrapolated_V_spinorbit}. The physically relevant parameters $\alpha_{LS}(\mu)$ and $\Lambda_{LS}(\mu)$ are collected in Tab.\ \ref{tab:v_ls_results}, together with the corresponding $\chi^2 / \text{DOF}$, which is of order 1 and indicates a good fit.

\begin{table}[htb]
	\centering
	\begin{tabular}{c|c|c}
		$\alpha_{LS}(\mu)$ & $\Lambda_{LS}(\mu)\,[\text{GeV}]$ & $\chi^2/\text{DOF}$
		\\
		\hline
		$0.192(13)$ & $-0.1218(43)$ & $1.33$
	\end{tabular}
	\caption{\label{tab:v_ls_results}Fit results for the parameterizations of the spin-orbit potentials (\ref{EQN_V_L2S1_para}) and (\ref{EQN_V_LS_para}) for $\overline{\text{MS}}$ scale $\mu = 3 \, \text{GeV}$. Correlations between all fit parameters in Tabs.~\ref{tab:v0_results} to \ref{tab:v_sq_results} are given in App.~\ref{apdx:correlations}.}
\end{table}


\subsection{\label{sec:result_spin_spin}Spin-spin potentials \texorpdfstring{$V_{S_{12}}^{(1, 1)}$}{VS12(1,1)} and \texorpdfstring{$V_{S^2}^{(1, 1)}$}{VS2(1,1)}}

The spin-spin potentials $V_{S_{12}}^{(1, 1)}$ and $V_{S^2}^{(1, 1)}$ are related to integrated field strength correlators containing two chromomagnetic insertions (see Eqs.\ (\ref{eq:V_ST}) and (\ref{eq:V_Ssq})). Thus, matching coefficients $c_F$ are required for the extrapolation to the continuum and zero flow time and one has to set $k = 2$ in the general equations provided in Sec.\ \ref{sec:basic_parameterization}.
As before, we provide continuum and flow time extrapolated potentials in the $\overline{\text{MS}}$ scheme at fixed scale $\mu = 3 \, \text{GeV}$. To convert the $\overline{\text{MS}}$ potentials to quark mass dependent potentials, which can be used for Born-Oppenheimer studies of quarkonium, i.e.\ potentials $V_{S_{12}}^{(1, 1)}(r,m_1,m_2)$ and $V_{S^2}^{(1, 1)}(r,m_1,m_2)$ appearing in Eqs.\ (\ref{eq:V20}) and (\ref{eq:V11}), one has to multiply the $\overline{\text{MS}}$ results with matching coefficients $c_F^{(m_1)}(\mu) c_F^{(m_2)}(\mu)$ (see Eq.\ (\ref{eq:cf_msbar})) and, for $V_{S^2}^{(1,1)}$, add the contact term (see Eq.\ (\ref{eq:cf_V_Ssq})).

For the potentials $V_{S_{12}}^{(1, 1)}(r,\mu)$ and $V_{S^2}^{(1, 1)}(r,\mu)$ in the continuum and at zero flow time we use the parameterizations
\begin{eqnarray}
\label{EQN_V_S12_para} & & \hspace{-0.7cm} V_{S_{12},\text{fit}}^{(1, 1)}(r,\mu) = \frac{\alpha_{S_{12}}(\mu) + \alpha_{S_{12}}^\prime(\mu) r}{4 r^3 + 90 \alpha_{S_{12}}^\prime(\mu) \Lambda_{S^2}(\mu) r^6 / \pi^3} \\
\label{EQN_V_S_square_para}  & & \hspace{-0.7cm} V_{S^2,\text{fit}}^{(1, 1)}(r,\mu) = \frac{\alpha_{S^2}(\mu) + \alpha_{S^2}^\prime(\mu) r}{r^3 + 45  \alpha_{S^2}^\prime(\mu) \Lambda_{S^2}(\mu) r^6 / 2 \pi^3} \, .
\end{eqnarray}
Both parameterizations are motivated by the corresponding perturbative results $\sim 1/r^3$ at small separations (see Refs.\ \cite{Peset:2015vvi,Peset:2018jkf,Anzai:2018eua}) and the effective string theory prediction
\begin{eqnarray}
\label{EQN955} & & \hspace{-0.7cm} V_{S_{12}}^{(1, 1)}(r,\mu) = \frac{1}{4} V_{S^2}^{(1, 1)} (r,\mu) \propto \frac{1}{r^5}
\end{eqnarray}
at large separations (see Refs.\ \cite{Perez-Nadal:2008wtr,Brambilla:2014eaa}) and chosen such that there is a smooth interpolation between the two regimes. 
Note that there is a common parameter $\Lambda_{S^2}(\mu)$, which determines the respective strength of the long-range $1/r^5$ falloff in the two parameterizations (\ref{EQN_V_S12_para}) and (\ref{EQN_V_S_square_para}) and ensures that their ratio is consistent with Eq.\ (\ref{EQN955}).
Because of this common parameter, we have to carry out a combined fit.

As discussed in Sec.\ \ref{sec:basic_parameterization}, we do not include tree level corrections $\Delta V_{S_{12}}^{(1, 1),\text{LO}}(r, t_f, a)$ and $\Delta V_{S^2}^{(1, 1),\text{LO}}(r, t_f, a)$ in the fit functions, but subtract them from the lattice data points so that we can perform a straightforward and consistent error propagation. 
Their calculation can be found in App.\ \ref{sec:apdx_tree_level}.

\begin{figure}[htb]
\centering
\includegraphics[width=0.495\textwidth, page=1]{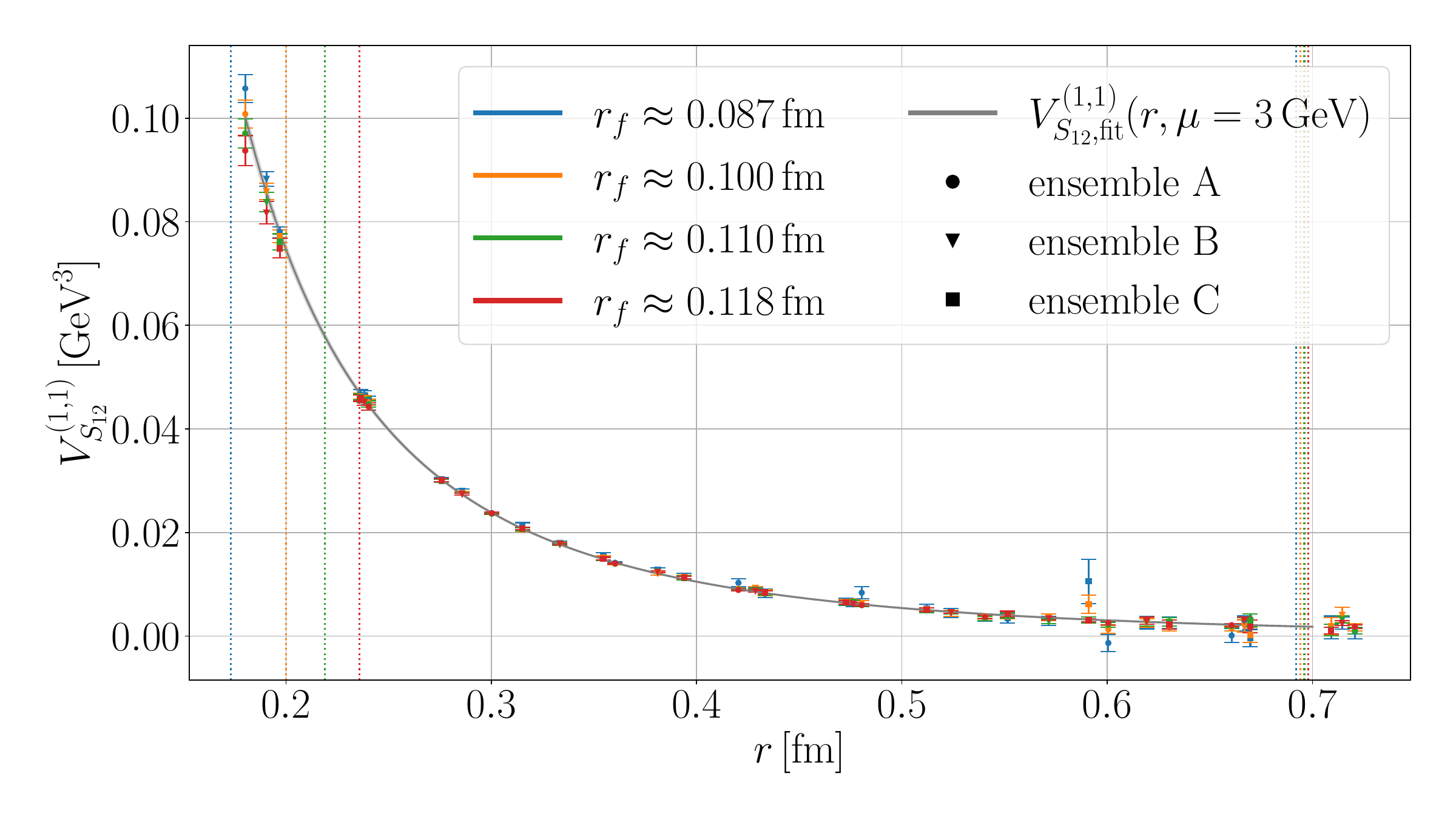}
\includegraphics[width=0.495\textwidth, page=2]{figs/spin_spin.pdf}
\caption{
\label{fig:extrapolated_V_spinspin}
Results for the spin-spin potentials $V_{S_{12}}^{(1, 1)}$ and $V_{S^2}^{(1, 1)}$ in the $\overline{\text{MS}}$ scheme at scale $\mu = 3 \, \text{GeV}$. The curves represent the resulting parameterizations in the continuum and at flow
time zero (\ref{EQN_V_S12_para}) and (\ref{EQN_V_S_square_para}). Data points have been corrected by subtracting all terms representing discretization errors and flow time effects and multiplied with $(c_F(\mu, t_f))^2$. Fit ranges for each flow time are indicated by vertical lines in the corresponding
color.}
\end{figure}

To summarize, we fit
\begin{eqnarray}
\nonumber & & \hspace{-0.7cm} \mathcal{U}_{S_{12}}^{(1,1)}(r,t_f,a) =
\Big(c_F(\mu, t_f)\Big)^{-2} \bigg(
\frac{\alpha_{S_{12}}(\mu) + \alpha_{S_{12}}^\prime(\mu) r}{4 r^3 + 90 \alpha_{S_{12}}^\prime(\mu) \Lambda_{S^2}(\mu) r^6 / \pi^3}
+ \frac{1}{r^3} \beta_{S_{12}, \text{flow}}^{(1,1)}(\mu) \frac{t_f}{r^2}
\bigg) \\
\label{eq:ansatz_v_t12_11} & &  \hspace{0.675cm} + \frac{1}{r^3} \bigg(\beta_{S_{12}, \text{lattice}}^{(1,1),1}(t_f/r^2) \frac{a^2}{r^2} + \beta_{S_{12}, \text{lattice}}^{(1,1),2}(t_f/r^2) \frac{a^2}{t_f} \bigg)
\end{eqnarray}
to data points
\begin{eqnarray}
& & \hspace{-0.7cm} V_{S_{12}}^{(1,1)}(r, t_f, a) - \Delta V_{S_{12}}^{(1, 1), \text{LO}}(r, t_f, a)
\end{eqnarray}
and
\begin{eqnarray}
\nonumber & & \hspace{-0.7cm} \mathcal{U}_{S^2}^{(1,1)}(r,t_f,a) =
\Big(c_F(\mu, t_f)\Big)^{-2} \bigg(
\frac{\alpha_{S^2}(\mu) + \alpha_{S^2}^\prime(\mu) r}{r^3 + 45  \alpha_{S^2}^\prime(\mu) \Lambda_{S^2}(\mu) r^6 / 2 \pi^3}
+ \frac{1}{r^3} \beta_{S^2, \text{flow}}^{(1,1)}(\mu) \frac{t_f}{r^2}
\bigg) \\
\label{eq:ansatz_v_s1s2_11} & &  \hspace{0.675cm} + \frac{1}{r^3} \bigg(\beta_{S^2, \text{lattice}}^{(1,1),1}(t_f/r^2) \frac{a^2}{r^2} + \beta_{S^2, \text{lattice}}^{(1,1),2}(t_f/r^2) \frac{a^2}{t_f} \bigg)
\end{eqnarray}
to data points
\begin{eqnarray}
& & \hspace{-0.7cm} V_{S^2}^{(1,1)}(r, t_f, a) - \Delta V_{S^2}^{(1, 1), \text{LO}}(r, t_f, a) \, .
\end{eqnarray}
The fit parameters are $\alpha_{S_{12}}(\mu)$, $\alpha_{S_{12}}^\prime(\mu)$,
$\alpha_{S^2}(\mu)$, $\alpha_{S^2}^\prime(\mu)$ and $\Lambda_{S^2}(\mu)$ as well as five $\beta$ parameters for each potential, describing discretization errors and flow effects.
Results from an uncorrelated fit across all ensembles and flow times in the range $2 r_f \leq r < L/2$ are shown in Fig.~\ref{fig:extrapolated_V_spinspin}. The physically relevant parameters $\alpha_{S_{12}}(\mu)$, $\alpha_{S_{12}}^\prime(\mu)$,
$\alpha_{S^2}(\mu)$, $\alpha_{S^2}^\prime(\mu)$ and $\Lambda_{S^2}(\mu)$ are collected in Tab.\ \ref{tab:v_sq_results}, together with the corresponding $\chi^2 / \text{DOF}$, which is of order 1 and indicates a good fit.

\begin{table}[htb]
	\centering
	\begin{tabular}{c|c|c|c|c|c}
		$\alpha_{S_{12}}(\mu)$ & $\alpha_{S_{12}}^\prime(\mu)\,[\text{GeV}]$ & $\alpha_{S^2}(\mu)$ & $\alpha_{S^2}^\prime(\mu)\,[\text{GeV}]$ & $\Lambda_{S^2}(\mu)\,[\text{GeV}^{2}]$ & $\chi^2/\text{DOF}$
		\\
		\hline
		$0.243(14)$ & $0.071(12)$ & $-0.052(15)$ & $0.101(10)$ & $0.240(43)$ & $1.50$
	\end{tabular}
	\caption{\label{tab:v_sq_results}Fit results for the parameterizations of the spin-spin potentials $V_{S_{12}}^{(1, 1)}$ and $V_{S^2}^{(1, 1)}$ for $\overline{\text{MS}}$ scale $\mu = 3 \, \text{GeV}$. Correlations between all fit parameters in Tabs.~\ref{tab:v0_results} to \ref{tab:v_sq_results} are given in App.~\ref{apdx:correlations}.}
\end{table}


\FloatBarrier

\subsection{\label{sec:discussion}Discussion of results}


\subsubsection{\label{sec:discussion_mu}\texorpdfstring{$\mu$}{mu} independence of spin-dependent potentials for physical heavy quark masses}

Our results for the spin dependent potentials $V_{LS}^{(2, 0)}$, $V_{L_2S_1}^{(1, 1)}$, $V_{S_{12}}^{(1, 1)}$ and $V_{S^2}^{(1, 1)}$ are provided in the $\overline{\text{MS}}$ scheme at the scale $\mu = 3 \, \text{GeV}$ (see Secs.\ \ref{SEC_spin_orbit} and \ref{sec:result_spin_spin}).
These $\overline{\text{MS}}$ potentials are, however, intermediate quantities without direct physical meaning.
The potentials entering a Born-Oppenheimer study of quarkonium, i.e.\ $V_{LS}^{(2, 0)}(r, m_j)$, $V_{L_2S_1}^{(1, 1)}(r, m_j)$, $V_{S_{12}}^{(1, 1)}(r, m_1, m_2)$ and $V_{S^2}^{(1, 1)}(r, m_1, m_2)$, are obtained via Eqs.\ (\ref{eq:cf_V_LS20}) to (\ref{eq:cf_V_Ssq}) and depend on the heavy quark masses, but should not depend on $\mu$.
In our analysis $\mu$ enters at two places, first via the gradient flow matching coefficient $c_F(\mu, t_f)$, Eq.\ (\ref{eq:cf_flow}), which is used in the extrapolation to zero flow time (see Eq.\ (\ref{EQN_lim_tf_with_cF})), and second via the NRQCD matching coefficient $c_F^{(m_j)}(\mu)$, Eq.\ (\ref{EQN753}) or Eq.\ (\ref{eq:cf_msbar}), which relates the $\overline{\text{MS}}$ potentials to the quark mass dependent potentials.
At order $\alpha_s$ the logarithms of $\mu$ in the two matching coefficients cancel, i.e.\ the product $c_F^{(m_j)}(\mu) \, c_F(\mu, t_f)$ is independent of $\mu$ up to corrections of order $\alpha_s^2$.
This holds both for the NLO expression (\ref{EQN753}) and for the NLL expression (\ref{eq:cf_msbar}) for $c_F^{(m_j)}(\mu)$, which differ from each other only at order $\alpha_s^2$.
Because the matching coefficients are known only to a finite order in perturbation theory, a residual $\mu$ dependence of order $\alpha_s^2$ of the quark mass dependent potentials remains.
The size of this residual dependence provides a consistency check of the perturbative matching. It cannot, however, reveal unknown $\mu$ independent contributions of higher order and, thus, does not provide a complete estimate of the perturbative uncertainty.

\begin{figure}[ht]
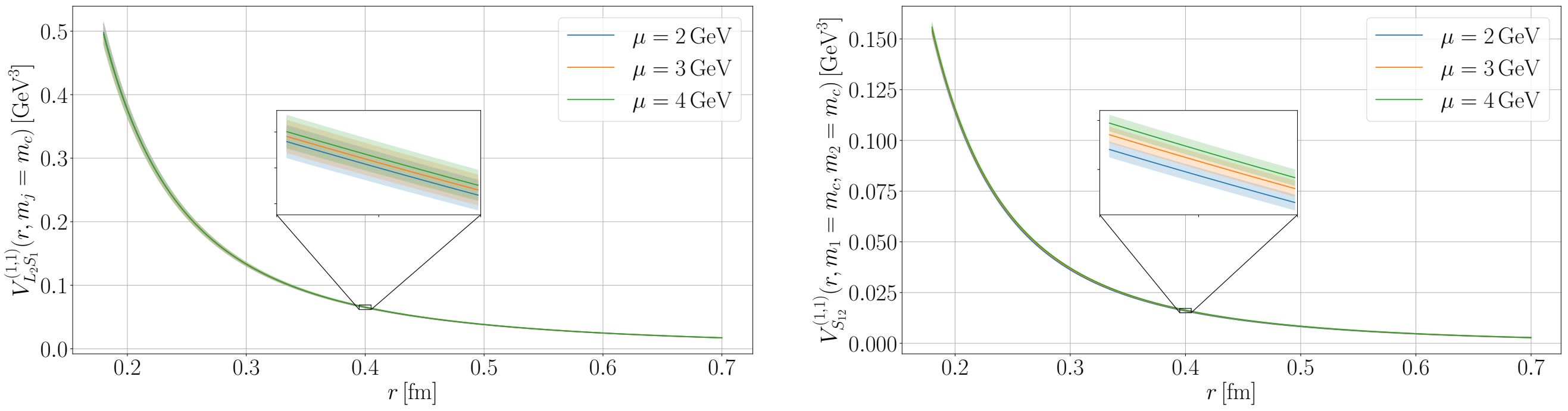

\centering
\includegraphics[width=0.495\textwidth, page=4]{figs/different_matching_mq_spin_orbit.pdf}
\includegraphics[width=0.495\textwidth, page=3]{figs/different_matching_mq_spin_spin.pdf}
\caption{\label{fig:mu_independence}Spin dependent potentials $V_{L_2S_1}^{(1, 1)}(r, m_c)$ (left) and $V_{S_{12}}^{(1, 1)}(r, m_c, m_c)$ (right) for the charm quark pole mass $m_c = 1.67\,\text{GeV}$, obtained from $\overline{\text{MS}}$ results at scales $\mu = 2, 3, 4\,\text{GeV}$. The insets show a magnification of the region around $r = 0.4 \, \text{fm}$.}
\end{figure}

To quantify this residual $\mu$ dependence, we have repeated the complete analysis, in particular the combined continuum and zero flow time extrapolation discussed in Sec.\ \ref{sec:basic_parameterization}, for $\mu = 2 \, \text{GeV}$ and $\mu = 4 \, \text{GeV}$ in addition to $\mu = 3 \, \text{GeV}$, which was used for the results presented in Secs.\ \ref{SEC_spin_orbit} and \ref{sec:result_spin_spin}.
We then converted the three sets of $\overline{\text{MS}}$ potentials to quark mass dependent potentials for the charm quark pole mass $m_c = 1.67 \, \text{GeV}$ and the bottom quark pole mass $m_b = 4.78 \, \text{GeV}$ \cite{ParticleDataGroup:2024cfk}, using the NLL expression (\ref{eq:cf_msbar}) for $c_F^{(m_j)}(\mu)$.
In Fig.\ \ref{fig:mu_independence} we show two examples of the resulting potentials for the charm quark, the spin-orbit potential $V_{L_2S_1}^{(1, 1)}(r, m_c)$, which contains a single chromomagnetic field and, thus, a single matching coefficient $c_F^{(m_c)}(\mu)$, and the tensor potential $V_{S_{12}}^{(1, 1)}(r, m_c, m_c)$, which contains two chromomagnetic fields and, thus, the square of the matching coefficient, $(c_F^{(m_c)}(\mu))^2$.
Both potentials have small statistical uncertainties, and the latter is particularly sensitive to a residual $\mu$ dependence.
We have chosen the charm quark, because it is both the more stringent and the phenomenologically more relevant case: $\alpha_s(m_c)$ is larger than $\alpha_s(m_b)$, i.e.\ perturbative matching is expected to be less precise, and the $1/m_Q^2$ corrections are less suppressed for charmonium than for bottomonium (by a factor $m_b^2/m_c^2 \approx 8$), so that a precise determination of the spin dependent potentials is particularly important for charmonium.

For both potentials the curves for $\mu = 2 \, \text{GeV}$, $3 \, \text{GeV}$ and $4 \, \text{GeV}$ are visually hard to distinguish over the whole range of separations $0.18 \, \text{fm} \ltapprox r \ltapprox 0.7 \, \text{fm}$.
A residual $\mu$ dependence is only visible in the magnified insets, where the three curves are ordered according to their value of $\mu$.
This residual dependence is of the same order as the statistical uncertainties of our results.
As expected, it is somewhat more pronounced for $V_{S_{12}}^{(1, 1)}$ with two matching coefficients than for $V_{L_2S_1}^{(1, 1)}$ with a single matching coefficient.
To estimate the impact of this residual $\mu$ dependence on quarkonium spectra, we consider the difference between the potentials obtained for $\mu = 2 \, \text{GeV}$ and $\mu = 4 \, \text{GeV}$ at $r = 0.4 \, \text{fm}$, a typical quark-antiquark separation for low-lying charmonium states, and multiply it by the suppression factor $1/m_c^2$ appearing in Eq.\ (\ref{eq:VpNRQCD}).
For both $V_{L_2S_1}^{(1, 1)}$ and $V_{S_{12}}^{(1, 1)}$ this yields values well below $1 \, \text{MeV}$, which illustrates that the corresponding effect on quarkonium masses computed within the Born-Oppenheimer approach using our potentials is expected to be small. A quantitative estimate would require computing the resulting changes of these quarkonium masses, i.e.\ solving the Schr\"odinger equation with the potentials obtained for different values of $\mu$, which is beyond the scope of this work.

For the bottom quark as well as for the remaining spin dependent potentials $V_{LS}^{(2, 0)}$ and $V_{S^2}^{(1, 1)}$ (the latter having larger statistical uncertainties), we find the same qualitative behavior and plots very similar to those shown in Fig.\ \ref{fig:mu_independence}.

We conclude that the residual $\mu$ dependence originating from the perturbative matching is small. It is of the same order as the statistical uncertainties of the two most precisely determined spin dependent potentials and, thus, expected to be smaller than the combined statistical and systematic uncertainties of Born-Oppenheimer predictions for quarkonium spectra based on the potentials computed in this work (which also include e.g.\ the larger statistical uncertainties of the other potentials and the truncation of the $1/m_Q$ expansion).

\subsubsection{\label{sec:discussion_literature}Comparison to existing lattice gauge theory studies}

Lattice computations of the potential corrections have been carried out in the past mainly by two groups, by Bali and collaborators \cite{Bali:1997am} and by Koma and collaborators \cite{Koma:2006fw,Koma:2006si,Koma:2007jq,Koma:2009ws,Koma:2012bc}.
Both are pioneering studies, which complement each other.
Ref.\ \cite{Bali:1997am} performs a matching of the spin dependent potentials to QCD, but has comparatively large statistical uncertainties.
Refs.\ \cite{Koma:2006fw,Koma:2006si,Koma:2007jq,Koma:2009ws,Koma:2012bc} achieve very small statistical uncertainties, comparable to ours and in some cases even smaller, using the multilevel algorithm, but do not carry out any matching.
The present work combines the advantages of both approaches:
the spin dependent potentials are matched to QCD, which avoids the systematic error inherent in unmatched lattice results, and this matching is even carried out at higher order than in Ref.\ \cite{Bali:1997am}, while the use of gradient flow leads to small statistical uncertainties, comparable to those obtained with the multilevel algorithm.
Moreover, in contrast to Refs.\ \cite{Bali:1997am,Koma:2006fw,Koma:2006si,Koma:2007jq,Koma:2009ws,Koma:2012bc}, our results are extrapolated to the continuum.

Ref.\ \cite{Bali:1997am} (together with the preceding SU(2) study \cite{Bali:1996cj}) provided the first comprehensive lattice computation of the spin dependent and spin independent $1/m_Q^2$ corrections to the static potential in SU(3) gauge theory (the $1/m_Q$ potential $V^{(1)}$ was not considered) and developed many fundamental methods, which are also used in this work, e.g.\ the spectral decomposition of field strength correlators to compute the large-$t$ integrals and tree level improvement.
For the spin dependent potentials Ref.\ \cite{Bali:1997am} employs matching coefficients with leading logarithmic (LL) accuracy, whereas in this work the gradient flow matching coefficient $c_F(\mu, t_f)$ is used at NLO (Eq.\ (\ref{eq:cf_flow})) and the NRQCD matching coefficient $c_F^{(m_j)}(\mu)$ at NLL (Eq.\ (\ref{eq:cf_msbar})), i.e.\ one order higher in $\alpha_s$.
Refs.\ \cite{Bali:1997am,Koma:2006fw} cover a similar range of separations as this work, up to about $0.7 \, \text{fm}$, while Refs.\ \cite{Koma:2007jq,Koma:2009ws} extend to larger separations of up to $0.9 \, \text{fm}$ and $1.2 \, \text{fm}$, respectively.
Both groups provide results at several lattice spacings (two in Refs.\ \cite{Bali:1997am,Koma:2006fw,Koma:2007jq} and up to four in the scaling study of Ref.\ \cite{Koma:2009ws}), which are, however, used to estimate the size of discretization errors, but not for a continuum extrapolation.
In this work, the continuum extrapolation is part of a combined fit with ans\"atze depending both on the lattice spacing and the flow time (see Sec.\ \ref{sec:basic_parameterization}), i.e.\ discretization errors, an important source of systematic error, are largely eliminated.

Our parameterizations of the potential corrections are motivated by perturbation theory at small separations and by effective string theory at large separations.
It is, however, not our aim to provide parameterizations, which are valid beyond the range of separations covered by our lattice data points; in principle, any other functional form leading to a small $\chi^2/\text{DOF}$ could have been used as well.
As emphasized in Ref.\ \cite{Koma:2006fw}, the same is true for the partly different parameterizations of Refs.\ \cite{Bali:1997am,Koma:2006fw,Koma:2007jq}, which are at large separations guided by the minimal area law \cite{Barchielli:1986zs,Barchielli:1988zp}, i.e.\ by the leading large-distance contribution of effective string theory.
What matters is, thus, not the choice of the functional form, but the comparison of the resulting curves.
Since Refs.\ \cite{Bali:1997am,Koma:2006fw,Koma:2007jq} provide parameterizations only at fixed lattice spacing, we compare our continuum extrapolated results to their parameterizations for $\beta = 6.0$, i.e.\ for lattice spacing $a \approx 0.09 \, \text{fm}$.
For the spin dependent potentials, the comparison is carried out at fixed heavy quark masses, using the quark masses of Ref.\ \cite{Bali:1997am}.
Since Ref.\ \cite{Koma:2006fw} does not perform a matching, we have applied the matching coefficients of Ref.\ \cite{Bali:1997am}, which were derived for lattice results without gradient flow, to the results of Ref.\ \cite{Koma:2006fw} to allow for a meaningful comparison.
This comparison shows qualitative agreement for most of the potentials, i.e.\ the same sign, a similar curvature and a similar behavior at large separations.
Quantitatively, there are deviations ranging from a few percent to a factor of about two, which are typically largest at small separations $r \ltapprox 0.3 \, \text{fm}$.
The most prominent example is $V_{S_{12}}^{(1,1)}$, where the results of Refs.\ \cite{Bali:1997am} and \cite{Koma:2006fw} agree with each other, but are smaller than our result by a factor of around $1.7$.
Since $V_{S_{12}}^{(1,1)}$ is also the potential, for which we observe particularly large discretization errors, and since discretization errors are expected to be largest at small $r/a$, a significant part of these deviations might be caused by the missing continuum extrapolation in Refs.\ \cite{Bali:1997am,Koma:2006fw}.
We conclude that our results are affected by fewer sources of systematic error, because they are extrapolated to the continuum and the matching is carried out at higher order.


\newpage

\section{\label{SEC_conclusions}Conclusions and outlook}

In this work we have computed the $1/m_Q$ potential $V^{(1)}$, the spin independent $1/m_Q^2$ potentials $V_{p^2}^{(2,0)}$, $V_{p^2}^{(1,1)}$, $V_{L^2}^{(2,0)}$ and $V_{L^2}^{(1,1)}$ as well as the spin dependent $1/m_Q^2$ potentials $V_{LS}^{(2,0)}$, $V_{L_2S_1}^{(1,1)}$, $V_{S_{12}}^{(1,1)}$ and $V_{S^2}^{(1,1)}$ of pNRQCD in SU(3) gauge theory, where the latter are provided in the $\overline{\text{MS}}$ scheme at the scale $\mu = 3 \, \text{GeV}$.
The field strength insertions in the generalized Wilson loops were renormalized using gradient flow, which is particularly suited for this purpose:
it leads to small statistical uncertainties and, since the matching coefficient from the gradient flow scheme to the $\overline{\text{MS}}$ scheme is known to NLO \cite{Brambilla:2023vwm} (and recently also to NNLO \cite{Wang:2026toy}), it allows a rigorous matching of the spin dependent potentials.
Based on three lattice spacings and four flow times, all potentials were extrapolated to the continuum and to zero flow time.
Consequently, our results combine small statistical uncertainties with a good control of systematic errors: discretization errors and gradient flow effects are removed by the extrapolations, the results are stable under variations of the fit ranges, and the residual uncertainty associated with the perturbative matching is small (see Sec.\ \ref{sec:discussion_mu}).
As discussed in Sec.\ \ref{sec:discussion_literature}, they thus improve on existing lattice results \cite{Bali:1997am,Koma:2006fw,Koma:2006si,Koma:2007jq,Koma:2009ws,Koma:2012bc}.
They are provided in terms of simple parameterizations (see Eqs.\ (\ref{EQN_Cornell}), (\ref{EQN_V_1_para}), (\ref{EQN_V_p_square_11_para}), (\ref{EQN_V_p_square_20_para}), (\ref{EQN_V_L_square_11_para}), (\ref{EQN_V_L_square_20_para}), (\ref{EQN_V_L2S1_para}), (\ref{EQN_V_LS_para}), (\ref{EQN_V_S12_para}) and (\ref{EQN_V_S_square_para}) with parameters listed in Tabs.\ \ref{tab:v0_results} to \ref{tab:v_sq_results}), which can directly be used in Born-Oppenheimer studies of quarkonium, where the spin dependent potentials can easily be converted to arbitrary heavy quark masses via Eqs.\ (\ref{eq:cf_V_LS20}) to (\ref{eq:cf_V_Ssq}).

Both the method developed in this work and the resulting quarkonium potentials are also relevant for hybrid mesons.
In the Born-Oppenheimer effective field theory \cite{Brambilla:2017uyf} hybrid mesons are described by hybrid static potentials and corresponding corrections proportional to powers of $1/m_Q$, including hybrid spin dependent potentials, as well as by potentials describing the mixing of hybrid mesons and quarkonium.
These hybrid potentials are computed in an ongoing long-term project in our group using exactly the same lattice setup as in this work (first results were recently presented at a conference \cite{deJonge:2026htb}).
Since hybrid mesons and quarkonium states can have the same quantum numbers and mix, the quarkonium potentials presented here are also an essential ingredient for a consistent lattice based Born-Oppenheimer study of hybrid mesons and their mixing with quarkonium.


\newpage

\appendix

\section{\label{sec:apdx_tree_level}Tree level expressions for the static potential and the potential corrections}

In this appendix we collect the tree level expressions for the static potential and the potential corrections defined in Eqs.\ (\ref{eq:V_1}) to (\ref{eq:V_Ssq}) in Sec.\ \ref{sec:potential_definitions}.
At tree level, they are given by one gluon exchange diagrams (see Refs.\ \cite{Bali:1996cj,Bali:1997am}).

\subsection{\label{sec:apdx_tree_level_continuum}Tree level expressions in the continuum}

In the continuum and at zero flow time, the one gluon exchange between a static quark and a static antiquark can be described using e.g.\ the gluon propagator in Feynman gauge,
\begin{eqnarray}
& & \hspace{-0.7cm} \Big\langle A_\mu^a(x) A_\nu^b(y) \Big\rangle^\text{LO}
 =
\delta_{\mu\nu} \delta^{ab}
\frac{1}{4 \pi^2 (x - y)^2}
\, .
\label{eq:propagator_continuum}
\end{eqnarray}

The static potential at separation $\bm{r} = r\bm{e}_z$ is obtained by integrating the temporal component of the propagator (i.e.\ $\mu = \nu = 0$) over the temporal separation $t$ of its two endpoints, which are located on the two temporal Wilson lines representing the static quark and the static antiquark \cite{Bali:1996cj,Bali:1997am},
\begin{eqnarray}
& & \hspace{-0.7cm} V^{(0),\text{LO}}(r)
 =
- 4 \pi C_F \alpha_s \int_{-\infty}^{+\infty} \mathrm{d}t \, \frac{1}{4 \pi^2 (t^2 + r^2)}
=
- \frac{C_F \alpha_s}{r}
\, .
\end{eqnarray}
Contributions, where both endpoints of the gluon propagator are located on the same temporal Wilson line (self energy contributions), are $r$ independent and are absorbed by $\mu^{(0)}(t_f, a)$ (see Sec.~\ref{sec:result_static_potential}), i.e.\ we omit them here.

The potential corrections are given by integrals over field strength correlators, Eqs.\ (\ref{eq:V_1}) to (\ref{eq:V_Ssq}). At tree level, the Wilson lines representing the static quark and antiquark do not contribute to these correlators, which then reduce to two-point functions of the field strength. In this way, one obtains the known tree level expressions (again omitting $r$ independent self energy contributions; see e.g.\ Refs.\ \cite{Pineda:2000sz,Brambilla:2004jw})
\begin{eqnarray}
& & \hspace{-0.7cm} V^{(1),\text{LO}}(r) = 0
\\
& & \hspace{-0.7cm} V_{p^2}^{(2,0),\text{LO}}(r) = 0
\\
& & \hspace{-0.7cm} V_{p^2}^{(1,1),\text{LO}}(r) = -\frac{C_F\alpha_s}{r}
\\
& & \hspace{-0.7cm} V_{L^2}^{(2,0),\text{LO}}(r) = 0
\\
& & \hspace{-0.7cm} V_{L^2}^{(1,1),\text{LO}}(r) = \frac{C_F\alpha_s}{2r}
\\
& & \hspace{-0.7cm} V_{LS}^{(2,0),\text{LO}}(r,m_j) = \frac{2c_F^{(m_j)}(\mu)-1}{2r} V^{(0)\prime,\text{LO}}(r) = \frac{C_F\alpha_s}{2r^3}
\\
& & \hspace{-0.7cm} V_{L_2S_1}^{(1,1),\text{LO}}(r,m_j) = c_F^{(m_j)}(\mu) \frac{C_F\alpha_s}{r^3} = \frac{C_F\alpha_s}{r^3}
\\
& & \hspace{-0.7cm} V_{S_{12}}^{(1,1),\text{LO}}(r,m_1,m_2) = c_F^{(m_1)}(\mu) c_F^{(m_2)}(\mu) \frac{C_F \alpha_s}{4 r^3} = \frac{C_F \alpha_s}{4 r^3}
\\
& & \hspace{-0.7cm} V_{S^2}^{(1,1),\text{LO}}(r,m_1,m_2) = c_F^{(m_1)}(\mu) c_F^{(m_2)}(\mu) \frac{8 \pi C_F\alpha_s}{3} \delta^{(3)}(\bm{r}) = \frac{8 \pi C_F\alpha_s}{3} \delta^{(3)}(\bm{r})
\,.
\end{eqnarray}
At tree level, $c_F^{(m_j)}(\mu)$ can be set to $1$, since $c_F^{(m_j)}(\mu) = 1 + \mathcal{O}(\alpha_s)$. The contact terms proportional to $d_{sv}$ and $d_{vv}$ in Eq.\ (\ref{eq:cf_V_Ssq}) are not considered.

\subsection{\label{sec:apdx_tree_level_lattice}Tree level expressions at finite flow time and lattice spacing}

At finite flow time $t_f$ and finite lattice spacing $a$, the analog of Eq.\ (\ref{eq:propagator_continuum}) for the flowed gauge field $B_\mu$ (see Sec.\ \ref{sec:gradient_flow}) is \cite{Fodor:2014cpa}
\begin{eqnarray}
& & \hspace{-0.7cm} \Big\langle B_\mu^a(x, t_f, a) B_\nu^b(y, t_f, a) \Big\rangle^\text{LO}
 =
\delta_{\mu\nu} \delta^{ab} \int_{[-\pi/a, \pi/a)^4} \frac{\mathrm{d}^4 p}{(2\pi)^4} e^{i p(x-y) - 2 t_f \mathcal{D}_\mathrm{GF}^{-1}(p, a)} \mathcal{D}_\mathrm{MC}(p, a)
\,,
\label{eq:flow_propagator_lattice}
\end{eqnarray}
where $\mathcal{D}_\mathrm{MC}(p, a)$ denotes the momentum space gluon propagator in Feynman gauge for the action used in the Monte Carlo simulation and $\mathcal{D}_\mathrm{GF}(p, a)$ the corresponding inverse lattice Laplacian for the action used in the flow equation.
In this work, we use the Wilson plaquette action in both cases, i.e.\ $\mathcal{D}_\mathrm{GF}^{-1}(p, a) = \mathcal{D}_\mathrm{MC}^{-1}(p, a) = (4/a^2) \sum_\mu \sin^2(a p_\mu / 2)$.

Analogously to App.\ \ref{sec:apdx_tree_level_continuum}, the tree level expression for the static potential at finite flow time $t_f$ and finite lattice spacing $a$ is obtained by summing the temporal component of Eq.\ (\ref{eq:flow_propagator_lattice}) over the temporal separation $t$ of its two endpoints, which restricts the momentum integration to $p_0 = 0$ (see also Ref.\ \cite{Brambilla:2023fsi}),
\begin{eqnarray}
& & \hspace{-0.7cm} V^{(0), \text{LO}}(r, t_f, a)
 =
- 4\pi C_F \alpha_s G(r, t_f, a)
\, , \quad
G(r, t_f, a)
=
\int_{[-\pi/a, \pi/a)^3} \frac{\mathrm{d}^3 p}{(2\pi)^3} e^{i \bm{p}\bm{r} - 2 t_f \mathcal{D}_\mathrm{GF}^{-1}(p, a)} \mathcal{D}_\mathrm{MC}(p, a) \bigg\vert_{p_0 = 0}
\, .
\label{eq:G_lattice}
\end{eqnarray}
Note that a straightforward evaluation of Eq.\ (\ref{eq:G_lattice}) is numerically inefficient due to the integrable singularity at the origin.
We avoid this issue as follows.
Consider the first octant of the spatial integration domain and the following decomposition:
\begin{eqnarray}
& & \hspace{-0.7cm} V = \lbrace \bm{p}: 0 \leq p_1, p_2, p_3 < \pi/a \rbrace
\nonumber
\\
& & \hspace{-0.7cm}\hphantom{V} =
\underbrace{\lbrace\bm{p}: 0 \leq p_2,p_3 \leq p_1 < \pi/a \rbrace}_{=V_1}
\cup
\underbrace{\lbrace\bm{p}: 0 \leq p_1,p_3 \leq p_2 < \pi/a \rbrace}_{=V_2}
\cup
\underbrace{\lbrace\bm{p}: 0 \leq p_1,p_2 \leq p_3 < \pi/a \rbrace}_{=V_3}
\,.
\end{eqnarray}
A transformation leading to a stable integral over the $V_3$ domain is
\begin{eqnarray}
& & \hspace{-0.7cm} p_1 = uw\,,\quad p_2 = vw\,, \quad p_3 = w\,,
\end{eqnarray}
with Jacobian $w^2$ and domain $\lbrace (u,v,w): 0\leq u,v \leq 1; 0\leq w < \pi/a \rbrace$.
In these variables and at $p_0 = 0$,
\begin{eqnarray}
& & \hspace{-0.7cm} \mathcal{D}_\mathrm{GF}^{-1}(u,v,w,a) = \mathcal{D}_\mathrm{MC}^{-1}(u,v,w,a) = (4/a^2) \big(\sin^2(auw/2) + \sin^2(avw/2) + \sin^2(aw/2)\big)
\, ,
\end{eqnarray}
and the factor $w^2 \mathcal{D}_\mathrm{MC}$ in the integrand of Eq.\ (\ref{eq:G_lattice}) becomes finite,
\begin{eqnarray}
& & \hspace{-0.7cm} \frac{w^2}{\mathcal{D}_\mathrm{MC}^{-1}(u,v,w,a)}
=\frac{1}{u^2\operatorname{sinc}^2(auw/2) + v^2\operatorname{sinc}^2(avw/2) + \operatorname{sinc}^2(aw/2)}
\end{eqnarray}
with $\operatorname{sinc}(x) = \sin(x)/x$.
Finite integrands for the domains $V_1$ and $V_2$ are obtained analogously.
Finally, reflection properties of Eq.\ (\ref{eq:G_lattice}) are used to reduce the integration domain to the first octant $V$.
One obtains
\begin{eqnarray}
& & \hspace{-0.7cm} G(r, t_f, a)
=
\frac{1}{\pi^3} \int_0^{\pi/a} \mathrm{d}w \int_0^1 \mathrm{d}v \int_0^1 \mathrm{d}u
\frac{\cos(w r) + \cos(vw r) + \cos(uw r)}{u^2\operatorname{sinc}^2(auw/2) + v^2\operatorname{sinc}^2(avw/2) + \operatorname{sinc}^2(aw/2)} e^{-2t_f \mathcal{D}_\text{GF}^{-1}(u,v,w,a)}
\,.
\label{eq:flow_propagator}
\end{eqnarray}
We compute this integral numerically using multi-quadrature methods implemented in the \texttt{SciPy} library \cite{Virtanen:2019joe}.

To our knowledge, tree level expressions for the potential corrections at finite flow time and finite lattice spacing have not been computed before. Their computation is, however, straightforward. They are given by integrals of the type (\ref{eq:general_correlator_integral}) over tree level field strength correlators. Since the latter are not affected by statistical errors, we do not use the splitting (\ref{EQN_I_split}), but compute these integrals in the full range $0 \leq t < \infty$ using the composite Simpson's rule (\ref{eq:integral_simpson}), i.e.\ in the same way as $I^{(s)}(0,\tau)$ for the lattice data. In practice, the upper integration limit $\infty$ is replaced by $t_\text{max} = 16a, 20a, 24a$ for the ensembles A, B and C, respectively. The remaining contribution from $t_\text{max}$ to $\infty$ is not neglected, but approximated by the corresponding contribution in the continuum at the same flow time, which is computed from the continuum analog of Eq.\ (\ref{eq:GG_lattice}). In this way, only the difference between the lattice and the continuum contributions for $t > t_\text{max}$ is neglected, which we have checked to be negligible compared to the statistical uncertainties of the lattice data points $V_X(r,t_f,a)$ for the potential corrections and of the parameters of their parameterizations determined in Secs.~\ref{SEC_spin_independent} to \ref{sec:result_spin_spin}.
As in the continuum, the field strength correlators reduce at tree level to two-point functions of the field strength \cite{Bali:1996cj,Bali:1997am},
\begin{eqnarray}
& & \hspace{-0.7cm} \bra{0, \Sigma_g^+} gF_{\mu\nu}^{0,r}(t) gF_{\rho\sigma}^0(0) \ket{0, \Sigma_g^+}^{\text{LO}}(r,t_f,a)
=
\Big\langle gG_{\mu\nu}(x, t_f, a) gG_{\rho\sigma}(y, t_f, a) \Big\rangle^\text{LO}
\, ,
\end{eqnarray}
where $x=(t, \bm{0})$ or $x=(t, \bm{r})$, depending on the potential, and $y=0$.
For the clover leaf definition of the field strength (see Sec.\ \ref{sec:analysis_details}), linearized in the flowed gauge field $B_\mu$,
\begin{eqnarray}
\nonumber & & \hspace{-0.7cm} G_{\mu\nu}(x, t_f, a) = \frac{1}{4a} \Big(B_\nu(x+ae_\mu) + B_\nu(x+ae_\mu-ae_\nu) - B_\nu(x-ae_\mu) - B_\nu(x-ae_\mu-ae_\nu)
\\
& & \hspace{0.675cm} - B_\mu(x+ae_\nu) - B_\mu(x+ae_\nu-ae_\mu) + B_\mu(x-ae_\nu) + B_\mu(x-ae_\nu-ae_\mu)\Big)
\end{eqnarray}
(where $B_\nu(x) \equiv B_\nu(x, t_f, a)$ is associated with the link from $x$ to $x + a e_\nu$), this two-point function can be expressed in terms of Eq.\ (\ref{eq:flow_propagator_lattice}),
\begin{eqnarray}
& & \hspace{-0.7cm} \Big\langle gG_{\mu\nu}(x, t_f, a) gG_{\rho\sigma}(y, t_f, a) \Big\rangle^\text{LO}
=
4\pi C_F \alpha_s \int_{[-\pi/a,\pi/a)^4} \frac{\mathrm{d}^4p}{(2\pi)^4} e^{ip(x-y) -2t_f \mathcal{D}_\text{GF}^{-1}(p,a)}  \mathcal{D}_\text{MC}(p,a) \frac{1}{a^2} K_{\mu\nu\rho\sigma}(pa)\,,
\label{eq:GG_lattice}
\end{eqnarray}
where
\begin{eqnarray}
\nonumber & & \hspace{-0.7cm} K_{\mu\nu\rho\sigma}(n) =
\sin(n_\mu)\sin(n_\rho) \cos(n_\nu/2)\cos(n_\sigma/2) \delta_{\nu\sigma}
-\sin(n_\mu)\sin(n_\sigma) \cos(n_\nu/2)\cos(n_\rho/2) \delta_{\nu\rho}
\\
& & \hspace{0.675cm} +\sin(n_\nu)\sin(n_\sigma) \cos(n_\mu/2)\cos(n_\rho/2) \delta_{\mu\rho}
-\sin(n_\nu)\sin(n_\rho) \cos(n_\mu/2)\cos(n_\sigma/2) \delta_{\mu\sigma}
\, .
\end{eqnarray}
In contrast to Eq.\ (\ref{eq:G_lattice}), the integrand of Eq.\ (\ref{eq:GG_lattice}) has no singularity at $p = 0$, since $K_{\mu\nu\rho\sigma}(pa) = \mathcal{O}(a^2 p^2)$. Close to $p = 0$, however, the integrand is a ratio of two small quantities, which renders a direct evaluation unstable. We therefore use the same coordinate transformation as for the integral in Eq.\ (\ref{eq:G_lattice}).


\section{\label{apdx:correlator_fits}Selection of fit ranges for the correlated fits of the spectral decompositions}

As discussed in Sec.~\ref{SEC_potential_corrections}, we determine the
parameters $D_{n,\Lambda_\eta^\epsilon}$ and $\Delta V_{n,\Lambda_\eta^\epsilon}$
of the truncated spectral decompositions $I_\text{spectral}^{(s)}(t_1,t_2)$ (Eqs.~(\ref{eq:rootfct_t0}) to (\ref{eq:rootfct_t2}))
by correlated fits to
numerical integrals of field strength correlators. The data points
entering a fit are
\begin{eqnarray}
& & \hspace{-0.7cm} I_\text{Simpson}^{(s)}(t_1,t_2) \, , \quad t_{2,\text{min}} \leq t_2 \leq t_{2,\text{max}} \, ,
\end{eqnarray}
where the lower integration bound $t_1$ is fixed, while the upper
integration bound $t_2$ runs over all values in the fit range. In the following, we discuss how we select $t_1$ and the fit
range $t_{2,\text{min}} \leq t_2 \leq t_{2,\text{max}}$.
We first restrict the candidate fit ranges, which are specified by $t_1$, $t_{2,\text{min}}$ and $t_{2,\text{max}}$, by the following conditions, which do not depend on the lattice data:

{\begin{itemize}
\item The minimum upper integration bound, i.e.\ the start of the fit range, is $t_{2,\text{min}} = t_1 + 2a$, such that each data point $I_\text{Simpson}^{(s)}(t_1,t_2)$ is computed from at least three lattice data points of the field strength correlator, as required for Simpson's rule.

\item $t_{2,\text{max}} \geq t_{2,\text{min}} + 7a$, i.e.\ at least eight data points enter the fit, twice as many as there are fit parameters ($D_{n,\Lambda_\eta^\epsilon}$ and $\Delta V_{n,\Lambda_\eta^\epsilon}$ for two states).

\item $t_{2,\text{min}} \leq \tau$ and $t_{2,\text{max}} \geq \tau$, where $\tau$ is the boundary between the small and the large $t$ integrals in Eq.~(\ref{EQN_I_split}) ($\tau / a = 8, 10, 12$ for ensembles A, B, C, see Sec.~\ref{SEC_potential_corrections}). In this way, the fit range contains data points with $t_2 \leq \tau$, i.e.\ in the region already covered by the Simpson integration of $I^{(s)}(0,\tau)$, which ensures that the spectral decomposition connects smoothly to the small $t$ integral, as well as data points with $t_2 \geq \tau$, which constrain the behavior of the spectral decomposition in the region relevant for $I^{(s)}(\tau,\infty)$.

\item For field strength correlators containing one or two chromoelectric fields, we require $t_1 \geq a$ or $t_1 \geq 2a$, respectively. Since the clover leaf discretization of a chromoelectric field extends over $\pm a$ in temporal direction, this ensures that the two field strength insertions do not overlap in time, i.e.\ are time ordered, as required for the identification of path integral expectation values with time-ordered matrix elements and, thus, for the spectral decomposition.

\item We exclude lattice data points, for which the two field strength insertions, smeared by the application of gradient flow, have a sizable overlap (similar to the static potential, where we require that the two spatial parallel transporters of the Wilson loop are separated in time by at least $2 r_f$, see Sec.~\ref{SEC467}). The corresponding criterion depends on whether both field strengths are inserted on the same temporal line or on opposite temporal lines of the Wilson loop (see Fig.~\ref{fig:skizze}), on whether they are chromoelectric or chromomagnetic fields and on their orientation. It can be written compactly as $d(t_1,r,a) \geq 2 r_f$, where the geometrical separation $d(t_1,r,a)$ is specified for each of the considered correlators in Table~\ref{tab:integral_fit_separations}. Since $d$ grows with the temporal separation of the field strength insertions, this condition is then not only satisfied at $t_1$, but for all correlators entering $I_\text{Simpson}^{(s)}(t_1,t_2)$.
\end{itemize}
}

\begin{table}[tbp]
\centering
\renewcommand{\arraystretch}{1.25}
\begin{tabular}{cc}
\hline
field strength insertions & $d(t_1,r,a)$ \\
\hline
$B_z^r B_z^0$
 & $\sqrt{r^2+t_1^2}$ \\
$B_x^r B_x^0$
 & $\sqrt{(r-2a)^2+t_1^2}$ \\
$E_x^r E_x^0$
 & $\sqrt{r^2+(t_1-2a)^2}$ \\
$E_z^r E_z^0$
 & $\sqrt{(r-2a)^2+(t_1-2a)^2}$ \\
$B_x^r E_y^0$
 & $\sqrt{(r-a)^2+(t_1-a)^2}$ \\
$E_x^0 E_x^0$, $E_z^0 E_z^0$
 & $t_1-2a$ \\
$B_x^0 E_y^0$
 & $t_1-a$ \\
\hline
\end{tabular}
\caption{
Geometrical separations $d(t_1,r,a)$ in the condition $d(t_1,r,a) \geq 2r_f$ for the considered field strength correlators.
}
\label{tab:integral_fit_separations}
\end{table}

For all candidate fit ranges satisfying these conditions, we perform fully correlated fits.
Taking correlations into account is particularly important here, since the data points $I_\text{Simpson}^{(s)}(t_1,t_2)$ are integrals with a common lower bound $t_1$: field strength correlators at $t \leq t_{2,\text{min}}$ enter all data points of a fit, while those at $t > t_{2,\text{min}}$ enter fewer and fewer data points with increasing $t$.
An uncorrelated fit would, therefore, assign an unnaturally large weight to the field strength correlators at small $t$.
We exclude fit ranges, for which the covariance matrix cannot be estimated in a numerically stable way.

The fits are stabilized by Gaussian priors for the potential
differences $\Delta V_{n,\Lambda_\eta^\epsilon}(r,t_f,a)$, as detailed in
App.~\ref{apdx:priors}.
The $\chi^2$ minimized in the fits consists of two parts,
\begin{eqnarray}
& & \hspace{-0.7cm} \chi^2 = \chi^2_\text{data} + \chi^2_\text{prior} \, .
\label{eq:chi2_total}
\end{eqnarray}
$\chi^2_\text{data}$ is the correlated $\chi^2$ quantifying the deviation of the fit function $I_\text{spectral}^{(s)}$ from the data points $I_\text{Simpson}^{(s)}$,
\begin{eqnarray}
& & \hspace{-0.7cm} \chi^2_\text{data} = \sum_{t_2, t_2^\prime} \delta I^{(s)}(t_1,t_2) \Big(\operatorname{Cov}^{-1}\Big)_{t_2 t_2^\prime} \delta I^{(s)}(t_1,t_2^\prime) \, ,
\label{eq:chi2_data}
\end{eqnarray}
where $\delta I^{(s)}(t_1,t_2) = I_\text{Simpson}^{(s)}(t_1,t_2) - I_\text{spectral}^{(s)}(t_1,t_2)$, $\operatorname{Cov}$ is the covariance matrix of the data points and $t_2, t_2^\prime$ run over the fit range.
$\chi^2_\text{prior}$ is the uncorrelated $\chi^2$ of the priors,
\begin{eqnarray}
& & \hspace{-0.7cm} \chi^2_\text{prior} = \sum_{n,\Lambda_\eta^\epsilon} \frac{\Big(\Delta V_{n,\Lambda_\eta^\epsilon} - \Delta V_{n,\Lambda_\eta^\epsilon}^\text{prior}\Big)^2}{\Big(\sigma_{n,\Lambda_\eta^\epsilon}^\text{prior}\Big)^2} \, ,
\label{eq:chi2_prior}
\end{eqnarray}
with mean values $\Delta V_{n,\Lambda_\eta^\epsilon}^\text{prior}$ and widths $\sigma_{n,\Lambda_\eta^\epsilon}^\text{prior}$ from App.~\ref{apdx:priors}. It contains two terms, one for each of the two potential differences.
To use $\chi^2$ as a quality criterion, we divide it by the number of degrees of freedom,
\begin{eqnarray}
& & \hspace{-0.7cm} \text{DOF} = \bigg(\frac{t_{2,\text{max}} - t_{2,\text{min}}}{a} + 1\bigg) + 2 - 4 \, .
\label{eq:integral_fit_dof}
\end{eqnarray}
The first term is the number of data points, $+2$ accounts for the two priors, which correspond to two additional data points, and $-4$ for the four fit parameters. For fits without priors (see App.~\ref{apdx:priors}), $\chi^2_\text{prior}$ in Eq.~(\ref{eq:chi2_total}) and the term $+2$ in Eq.~(\ref{eq:integral_fit_dof}) are absent.

Among all fits with $\chi^2/\text{DOF}<1$, we select the fit with the
largest number of data points. If several such fits have the same number of data points, we select the one with the smallest $t_{2,\text{min}}$ (or, equivalently, the smallest $t_1$).
If no fit has $\chi^2/\text{DOF}<1$, we select the fit with the
smallest $\chi^2/\text{DOF}$.

We apply this selection independently to each integral $I^{(s)}$ appearing in the potential corrections, i.e.\ separately for each combination of field strength correlator and power $s$, quark-antiquark separation, flow time and lattice spacing.


\section{\label{apdx:priors}Gaussian priors for static potential differences}

To stabilize the correlated fits described in
Sec.~\ref{SEC_potential_corrections} and App.~\ref{apdx:correlator_fits}, we use Gaussian priors for the potential
differences $\Delta V_{n,\Lambda_\eta^\epsilon}(r,t_f,a)$. The priors are
constructed from (hybrid) static potentials computed in
Ref.~\cite{Capitani:2018rox}. This reference provides the ground-state
potentials in the contributing $\Sigma$ and $\Pi$ sectors as well as
the first excited potential in the $\Sigma_g^+$ sector. Thus, all
potential differences needed for the two retained states
in each spectral fit are available, including the $n=1$, $\Sigma_g^+$ contribution
used for $E_z$ correlators.

These potentials were, however, computed at a different lattice
spacing and with HYP2-smeared temporal links rather than with gradient flow.
To obtain from them approximations of the corresponding potential differences
in our lattice setup, i.e.\ with gradient flow and at the lattice spacings of
the ensembles A, B and C, we proceed as follows.
First, we apply tree level improvement to the HYP2 results to obtain
approximate continuum results. Then we interpolate these to the separations
needed in our computations. Finally, we apply tree level improvement again,
but in the reverse direction, to supplement the results with the expected
discretization and flow effects of our setup.

The reference data were computed with HYP2-smeared temporal gauge
links at lattice spacing $a_\text{Ref.~\cite{Capitani:2018rox}}\approx 0.093\,\text{fm}$ and
on-axis quark-antiquark separations $r_\text{Ref.~\cite{Capitani:2018rox}} = \hat r_\text{Ref.~\cite{Capitani:2018rox}} a_\text{Ref.~\cite{Capitani:2018rox}}$,
where $\hat r_\text{Ref.~\cite{Capitani:2018rox}}$ is an integer. We first compute
\begin{eqnarray}
& & \hspace{-0.7cm} \Delta V^\text{HYP2}_{n,\Lambda_\eta^\epsilon}(r_\text{Ref.~\cite{Capitani:2018rox}})
 =
V^\text{HYP2}_{n,\Lambda_\eta^\epsilon}(r_\text{Ref.~\cite{Capitani:2018rox}})
-
V^\text{HYP2}_{0,\Sigma_g^+}(r_\text{Ref.~\cite{Capitani:2018rox}}) \, .
\label{eq:prior_reference_gaps}
\end{eqnarray}
The statistical uncertainties of these differences are determined
using linear error propagation, neglecting correlations between
$V^\text{HYP2}_{n,\Lambda_\eta^\epsilon}(r_\text{Ref.~\cite{Capitani:2018rox}})$ and $V^\text{HYP2}_{0,\Sigma_g^+}(r_\text{Ref.~\cite{Capitani:2018rox}})$,
i.e.\ by adding their uncertainties in quadrature.

For tree level improvement we use improved separations
$r_\text{improved}^\text{HYP2}$ defined through the corresponding tree
level propagators, following the approach of
Refs.~\cite{Sommer:1993ce,Necco:2001xg}. We define such improved separations $r_\text{improved}^\text{HYP2}$ for all reference separations $r_\text{Ref.~\cite{Capitani:2018rox}}$ via

\begin{eqnarray}
& & \hspace{-0.7cm} \frac{1}{4\pi r_\text{improved}^\text{HYP2}}
 =
\frac{G_\text{HYP2}(0,0,\hat r_\text{Ref.~\cite{Capitani:2018rox}})}{a_\text{Ref.~\cite{Capitani:2018rox}}} \, .
\label{eq:improved_r_hyp2}
\end{eqnarray}
Here, $G_\text{HYP2}$ denotes the tree level lattice propagator
appropriate to HYP2-smeared temporal links; detailed expressions are
given in Ref.~\cite{Schlosser:2021wnr}.

Up to corrections beyond tree level, the lattice potential differences
(\ref{eq:prior_reference_gaps}) at separation $r_\text{Ref.~\cite{Capitani:2018rox}}$ are equal to
the corresponding continuum potential differences at the improved separation
$r_\text{improved}^\text{HYP2}$ defined in Eq.~(\ref{eq:improved_r_hyp2}),
\begin{eqnarray}
& & \hspace{-0.7cm} \Delta V^\text{HYP2}_{n,\Lambda_\eta^\epsilon}(r_\text{Ref.~\cite{Capitani:2018rox}})
 \approx
\Delta V_{n,\Lambda_\eta^\epsilon}(r_\text{improved}^\text{HYP2}) \, .
\label{eq:prior_tree_level_matching}
\end{eqnarray}
To obtain potential differences at the separations determined by the
lattices of the ensembles A, B and C, these continuum-like results have to be interpolated.
Since the potential differences multiplied by the separation are smoother
functions than the potential differences themselves, we multiply
$\Delta V^\text{HYP2}_{n,\Lambda_\eta^\epsilon}(r_\text{Ref.~\cite{Capitani:2018rox}})$, which is
approximately the continuum result
$\Delta V_{n,\Lambda_\eta^\epsilon}(r_\text{improved}^\text{HYP2})$,
by the corresponding separation $r_\text{improved}^\text{HYP2}$ and
interpolate the data points
\begin{eqnarray}
& & \hspace{-0.7cm} \Big(r_\text{improved}^\text{HYP2} \, , \, r_\text{improved}^\text{HYP2} \, \Delta V^\text{HYP2}_{n,\Lambda_\eta^\epsilon}(r_\text{Ref.~\cite{Capitani:2018rox}})\Big)
\label{eq:prior_interpolated_quantity}
\end{eqnarray}
by a cubic spline, which we denote by
$r \, \Delta V_{n,\Lambda_\eta^\epsilon,\text{fit}}(r)$.
The statistical uncertainties of the data points are interpolated in the same way by a
second cubic spline. We only use potential data from Ref.~\cite{Capitani:2018rox}
with separations $\hat r_\text{Ref.~\cite{Capitani:2018rox}} \geq 3$. Since we do not extrapolate, priors are
not available at smaller separations, and in such cases we perform the fits
without Gaussian priors.
We exclude $\hat r_\text{Ref.~\cite{Capitani:2018rox}} = 2$ as a precaution: HYP2-smeared temporal links extend over one lattice spacing in each spatial direction and may, despite tree level improvement, still cause sizable discretization effects at such a small separation. Moreover, the corresponding fits at small separations are stable without priors, and priors do not reduce their statistical errors significantly.

To approximate lattice results for the potential differences in our setup,
i.e.\ at separations, which are integer multiples of one of the lattice
spacings $a$ of the ensembles A, B and C (see Table~\ref{tab:ensembles}),
and at a given flow time $t_f$, we invert the
procedure discussed above. For on-axis separations $r = \hat r a$, where
$\hat r$ is an integer, we define improved separations
$r_\text{improved}^\text{flow}$ analogous to Eq.~(\ref{eq:improved_r_hyp2}),
\begin{eqnarray}
& & \hspace{-0.7cm} \frac{1}{4\pi r_\text{improved}^\text{flow}}
 =
G(r,t_f,a) \, ,
\label{eq:improved_r_flow_new}
\end{eqnarray}
where $G$ is given in Eq.~(\ref{eq:G_lattice}). We then evaluate the cubic
spline at $r_\text{improved}^\text{flow}$ and divide by
$r_\text{improved}^\text{flow}$. Up to corrections beyond tree level, the
result corresponds to the lattice potential difference at separation $r$,
flow time $t_f$ and lattice spacing $a$,
\begin{eqnarray}
& & \hspace{-0.7cm} \Delta V_{n,\Lambda_\eta^\epsilon}(r,t_f,a)
 \approx
\frac{r_\text{improved}^\text{flow} \, \Delta V_{n,\Lambda_\eta^\epsilon,\text{fit}}(r_\text{improved}^\text{flow})}{r_\text{improved}^\text{flow}} \, ,
\label{eq:prior_mean_new}
\end{eqnarray}
where we have not cancelled $r_\text{improved}^\text{flow}$ to make explicit that
the numerator is the cubic spline evaluated at $r_\text{improved}^\text{flow}$.
The right-hand side of Eq.~(\ref{eq:prior_mean_new}) is the mean value $\Delta V_{n,\Lambda_\eta^\epsilon}^\text{prior}(r,t_f,a)$ of the Gaussian prior. Its width $\sigma_{n,\Lambda_\eta^\epsilon}^\text{prior}(r,t_f,a)$ is obtained in the
same way from the second cubic spline, which represents the statistical
uncertainties of the potential differences.


\section{\label{apdx:correlations}Correlations between fit parameters}

In this appendix we provide the statistical correlations between the fit parameters appearing in the parameterizations of the static potential, Eq.\ (\ref{EQN_Cornell}), and of the potential corrections, Eqs.\ (\ref{EQN_V_1_para}), (\ref{EQN_V_p_square_11_para}), (\ref{EQN_V_p_square_20_para}), (\ref{EQN_V_L_square_11_para}), (\ref{EQN_V_L_square_20_para}), (\ref{EQN_V_L2S1_para}), (\ref{EQN_V_LS_para}), (\ref{EQN_V_S12_para}) and (\ref{EQN_V_S_square_para}).
Together with the central values and statistical uncertainties quoted in Tabs.~\ref{tab:v0_results} to \ref{tab:v_sq_results}, these correlations provide the full information on the parameter covariance matrices.
This makes it possible to use our parameterizations in further studies, e.g.\ in Born-Oppenheimer predictions of quarkonium spectra, with the complete statistical information.

We list the correlation coefficients
\begin{eqnarray}
& & \hspace{-0.7cm} \rho_{jk}
 =
\frac{\operatorname{Cov}(p_j,p_k)}
{\sigma_j \sigma_k}
\label{eq:fit_parameter_correlations}
\end{eqnarray}
in Tab.~\ref{tab:parameter_correlations}, where $p_j$ and $p_k$ denote two arbitrary fit parameters and $\sigma_j^2$ and $\sigma_k^2$ their squared statistical uncertainties.
Note that $\sigma_j^2 = \operatorname{Cov}(p_j,p_j)$, which leads to $\rho_{jj} = 1$, i.e.\ to the entries $1$ on the diagonal of the matrix in Tab.~\ref{tab:parameter_correlations}.
Since $\rho_{jk} = \rho_{kj}$, only the lower triangular part is shown.
The correlation coefficients play a similar role as the statistical uncertainties, for which we do not compute errors either. Therefore, we quote them, which by definition satisfy $-1 \leq \rho_{jk} \leq 1$, with two decimal places only.

\begin{table}[h]
\centering
\begin{tabular}{c|c|c|c|c|c|c|c|c|c|c|c|c|c|c}
& $\alpha^{(0)}$ & $\sigma$ & $\alpha^{(1)}$ & $\Lambda^{(1)}$ & $\alpha_{p^2}$ & $\alpha_{L^2}$ & $\Lambda_{L^2}$ & $\alpha_{LS}$ & $\Lambda_{LS}$ & $\alpha_{S_{12}}$ & $\alpha_{S_{12}}^\prime$ & $\alpha_{S^2}$ & $\alpha_{S^2}^\prime$ & $\Lambda_{S^2}$ \\
\hline
$\alpha^{(0)}$ & $1$ & & & & & & & & & & & & & \\
$\sigma$ & $-0.62$ & $1$ & & & & & & & & & & & & \\
$\alpha^{(1)}$ & $-0.04$ & $0.06$ & $1$ & & & & & & & & & & & \\
$\Lambda^{(1)}$ & $\hphantom{-}0.03$ & $\hphantom{-}0.06$ & $-0.94$ & $1$ & & & & & & & & & & \\
$\alpha_{p^2}$ & $\hphantom{-}0.03$ & $-0.10$ & $\hphantom{-}0.03$ & $\hphantom{-}0.05$ & $1$ & & & & & & & & & \\
$\alpha_{L^2}$ & $-0.04$ & $\hphantom{-}0.01$ & $\hphantom{-}0.09$ & $\hphantom{-}0.14$ & $\hphantom{-}0.41$ & $1$ & & & & & & & & \\
$\Lambda_{L^2}$ & $\hphantom{-}0.02$ & $-0.05$ & $-0.20$ & $\hphantom{-}0.28$ & $-0.03$ & $-0.69$ & $1$ & & & & & & & \\
$\alpha_{LS}$ & $\hphantom{-}0.08$ & $-0.06$ & $\hphantom{-}0.00$ & $\hphantom{-}0.00$ & $-0.01$ & $-0.01$ & $\hphantom{-}0.02$ & $1$ & & & & & & \\
$\Lambda_{LS}$ & $\hphantom{-}0.07$ & $-0.09$ & $\hphantom{-}0.00$ & $\hphantom{-}0.00$ & $0.00$ & $-0.01$ & $\hphantom{-}0.02$ & $\hphantom{-}0.86$ & $1$ & & & & & \\
$\alpha_{S_{12}}$ & $-0.29$ & $\hphantom{-}0.09$ & $\hphantom{-}0.01$ & $\hphantom{-}0.00$ & $\hphantom{-}0.00$ & $\hphantom{-}0.00$ & $\hphantom{-}0.02$ & $-0.02$ & $-0.01$ & $1$ & & & & \\
$\alpha_{S_{12}}^\prime$ & $\hphantom{-}0.26$ & $-0.08$ & $-0.01$ & $\hphantom{-}0.00$ & $\hphantom{-}0.00$ & $\hphantom{-}0.00$ & $-0.01$ & $\hphantom{-}0.02$ & $\hphantom{-}0.01$ & $-0.97$ & $1$ & & & \\
$\alpha_{S^2}$ & $\hphantom{-}0.25$ & $-0.09$ & $-0.01$ & $\hphantom{-}0.00$ & $-0.01$ & $-0.01$ & $\hphantom{-}0.00$ & $\hphantom{-}0.02$ & $\hphantom{-}0.00$ & $\hphantom{-}0.15$ & $\hphantom{-}0.05$ & $1$ & & \\
$\alpha_{S^2}^\prime$ & $-0.15$ & $\hphantom{-}0.06$ & $\hphantom{-}0.00$ & $\hphantom{-}0.00$ & $\hphantom{-}0.01$ & $\hphantom{-}0.02$ & $\hphantom{-}0.00$ & $-0.01$ & $\hphantom{-}0.00$ & $\hphantom{-}0.06$ & $\hphantom{-}0.04$ & $-0.96$ & $1$ & \\
$\Lambda_{S^2}$ & $\hphantom{-}0.09$ & $-0.03$ & $\hphantom{-}0.00$ & $\hphantom{-}0.00$ & $\hphantom{-}0.01$ & $\hphantom{-}0.00$ & $\hphantom{-}0.00$ & $\hphantom{-}0.01$ & $\hphantom{-}0.01$ & $-0.58$ & $\hphantom{-}0.71$ & $-0.33$ & $\hphantom{-}0.41$ & $1$
\end{tabular}
\caption{
{Correlation coefficients $\rho_{jk}$ as defined in Eq.~(\ref{eq:fit_parameter_correlations}) (we omit the argument $\mu$ of the spin dependent parameters).}
}
\label{tab:parameter_correlations}
\end{table}


\FloatBarrier

\clearpage

\section*{Acknowledgements}

We are particularly grateful to Nora Brambilla for her exceptional effort in carefully reading the nearly final manuscript and for her many helpful comments, which significantly improved this paper.
We also acknowledge useful discussions with Vilija de Jonge, Viljami Leino, Julian Mayer-Steudte, Paul P\"utz, Carolin Schlosser, Antonio Vairo and Xiang-Peng Wang.

The authors gratefully acknowledge the computing time provided to them at the NHR Center NHR@SW at Goethe University Frankfurt (project number 26459).
This is funded by the Federal Ministry of Education and Research, and the state governments participating on the basis of the resolutions of the GWK for national high performance computing at universities (\href{https://www.nhr-verein.de/unsere-partner}{www.nhr-verein.de/unsere-partner}).


\FloatBarrier


\bibliographystyle{utphys.bst}
\providecommand{\href}[2]{#2}\begingroup\raggedright\endgroup


\end{document}